\RequirePackage{fix-cm}
\documentclass[twocolumn]{svjour3}          %
\smartqed  %
\usepackage{cite}
\usepackage{multirow}
\usepackage{graphicx}
\graphicspath{{./}{./figures/}}
\usepackage{amsmath}
\usepackage{array}
\usepackage{xspace}
\usepackage{xcolor}
\usepackage{tablefootnote}

\begin{document}

\title{Dense Reconstruction of Transparent Objects by Altering Incident Light Paths Through Refraction
}

\author{Kai Han         \and
        Kwan-Yee K. Wong \and
        Miaomiao Liu
}

\institute{Kai Han \at
           The University of Hong Kong, Hong Kong, China \\
              \email{khan@cs.hku.hk}           %
           \and
           Kwan-Yee K. Wong \at
           The University of Hong Kong, Hong Kong, China
           \and
           Miaomiao Liu \at
           Data61, CSIRO and CECS, Australian National University, Canberra, Australia
}

\date{Received: date / Accepted: date}

\maketitle

\begin{abstract}
This paper addresses the problem of reconstructing the surface shape of transparent objects. The difficulty of this problem originates from the viewpoint dependent appearance of a transparent object, which quickly makes reconstruction methods tailored for diffuse surfaces fail disgracefully. In this paper, we introduce a fixed viewpoint approach to dense surface reconstruction of transparent objects based on refraction of light. We present a simple setup that allows us to alter the incident light paths before light rays enter the object by immersing the object partially in a liquid, and develop a method for recovering the object surface through reconstructing and triangulating such incident light paths. Our proposed approach does not need to model the complex interactions of light as it travels through the object, neither does it assume any parametric form for the  object shape nor the exact number of refractions and reflections taken place along the light paths. It can therefore handle transparent objects with a relatively complex shape and structure, with unknown and inhomogeneous refractive index. We also show that for thin transparent objects, our proposed acquisition setup can be further simplified by adopting a single refraction approximation. Experimental results on both synthetic and real data demonstrate the feasibility and accuracy of our proposed approach.
\end{abstract}

\section{Introduction}
\label{sec:intro}
Reconstructing a 3D model of an object from its 2D images has always been a hot topic in the field of computer vision. It has many important applications in robotics, augmented reality, video games, movie production, reverse engineering, etc. Despite the problem of 3D model reconstruction has virtually been solved for opaque objects with a diffuse surface, the literature is relatively sparse when it comes to shape recovery of transparent objects. It is still very challenging and remains an open problem. The viewpoint dependent appearance of a transparent object quickly renders reconstruction methods tailored for diffuse surfaces useless, and most of the existing methods for transparent object reconstruction are still highly theoretical. In fact, even with restrictive assumptions and special hardware setups, state-of-the-art methods can only handle transparent objects with a very simple shape. Meanwhile, it is not difficult to see that there exist many transparent objects in our world (e.g., glasses, plastics, crystals and diamonds). Hence, the study of 3D model reconstruction cannot be considered completed without taking transparent objects into account.

\begin{figure}
\begin{center}
\includegraphics[width=0.9\linewidth]{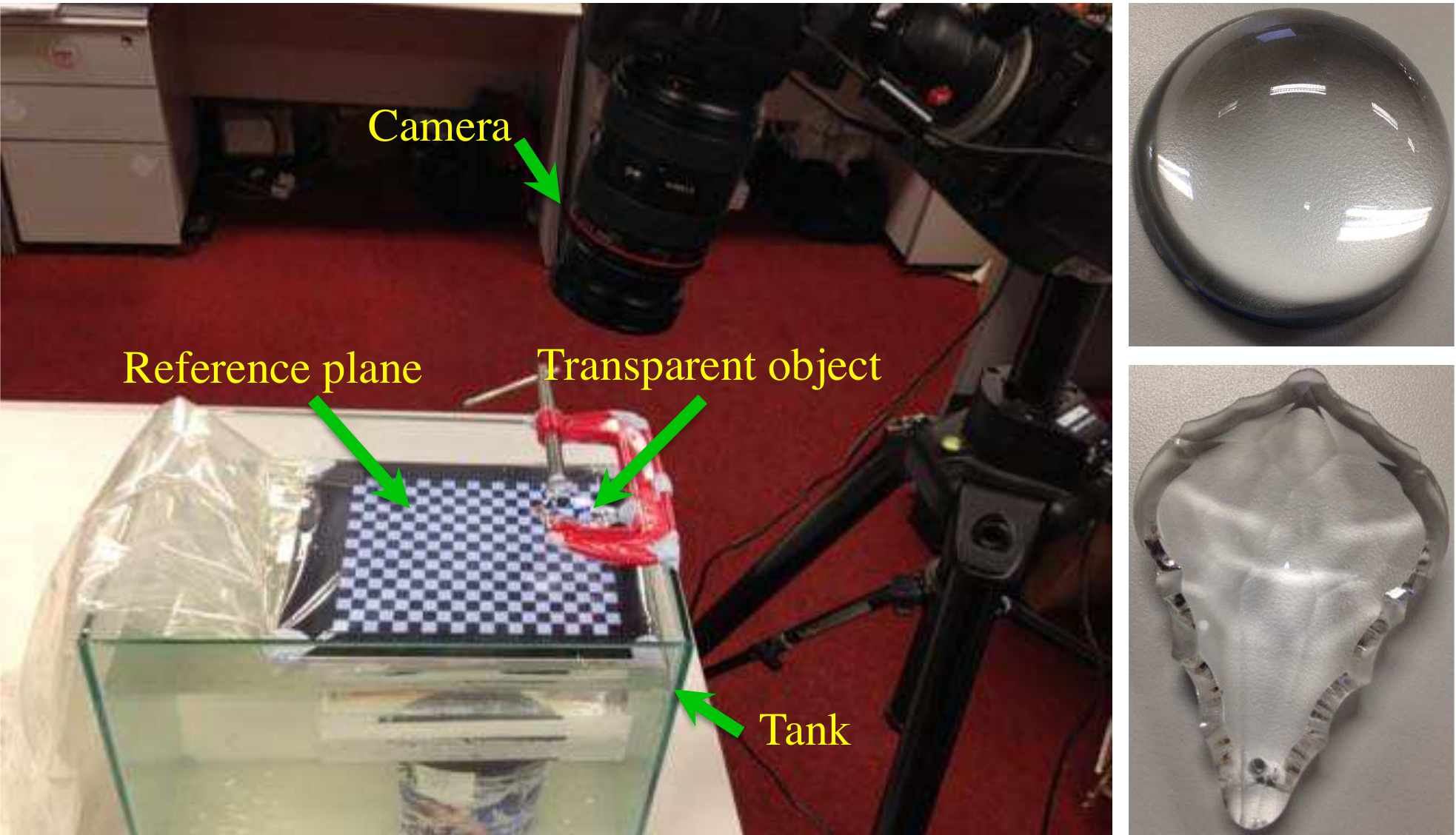}
\end{center}
   \caption{Real reconstruction setup and examples of transparent objects.}
\label{fig:setup}
\end{figure}
As mentioned previously, the difficulty of reconstructing a transparent object originates from its viewpoint dependent appearance. A transparent object may alter a light path by reflection, refraction, absorption and scattering at both its exterior surface as well as its interior structure. A number of existing work attempted to reconstruct a transparent object by exploiting specular highlights produced on the object surface \cite{Ihrke2008eurographics, Ihrke2010cgf}. This approach considers only reflection of light taken place at the object surface, and greatly simplifies the problem by making it not necessary to consider the complex interactions of light as it travels through the object. However, refraction of light is indeed an important and unique characteristic of transparent objects. It provides information on surface shape and should not be ignored. On the other hand, methods based on reflection of light often work only under very restrictive assumptions and precisely controlled environments, making them not very practical. 

In this paper, we focus our study in dense surface reconstruction of transparent objects. We introduce a fixed viewpoint approach to recovering the surface of a transparent object based on refraction of light. Like those methods that are based on specular highlights, our fixed viewpoint approach does not need to explicitly model the complex interactions of light as it travels through the object. We present a simple setup (see Fig.~\ref{fig:setup}) that allows us to alter the incident light paths before light rays enter the object by immersing the object partially in a liquid, and develop a method for recovering the surface of a transparent object through reconstructing and triangulating these incident light paths. We also show that for thin transparent objects, the acquisition setup can be further simplified by adopting a single refraction approximation. Compared with existing methods, our proposed method has the following benefits:

\begin{itemize}
\item[$\bullet$] It does not assume any parametric form for the shape of a transparent object.
\item[$\bullet$] It can handle a transparent object with a complex structure, with an unknown and inhomogeneous refractive index.
\item[$\bullet$] It considers only the incident light paths before light rays enter a transparent object, and makes no assumption on the exact number of refractions and reflections taken place as light travels through the object. 
\item[$\bullet$] The proposed setup is simple and inexpensive.
\end{itemize}

A preliminary version of this work appeared in \cite{Han2015cvpr}. This presentation extends \cite{Han2015cvpr} as follows: 1) We further extend our theory to the reconstruction of thin transparent objects. In particular, we adopt the single refraction approximation \cite{Wetzstein2011iccv}. It leads to a simplified setup and robust reconstruction. 2) We provide new experiments on both synthetic and real data to evaluate our method, including an experiment of reconstructing a real transparent hollow object. 3) We study the cases where total internal reflection happens inside the transparent object, and provide more comprehensive discussions on the shape of transparent objects that can be handled by our method.

The rest of the paper is organized as follows. Section~\ref{sec:related} briefly reviews existing techniques in the literature for shape recovery of transparent objects. Section~\ref{sec:method} describes our proposed approach to dense surface reconstruction of transparent objects in detail. Section~\ref{sec:thin_method} introduces our simplified approach to thin transparent object reconstruction. Section~\ref{sec:discussions} discusses the problem of total internal reflection and objects that are suitable for our approach. Experimental results on both synthetic and real data are presented in Section~\ref{sec:experiments}, followed by conclusions in Section~\ref{sec:conclusion}.

\section{Related Work}
\label{sec:related}
Great efforts have been devoted to the problem of transparent object reconstruction in the past two decades. To formulate this problem, existing methods often make assumptions such as orthographic projection~\cite{Murase1990iccv, Murase1992pami,QiShan2012cvpr}, $C^n$ continuity of the surface~\cite{QiShan2012cvpr}, known exact number of refractions along each light path~\cite{Kutulakos2005iccv, Kutulakos2008ijcv,Tsai2015icip}, etc. %
In \cite{Murase1990iccv, Murase1992pami}, Murase reconstructed a rippling water surface from the average observed coordinates of an underwater pattern under orthographic projection. Morris and Kutulakos \cite{Morris2011pami} solved a similar problem with an unknown refractive index of the liquid using two calibrated cameras and a known reference pattern. In \cite{QiShan2012cvpr}, Shan {\em et al.} introduced a framework for optimizing a refractive height field from a single image under the assumptions of an orthographic camera, known background, single refractive material and differentiable height field. In \cite{Hata1996icpr}, Hata {\em et al.}~used structured light and genetic algorithm to estimate the shape of a transparent paste drop on a board. Ben-Ezra and Nayar \cite{Ben-Ezra2003iccv} assumed a parametric form for the shape of a transparent object and estimated the shape parameters under the assumptions of a known camera motion and a distant background. In \cite{Kutulakos2005iccv, Kutulakos2008ijcv}, Kutulakos and Steger categorized reconstructible specular scenes, and developed algorithms for depth map computation in the cases where refraction/reflection of light occurs exactly once and twice respectively. Following the same fashion, Tsai {\em et al.}~\cite{Tsai2015icip} demonstrated two depth-normal ambiguities for transparent object recovery assuming the light path refracts exactly twice. In \cite{zuo2015iccv}, Zuo {\em et al.}~developed an interactive specular and transparent object reconstruction system based on visual hull refinement given the silhouettes under multiple views and labeled contours of the object in sparse key frames. In \cite{Qian2016cvpr}, Qian {\em et al.}~introduced a method to recover transparent objects by solving an optimization function with a position-normal consistency constraint, under the assumption of two refractions along each light path. Their system consists of two cameras and one display serving as a light source for correspondence estimation.

Many hardware setups have also been designed to recover the surfaces of transparent objects.
In \cite{Wetzstein2011iccv}, Wetzstein {\em et al.}~proposed a single image approach to reconstructing thin refractive surfaces using light field probes. In \cite{ding11iccv}, Ding {\em et al.}~introduced a $3 \times 3$ camera array to acquire correspondences for fluid surface recovery. In \cite{Eren2009optics_express}, Eren {\em et al.}~determined the surface shape of a glass object using laser surface heating and thermal imaging. In \cite{Ihrke2005iccv}, Ihrke {\em et al.} dyed water with a fluorescent chemical and presented a level set method for reconstructing a free flowing water surface from multi-video input data by minimizing a photo-consistency error computed using raytracing. Miyazaki and Ikeuchi \cite{Miyazaki2005cvpr} proposed an iterative method to estimate the front surface shape of a transparent object by minimizing the difference between observed polarization data and polarization raytracing result under the assumptions of a known refractive index, a known illumination distribution and a known back surface shape. In \cite{Trifonov2006siggraphsketches}, Trifonov {\em et al.}~introduced a visible light tomographic reconstruction method by immersing a transparent object into a fluid with a similar refractive index. The 3D shape was recovered by building the light paths within the fluid and the object. In \cite{Hullin2008SIGGRAPH}, Hullin {\em et al.}~embedded a transparent object into fluorescence and reconstructed the object surface by detecting the intersections of the visible laser sheets with the visual rays. A similar light sheet range scanning approach was introduced by Narasimhan {\em et al.} in \cite{Narasimhan2005iccv} for acquiring object geometry in the presence of a scattering medium. In \cite{Kutulakos2014cvpr}, O'Toole {\em et al.}~developed the structured light transport (SLT) technique. Based on SLT, they implemented an imaging device that allows one-shot indirect-invariant imaging for reconstructing transparent and mirror surfaces using structured light. In \cite{Wetzstein2014cvpr}, Ma {\em et al.} reformulated the intensity transport equation in terms of light fields, and presented a technique for refractive index field reconstruction using coded illumination. In \cite{Yuji2013cvpr}, Ji {\em et al.}~estimated the refractive index field of a gas volume by establishing ray-to-ray correspondences using a light field probe, and reconstructed the light paths through the refractive index field using a variational method based on Fermat's Principle.

Like specular surfaces, transparent objects also exhibit reflection properties. Hence, reflection correspondences designed for specular surface reconstruction (e.g., \cite{Balzer2010measurement, Reshetouski2013}) can also be adopted for the reconstruction of transparent objects. 
In \cite{Morris2007iccv}, Morris and Kutulakos introduced {\em scatter-trace} of a pixel and recovered the exterior surface of a transparent object using the non-negligible specular reflection component. Similarly, Yeung {\em et al.} \cite{Yeung2011cvpr} exploited specular highlights and proposed a dual-layered graph-cut method to reconstruct the surface of a solid transparent object. In \cite{Sturm2013cvpr}, Chari and Sturm introduced a method that integrates radiometric information into light path triangulation for reconstruction of transparent objects from a single image. In \cite{Yeung2014cvpr}, Liu {\em et al.}~proposed a frequency based method for establishing correspondences on transparent and mirror surfaces, and reconstruction can then be done using any stereo methods. 

Note that existing solutions for surface reconstruction of transparent objects often work only under restrictive assumptions (e.g., known refractive index, single refractive material, known exact number of refractions, non-negligible reflection of light, orthographic projection), using special hardware setups (e.g., light field probes, laser surface heating with thermal imaging, dying liquids with fluorescent chemical, immersing objects into liquids with similar refractive indexes), or for a particular class of objects (e.g., with known parametric model/average shape). There exists no general solution to this challenging and open problem. 

\begin{figure*}[htbp]
\tabcolsep=0.05\linewidth
  \centering
  \begin{tabular}{cc}
    \includegraphics[width=0.25\linewidth]{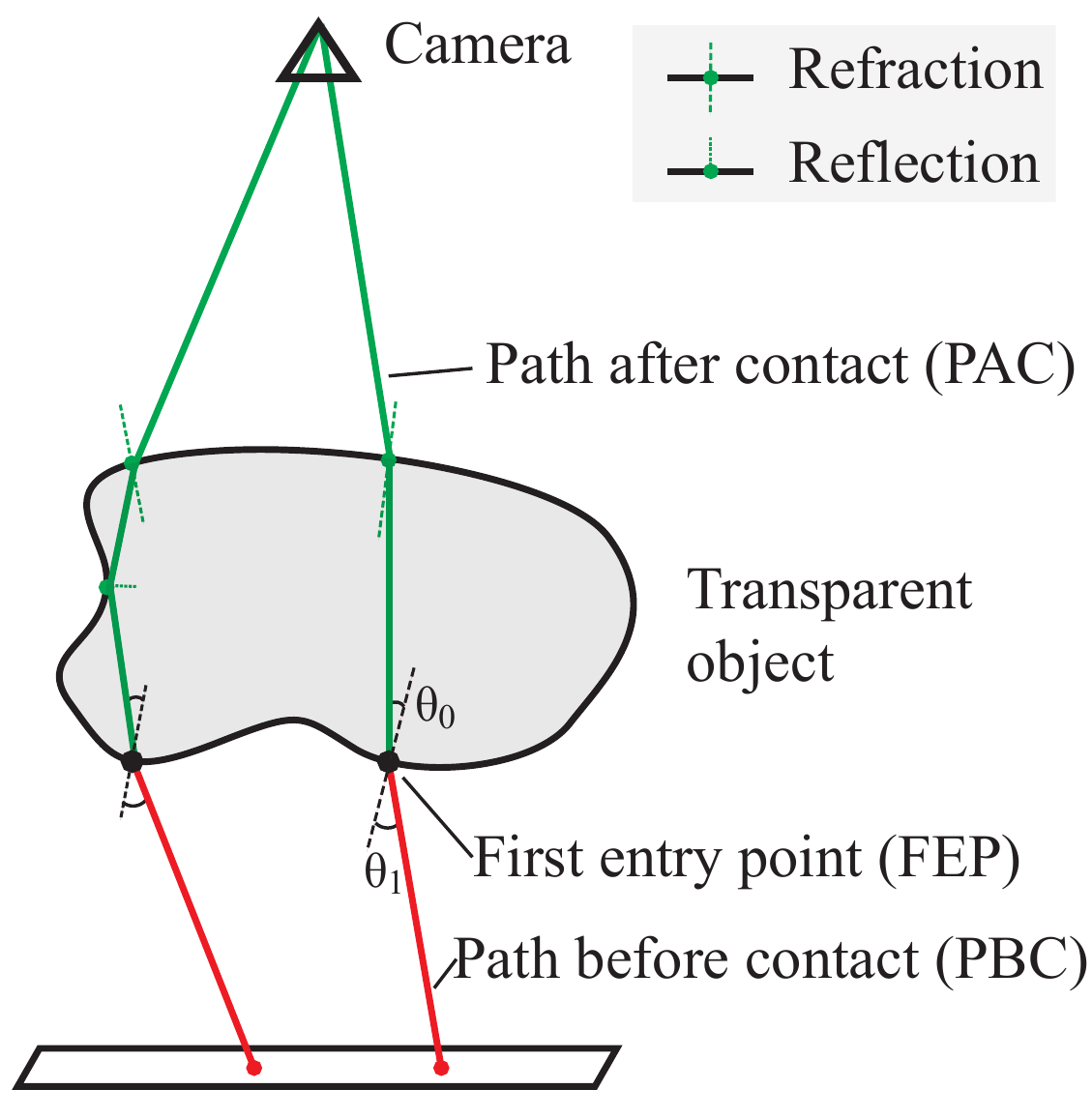} &    \includegraphics[width=0.32\linewidth]{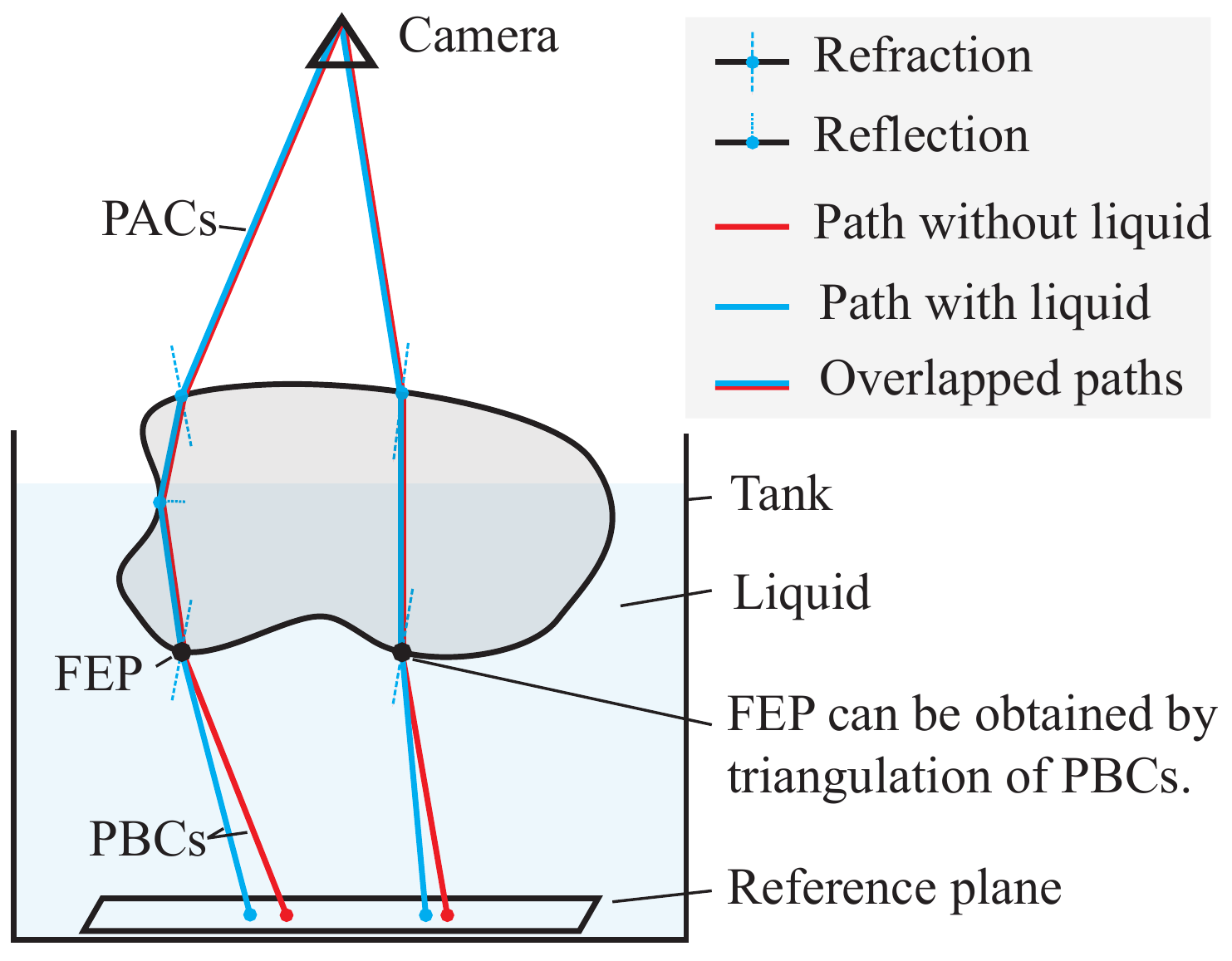}\\
    \small (a)\label{fig:fig_lightpath:a} %
     &
    \small (b)
    \label{fig:fig_lightpath:b} %
  \end{tabular}
 \caption{(a) A light path through an object is partitioned into two parts, namely i) the {\em path before contact} (PBC) which originates from the reference pattern to the {\em first entry point} (FEP) on the object surface (i.e., the red paths) and ii) the {\em path after contact} (PAC) that originates from FEP, passes through the interior of the object and terminates at the optical center of the camera (i.e., the green paths). (b) The PBC can be altered by filling the tank with a liquid, and the FEP can be recovered by triangulating two PBCs.}
\label{fig:fig_lightpath}
\end{figure*}

In this paper, we develop a fixed viewpoint approach to dense surface reconstruction of transparent objects based on altering and triangulating the incident light paths before light rays enter the object. We present a simple setup that allows us to alter the incident light paths by means of refraction of light. Under this proposed setup, the segment of a light path between the first entry point on the object surface and the optical center of the camera remains fixed. This allows us to ignore the details of the complex interactions of light inside the object. Compared with existing methods, our proposed approach (1) assumes neither a known nor homogeneous refractive index of the object; (2) places no restriction on the exact number of refractions and reflections taken place along a light path; and (3) assumes no parametric form for the object shape. This allows our approach to handle transparent objects with a relatively complex structure.

For thin transparent objects, we show that our acquisition setup can be further simplified by adopting a single refraction approximation. Such an approximation has been used by existing methods for recovering liquid surfaces (e.g., \cite{Morris2011pami, ding11iccv}), one-side flatten (e.g., \cite{QiShan2012cvpr}) or thin transparent surfaces (e.g., \cite{Wetzstein2011iccv}). The altering of the incident light paths can be achieved by the object itself without using any extra medium, and the surface can be recovered using the same formulation as the general approach.

\section{Shape Recovery of Transparent Objects}
\label{sec:method}
\subsection{Notations and Problem Formulation}
To solve the surface reconstruction problem, we consider a set of light paths originating from a reference pattern placed behind a transparent object, passing through the object and eventually reaching the image plane. We partition every such light path into two parts, namely (i) the {\em path before contact} (PBC) which originates from the reference pattern and ends at the {\em first entry point} (FEP) on the object surface (see the red paths in Fig.~\ref{fig:fig_lightpath:a}(a)) and (ii) the {\em path after contact} (PAC) which originates from the FEP, passes through the interior of the object and finally terminates at the optical center of the camera (see the green paths in Fig.~\ref{fig:fig_lightpath:a}(a)). We can now reformulate the surface reconstruction problem into estimating the FEP. The approach we take to tackle this problem is by altering the PBC while fixing the PAC for each light path. This enables us to ignore the details of the complex interactions of light inside the object, and recover the FEP by triangulating the PBCs. In the next section, we present a simple setup that allows us to alter the PBCs by means of refraction of light.

\subsection{Setup and Assumptions}
In our proposed setup, a camera is used to capture images of a transparent object in front of a reference pattern. The camera and the object are kept fixed with respect to each other to ensure the PACs remain unchanged for all the image points. The reference pattern is placed at two distinct positions and is used for reconstructing the PBCs. As mentioned before, our approach is based on altering and triangulating the PBCs. To achieve this, we employ a water tank and immerse the object partially into a liquid so as to alter the PBCs by means of refraction of light (see Fig.~\ref{fig:fig_lightpath:b}(b)). Two images of the transparent object are acquired for each position of the reference pattern, one without liquid in the tank and one with liquid in the tank. By calibrating the positions of the reference pattern and establishing correspondences between image points and points on the reference pattern, we can reconstruct two PBCs for each image point, one in air and one in the liquid, respectively. The FEP can then be recovered by triangulating these two PBCs.

Note that our proposed approach does not require the prior knowledge of the refractive index of the object or that of the liquid. If, however, the refractive index of the liquid is known a priori, it is possible to also recover the surface normal at each FEP. The only assumption made in our approach is that the PACs remain unchanged when the object is immersed partially into the liquid. 

\begin{figure*}[htbp]
\centering
\tabcolsep=0.15cm
\begin{tabular}{
>{\centering\arraybackslash} m{0.16\textwidth}
>{\centering\arraybackslash} m{0.16\textwidth}
>{\centering\arraybackslash} m{0.16\textwidth}
>{\centering\arraybackslash} m{0.16\textwidth}}
\includegraphics[width=\linewidth]{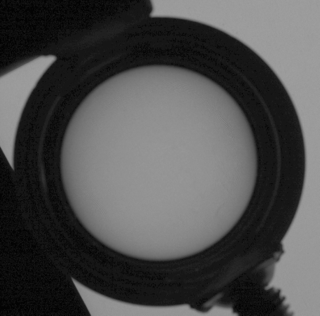} &
\includegraphics[width=\linewidth]{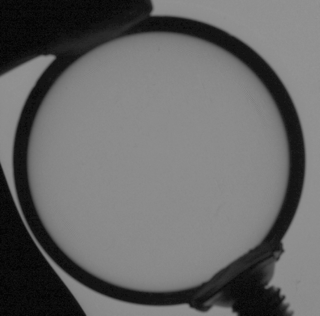}&
\includegraphics[width=\linewidth]{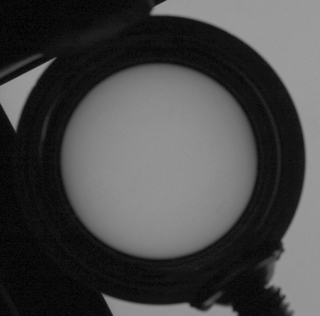}&
\includegraphics[width=\linewidth]{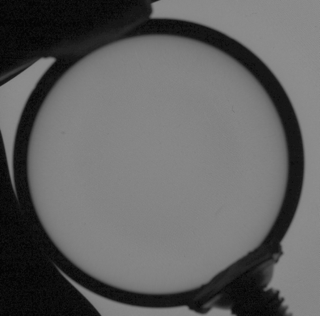}\\
\includegraphics[width=\linewidth]{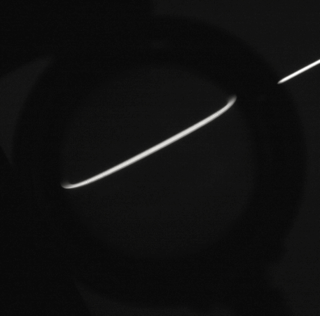} &
\includegraphics[width=\linewidth]{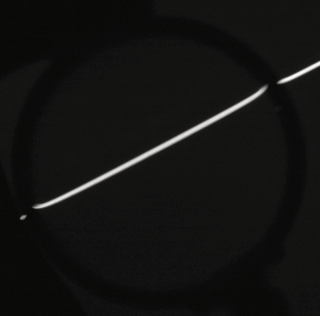} &
\includegraphics[width=\linewidth]{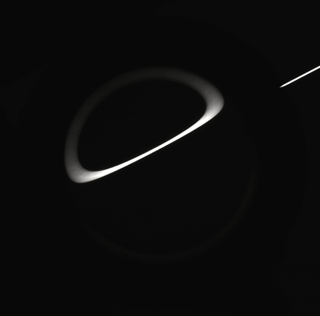} &
\includegraphics[width=\linewidth]{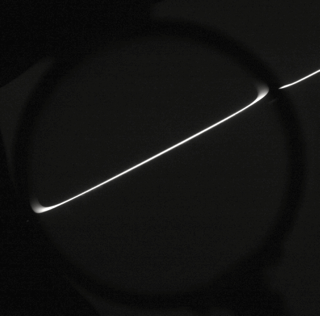}\\

\end{tabular}
\caption{The upper row shows images of a transparent hemisphere captured in front of a gray background (from left to right: reference pattern at a high position and without water, reference pattern at a high position and with water, reference pattern at a low position and without water, and reference pattern at a low position and with water). The lower row shows images of the hemisphere captured in front of a sweeping stripe (in the same order).} 
\label{fig:real_hemi_photo}
\end{figure*}

\subsection{Dense Refraction Correspondences} \label{sec:correspondences}
Before we can triangulate PBCs to recover the FEP, we first need to reconstruct the PBCs from the images. To achieve this, we first calibrate the two distinct positions of the reference pattern using \cite{MatlabCalibrationToolBox}. It is then straightforward to reconstruct the PBC for an image point by locating a correspondence point on the reference pattern under each of the two distinct positions in the same medium (i.e., with/without liquid in the tank). It is obvious that the quality of the correspondences will have a direct effect on the quality of the reconstruction. There exist many methods for establishing correspondences \cite{Batlle1998pr}, such as Gray Code \cite{Inokuchi1984icpr}, Phase Shift \cite{Wust1991MVA}, etc. However, these methods often can only provide sparse correspondences with limited precisions (e.g., a small patch of pixels is mapped to a small region on a reference plane due to finite discretization). In this work, we would like to establish quasi-point-to-point correspondences between the image and the reference pattern. We employ a portable display screen (e.g., an iPad) to serve as the reference pattern, and show a sequence of a thin stripe sweeping across the screen in vertical direction and then in horizontal direction \cite{Kutulakos2005iccv, Kutulakos2008ijcv}. We capture an image for each of the positions of the sweeping stripe (see Fig.~\ref{fig:real_hemi_photo}). For each image point, its correspondence on the reference pattern can then be solved by examining the sequence of intensity values of the image point for each sweeping direction and locating the peak intensity value. The position of the stripe that produces the peak intensity value in each sweeping direction then gives us the position of the correspondence on the reference pattern. In order to improve the accuracy of the peak localization, we fit a quadratic curve to the intensity profile in the neighborhood of the sampled peak value, and solve for the exact peak analytically.

\subsection{Light Path Triangulation} \label{sec:lightpathtriangulation}
Suppose high quality correspondences have been established between the images and the reference pattern under each of the two distinct positions and in each of the two media (i.e., with and without liquid in the tank). We can reconstruct two PBCs for each image point using the calibrated positions of the reference pattern. The FEP can then be recovered as the point of intersection between the two PBCs. Below we derive a simple solution for the FEP based on the established correspondences of an image point. 

\begin{figure}[htbp]
\begin{center}
\includegraphics[width=0.6\linewidth]{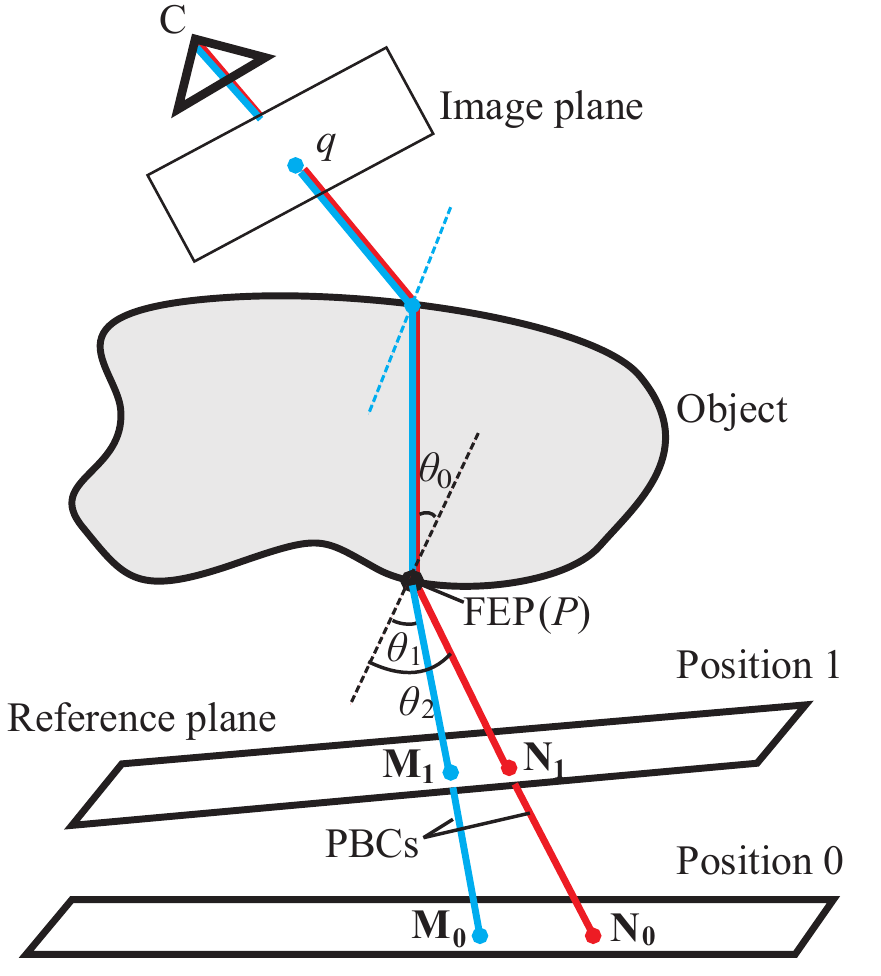}
\end{center}
 \caption{PBC reconstruction and FEP estimation. The correspondences of an image point $q$ on the reference pattern under position 0 and position 1 define a PBC. Given two PBCs in two different media, the FEP for $q$ can be obtained by triangulating the PBCs.}
\label{fig:fig_triangulation}
\end{figure}

Consider an image point $q$ (see Fig.~\ref{fig:fig_triangulation}). Suppose ${\bf M}_0$ and ${\bf M}_1$ denote, respectively, its correspondences on the reference pattern under position $0$ and position $1$ with liquid in the tank. Similarly, let ${\bf N}_0$ and ${\bf N}_1$ denote, respectively, its correspondences on the reference pattern under position $0$ and position $1$ without liquid in the tank. The PBCs for $q$ in liquid and in air can be expressed as
\begin{eqnarray}
L_M:&{\bf M}(s) = {\bf M}_0 + s{\bf U},\label{eq:M(s)}\\
L_N:&{\bf N}(t) = {\bf N}_0 + t{\bf V}, \label{eq:N(t)}
\end{eqnarray}
where ${\bf U} = \frac{{\bf M}_1 - {\bf M}_0}{\|{\bf M}_1 - {\bf M}_0\|}$ and ${\bf V} = \frac{{\bf N}_1 - {\bf N}_0}{\|{\bf N}_1 - {\bf N}_0\|}$. Under a perfect situation, the FEP for $q$ is given by the point of intersection between $L_M$ and $L_N$.

Due to noise, however, $L_M$ and $L_N$ often may not intersect with each other exactly at a point. In this situation, we seek the point ${\bf M}_c = {\bf M}(s_c)$ on $L_M$ and the point ${\bf N}_c = {\bf N}(t_c)$ on $L_N$ such that the Euclidean distance between ${\bf M}_c$ and ${\bf N}_c$ is a minimum. The distance between ${\bf M}_c$ and ${\bf N}_c$ can be taken as a quality measure of the reconstruction. If the distance is below a specified threshold, the mid-point between ${\bf M}_c$ and ${\bf N}_c$ can be taken as the FEP for $q$. Note that if ${\bf U}$ and ${\bf V}$ are parallel, there will not be a unique solution. This corresponds to the case where the two PBCs overlap with each other. This is a degenerate case which happens only when the incident ray is parallel to the surface normal.

\subsection{Surface Normal Reconstruction} \label{sec:surefacenormalreconstruction}
Recall that for the purpose of surface reconstruction, neither the refractive index of the object nor that of the liquid is needed. If, however, the refractive index of the liquid is known a priori, it is possible to recover the surface normal at each FEP (see Fig.~\ref{fig:fig_triangulation}). Let $\theta_1$ and $\theta_2$ denote the incident angles of the PBCs in the liquid and air, respectively, at the surface point $P$, and $\theta_0$ denote the refracted angle. Suppose the refractive index of the object, liquid and air are given by $\lambda_0$, $\lambda_1$ and $\lambda_2$, respectively. By Snell's Law, we have
\begin{equation}
\lambda_0\sin\theta_0 = \lambda_1\sin\theta_1 = \lambda_2\sin\theta_2. \label{eq:Snell}
\end{equation}  
Let $\Delta\theta = \cos^{-1}({\bf U}\cdot{\bf V})$ denote the angle between the two PBCs. Substituting this into (\ref{eq:Snell}) gives
\begin{equation}
\lambda_1\sin\theta_1 = \lambda_2\sin(\theta_1+\Delta\theta). \label{eq:DeltaTheta}
\end{equation}
With known refractive indices $\lambda_1$ and $\lambda_2$ for the liquid and air, respectively, the incident angle $\theta_1$ can be recovered by
\begin{equation}
\theta_1 = \tan^{-1}\left(\frac{\lambda_2\sin\Delta\theta}{\lambda_1-\lambda_2\cos\Delta\theta}\right). \label{eq:theta_1}
\end{equation}
The surface normal ${\bf n}_p$ at $P$ is then given by
\begin{equation}
{\bf n}_p = {\bf R}(\theta_1, {\bf V}\times{\bf U}){\bf U}, \label{eq:normal}
\end{equation}
where ${\bf R}(\theta, {\bf a})$ denotes a Rodrigues rotation matrix for rotating about the axis ${\bf a}$ by the angle ${\theta}$.

\section{Recovery of Thin Transparent Objects}
\label{sec:thin_method}
The method proposed in Section~\ref{sec:method} requires immersing an object partially into a liquid. However, this is not an easy task for flat thin transparent objects (e.g., glass plates, thin lens). In this section, we show that for thin transparent objects, the requirement of immersing the object partially into a liquid can be removed by a single refraction approximation, resulting in a simplified setup.

\subsection{Setup and Assumptions}
\begin{figure}[htbp]
\begin{center}
\includegraphics[width=0.6\linewidth]{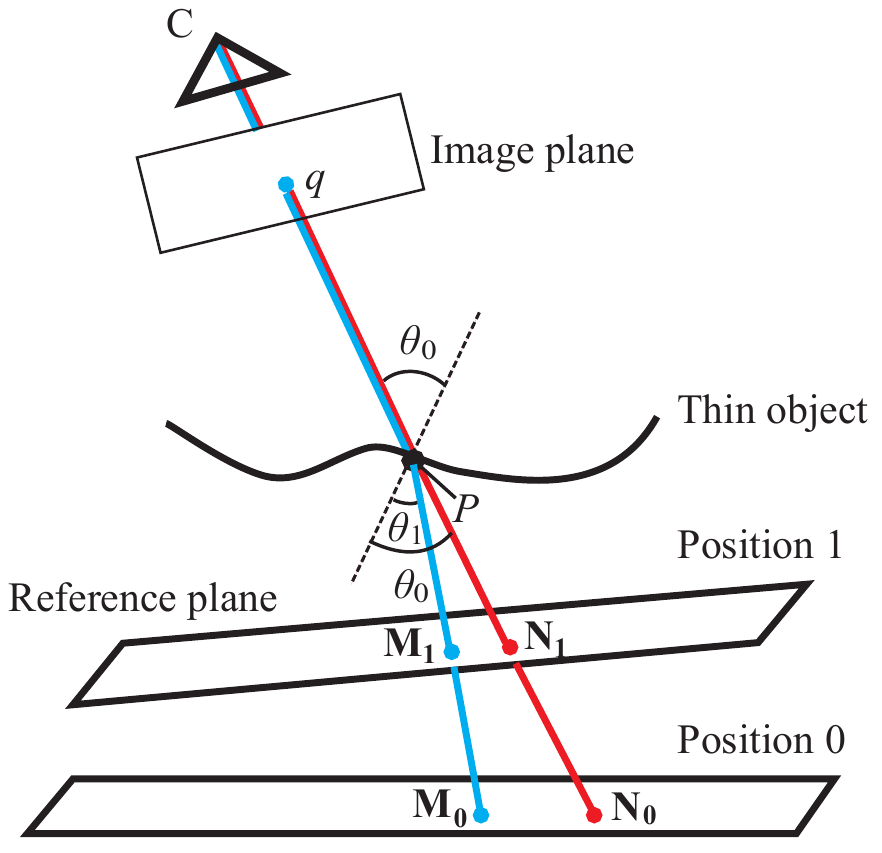}
\end{center}
 \caption{PBC and visual ray construction and surface estimation for thin objects. The correspondences of an image point $q$ on the reference pattern under position 0 and position 1 define a visual ray (red) and a PBC (blue). The visual ray and PBC are constructed from the direct view of the pattern and the refraction of the pattern caused by the object, respectively. Given these two rays, the surface point $P$ for $q$ can be obtained by ray triangulation, meanwhile the normal for $P$ can also be recovered using method described in Section ~\ref{sec:surefacenormalreconstruction}.}
\label{fig:fig_triangulation_thin}
\end{figure}

We follow the same notations used in Section~\ref{sec:method}. As discussed in \cite{Wetzstein2011iccv}, it is generally true for thin transparent objects to assume only one refraction occurs along each light path passing through the object. In general, a light path originates from the reference plane will be refracted (at least) twice at the surface of the object before it reaches the camera. However, if the object is very thin, the light path segment inside the object becomes negligible. In this case, we can assume only one single refraction along each light path. To reconstruct a PBC and a visual ray for each image point, the reference plane is placed at two distinct positions. For each position, two images of the reference pattern are captured, one with the object between the camera and the reference plane and the other without the object. After calibrating the positions of the reference plane and establishing correspondences between the image and the reference plane, we can reconstruct a PBC and a visual ray for each surface point (see Fig.~\ref{fig:fig_triangulation_thin}). The visual ray can be reconstructed from the direct view of the pattern (red path in Fig.~\ref{fig:fig_triangulation_thin})\footnote{If the camera is calibrated w.r.t the reference plane, it is straightforward to recover the visual ray of an image point, and two images are sufficient to construct the blue PBC. By using four images as described in the main text, the PBC and visual ray can be constructed even without calibrating the camera. We only need to calibrate the pattern poses, which is also required by the two-image method.}, and the PBC can be reconstructed from the refraction of the pattern caused by the thin surface (blue path in Fig.~\ref{fig:fig_triangulation_thin}).

\subsection{Surface Reconstruction}
Assuming a single refraction occurs along each light path passing through a thin transparent object, the FEP can be recovered by triangulating the visual ray and the PBC of each image point. Compared with the general setup discussed in Section \ref{sec:method}, the requirement of immersing the object partially into the liquid to alter the incident rays is removed along with the need for a water tank. However, the baseline between these two rays is quite narrow for a thin surface. It leads to noisy FEP cloud estimation. With a known refractive index of the object, the surface normal can be recovered using the method introduced in Section \ref{sec:method}. We therefore reconstruct the surface by integrating surface normals estimated from these rays, which proves to be more robust to noise.

\section{Discussions}
\label{sec:discussions}

\subsection{Total internal reflection}
It is well known that total internal reflection will occur if a light ray propagates from one medium with a larger refractive index to another medium with a smaller refractive index (e.g., from glass to air), but not for the opposite propagation direction (e.g., from air to glass). In the scenario of transparent surface reconstruction, as the refractive index of a solid object is generally larger than that of its surrounding environment (either air or liquid), total internal reflection will inevitably happen. Here, we discuss the potential total internal reflection situation when adopting our approach.

\begin{figure}[tbp]
\begin{center}
\includegraphics[width=0.6\linewidth]{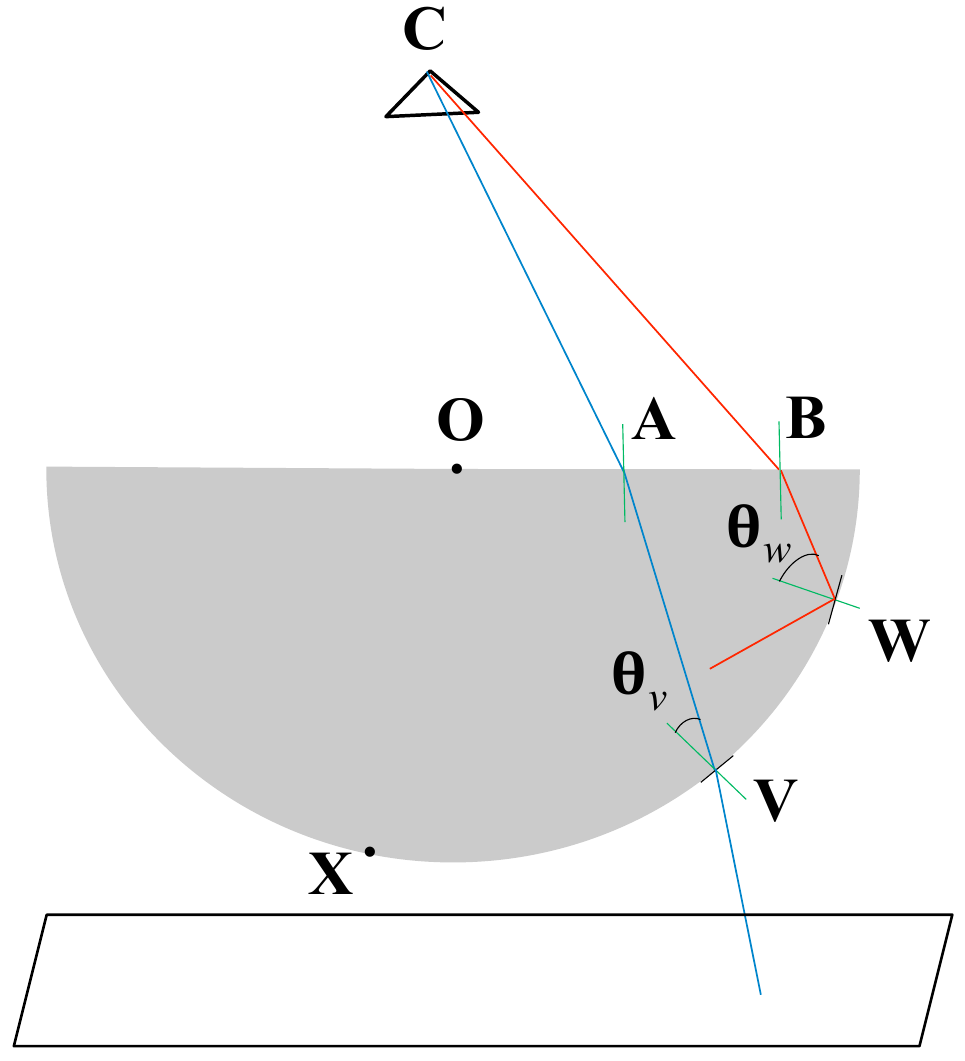}
\end{center}
   \caption{A total internal reflection example. A camera centered at $\mathbf{C}$ observes a transparent hemisphere centered at $\mathbf{O}$ and a reference plane is placed below the object. The blue light path traveling along $\mathbf{C}$, $\mathbf{A}$ and $\mathbf{V}$ has two refractions at $\mathbf{A}$ and $\mathbf{V}$ respectively. While the red light path traveling along $\mathbf{C}$, $\mathbf{B}$ and $\mathbf{W}$ first refracts at $\mathbf{B}$ and has a total internal reflection at $\mathbf{W}$. }
\label{fig:total_reflection}
\end{figure}

Consider a light path traveling from one medium with a refractive index $\lambda_1$ to another medium with a refractive index $\lambda_2$, where $\lambda_1 > \lambda_2$. Total internal reflection only happens when the incident angle is greater than the critical angle $\theta_c = \sin^{-1}(\frac{\lambda_2}{\lambda_1})$. Fig.~\ref{fig:total_reflection} depicts an example of total internal reflection, where $\theta_v < \theta_c$ and $\theta_w > \theta_c$. When total internal reflection happens, the light path may not reach the pattern and the refraction correspondences cannot be established. On the other hand, if after total internal reflection at $\mathbf{W}$, the light ray continues to propagate to anther surface point $\mathbf{X}$, and refracts at $\mathbf{X}$, and eventually reaches the pattern, our approach can still handle this case.

In practice, total internal reflection does not frequently happen, as the critical angle is normally very large (e.g., $\theta_c = 41.8^{\circ}$ for  glass to air). Only specially designed objects, like diamonds, will purposely make total internal reflection happen.

\subsection{Object analysis}

\begin{figure}[htbp]
\begin{center}
\includegraphics[width=0.65\linewidth]{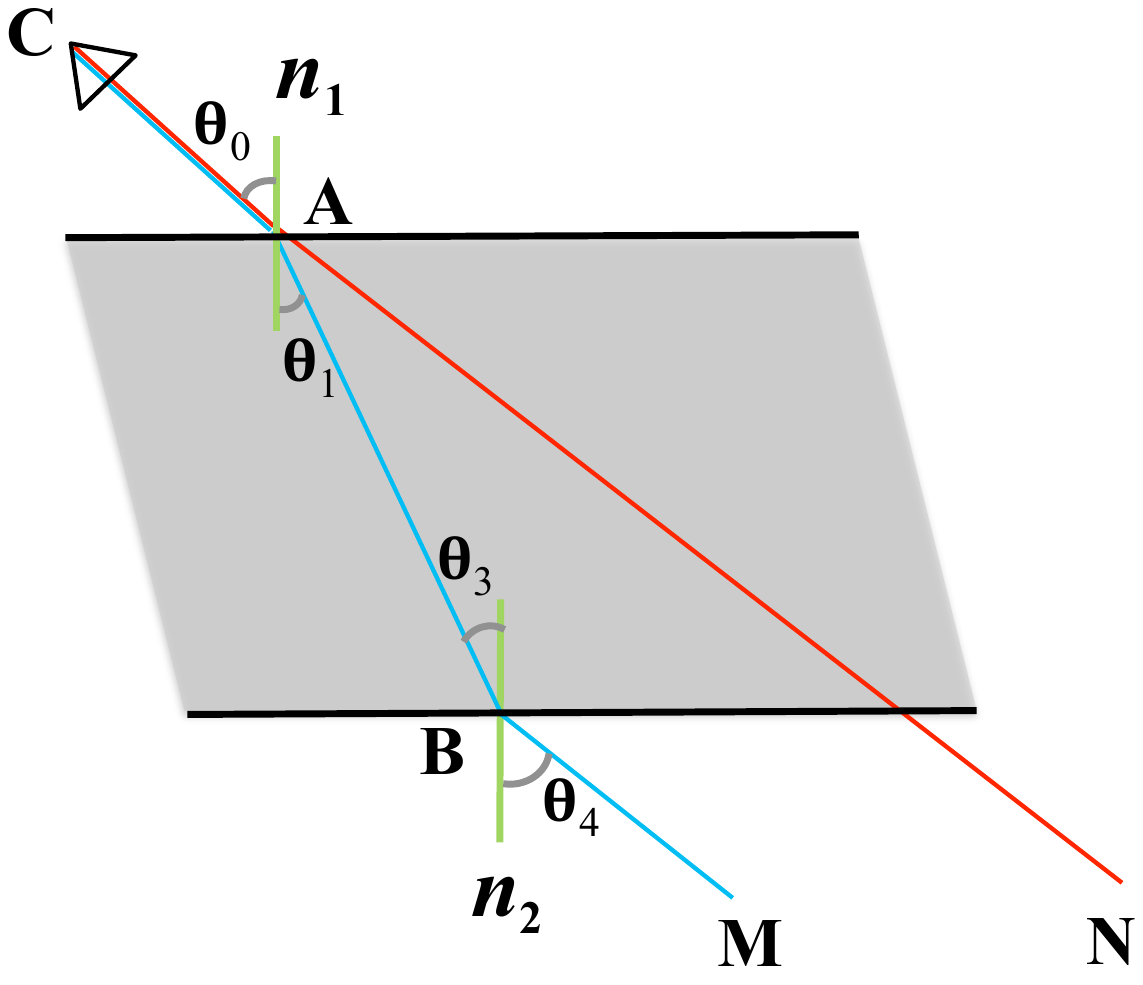}
\end{center}
   \caption{Refractions at the surfaces of a parallel planar plate. A camera centered at $\mathbf{C}$ observes a parallel planar surface. $\mathbf{A}$ is an upper surface point and $\mathbf{B}$ is a lower surface point along a light path. As their normals $\mathbf{n}_1$ and $\mathbf{n}_2$ are parallel, the PBC ($\mathbf{BM}$) is parallel to the visual ray ($\mathbf{CA}$). The angle between these two rays is $\Delta \theta = 0$.}
\label{fig:plane}
\end{figure}

Our general method only has the assumption that light paths (propagating from the pattern to the camera center) will not re-enter the liquid used for immersing the object once they enter the object. This assumption holds true for transparent objects with a convex shape, and for objects with holes completely enclosed inside the object. It allows us to handle object with an inhomogeneous refractive index. In practice, our method can also handle objects with shallow concavities as long as the previous assumption is satisfied.

Our thin object reconstruction method has the assumption of a single refraction. This is generally true for most thin transparent surfaces. The exception happens when the back and front sides of the surface are planar and parallel. For such surfaces (see Fig.~\ref{fig:plane}), the normals at the first and second refraction points are parallel and in opposite direction. In this case, the visual ray and PBC are parallel. This can be considered as no refraction occurs along the light path. Since this breaks the single refraction assumption, our thin transparent object reconstruction method cannot handle this case.

\subsection{Single refraction approximation}

Here we analyze the error induced by the single refraction approximation used in our second method, and demonstrate that such an approximation is appropriate for thin transparent objects. 

\begin{figure}[tbp]
\begin{center}
\includegraphics[width=0.8\linewidth]{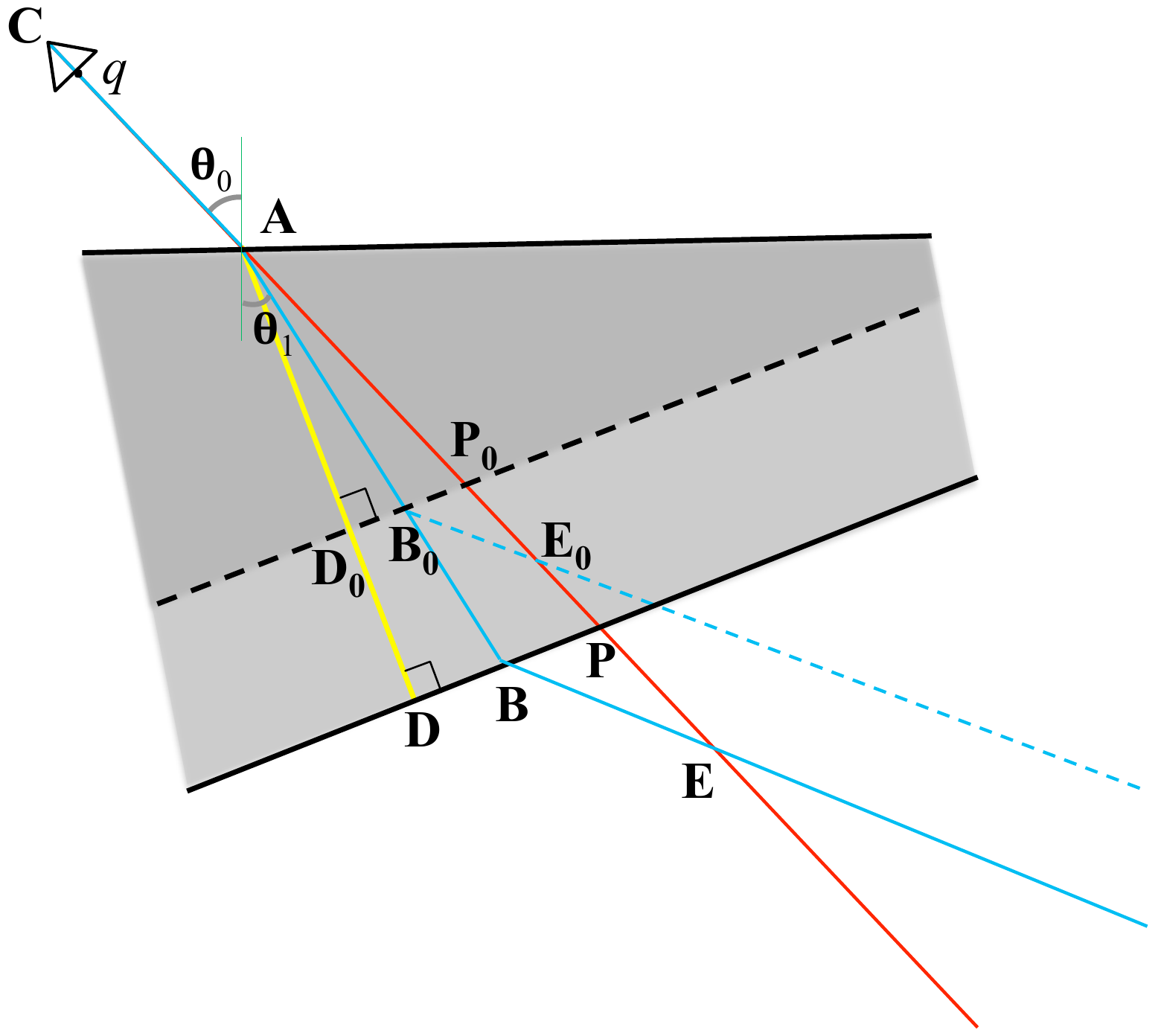}
\end{center}
\caption{Error induced by the single refraction approximation. It can be shown that the approximation error (i.e., distance between $\mathbf{B}$ and $\mathbf{E}$) is linearly proportional to the thickness of the transparent object (i.e., distance between $\mathbf{A}$ and $\mathbf{D}$. Please refer to the main text for details.}
\label{fig:error_single_refraction}
\end{figure}

Referring to Fig.~\ref{fig:error_single_refraction}, we have a camera centered at $\mathbf{C}$ observing a thin transparent object in front of a reference pattern. Let us consider the light path through an arbitrary image point $q$, and traverse this light path in reverse direction (i.e., beginning from the optical center of the camera, travelling through the thin transparent object, and eventually terminating at the reference pattern). After leaving the camera, this light path first refracts at point $\mathbf{A}$ on the upper surface of the transparent object. It continues to travel through the interior of the object, and refracts at point $\mathbf{B}$ on the lower surface of the object. After leaving the object, it continues to travel through the air and eventually terminates at a point on the reference pattern. If we apply the single refraction approximation, we will obtain point $\mathbf{E}$ from the intersection of the PBC and the visual ray of $q$. Note that the FEP for $q$ should be point $\mathbf{B}$, and therefore the distance between $\mathbf{B}$ and $\mathbf{E}$ is the approximation error.

Suppose we make the thin transparent object even thinner by moving its lower surface along its surface normal towards its upper surface, resulting in the new lower surface represented by the dotted line in Fig.~\ref{fig:error_single_refraction}. After leaving the camera, the light path for $q$ again first refracts at point $\mathbf{A}$ on the upper surface of the transparent object.  It continues to travel through the interior of the object, but this time refracts at point $\mathbf{B_0}$ on the lower surface of the thinner object. After leaving the object, it continues to travel through the air and eventually terminates at a point on the reference pattern. If we apply the single refraction approximation again, we will obtain point $\mathbf{E_0}$ from the intersection of the PBC and the visual ray of $q$. Similarly, the distance between $\mathbf{B_0}$ and $\mathbf{E_0}$ is the approximation error for this thinner object.

Consider the similar triangles $\triangle\mathbf{AB_0P_0}$ and $\triangle\mathbf{ABP}$. It is easy to see from  Fig.~\ref{fig:error_single_refraction} that $\mathbf{
\|AB_0\|}:\mathbf{\|AB\|} = \mathbf{\|AD_0\|}:\mathbf{\|AD\|}$. Hence, we have $\mathbf{\|B_0P_0\|}:\mathbf{\|BP\|} =   \mathbf{\|AB_0\|}:\mathbf{\|AB\|} = \mathbf{\|AD_0\|}:\mathbf{\|AD\|}$. Consider now the similar triangles $\triangle\mathbf{B_0P_0E_0}$ and $\triangle\mathbf{BPE}$. We have $\mathbf{\|B_0E_0\|}:\mathbf{\|BE\|} = \mathbf{\|B_0P_0\|}:\mathbf{\|BP\|} =  \mathbf{\|AD_0\|}:\mathbf{\|AD\|}$. Hence, we can conclude that the approximation error is linearly proportional to the thickness of the transparent object, and therefore the single refraction approximation is appropriate for thin transparent objects.

\begin{figure*}[htbp]
\centering
\tabcolsep=0.1cm
\begin{tabular}{
>{\centering\arraybackslash} m{0.27\textwidth}
>{\centering\arraybackslash} m{0.226\textwidth}
>{\centering\arraybackslash} m{0.220\textwidth}
>{\centering\arraybackslash} m{0.19\textwidth}}
\includegraphics[width=\linewidth]{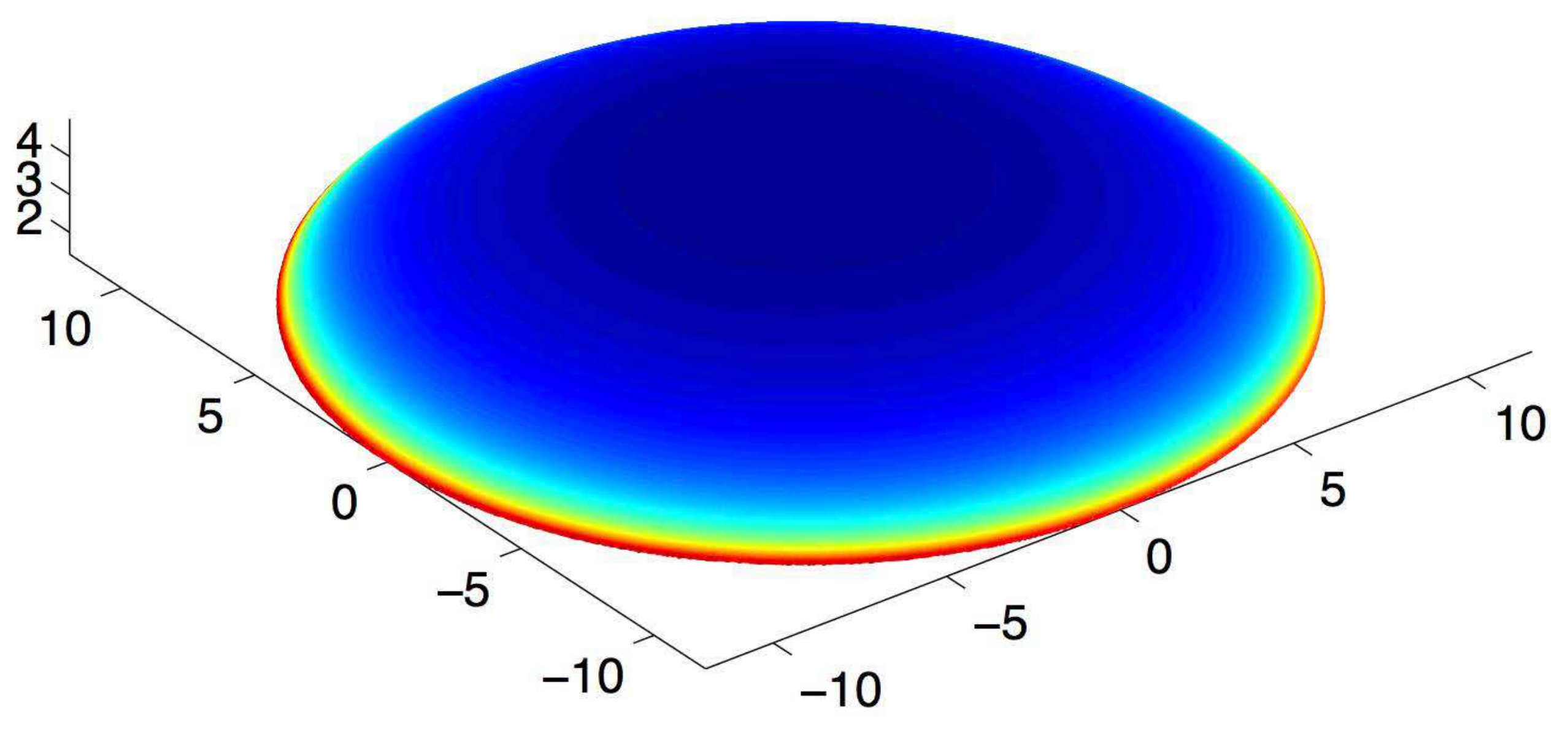} &
\includegraphics[width=\linewidth]{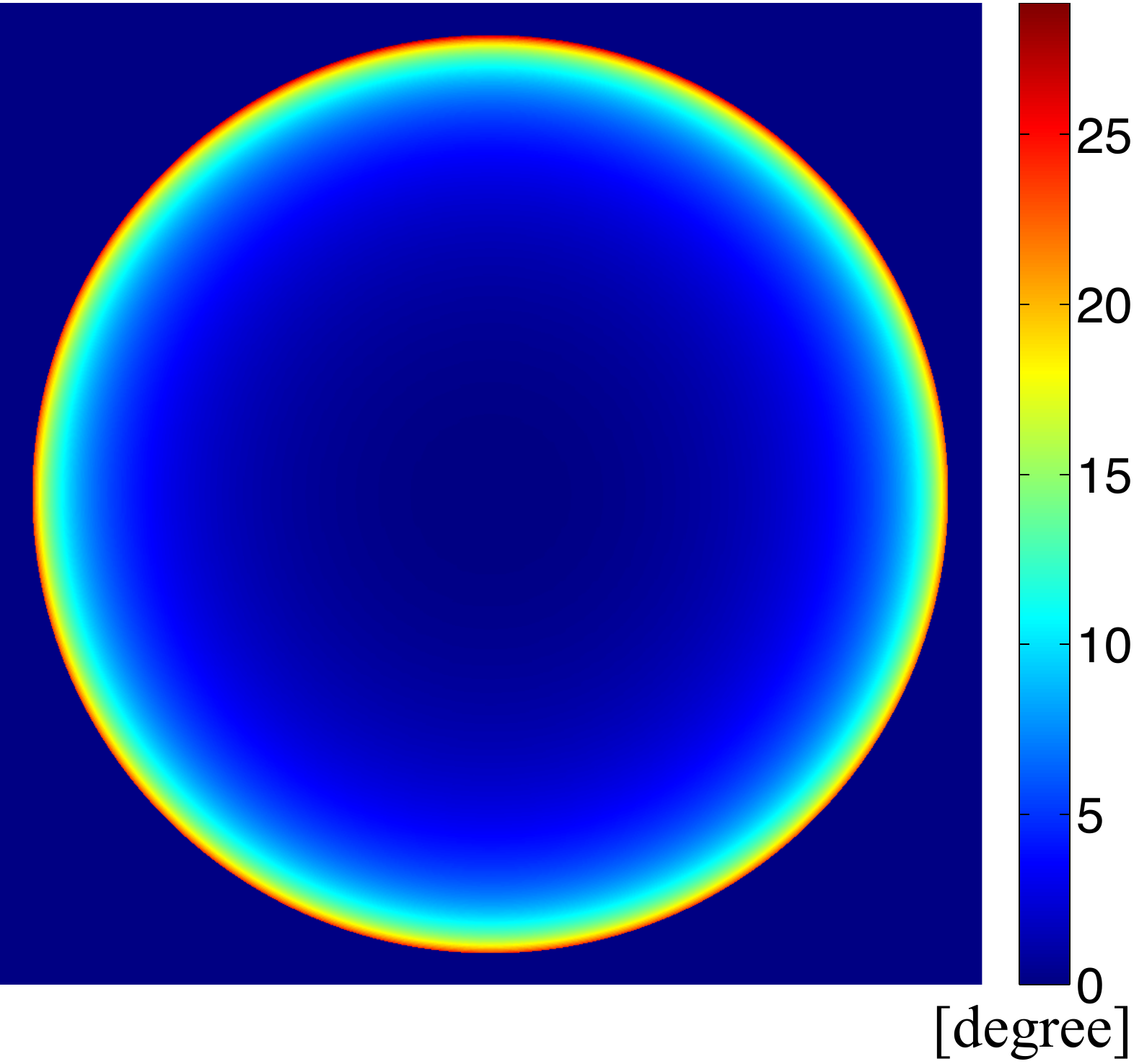}&
\includegraphics[width=\linewidth]{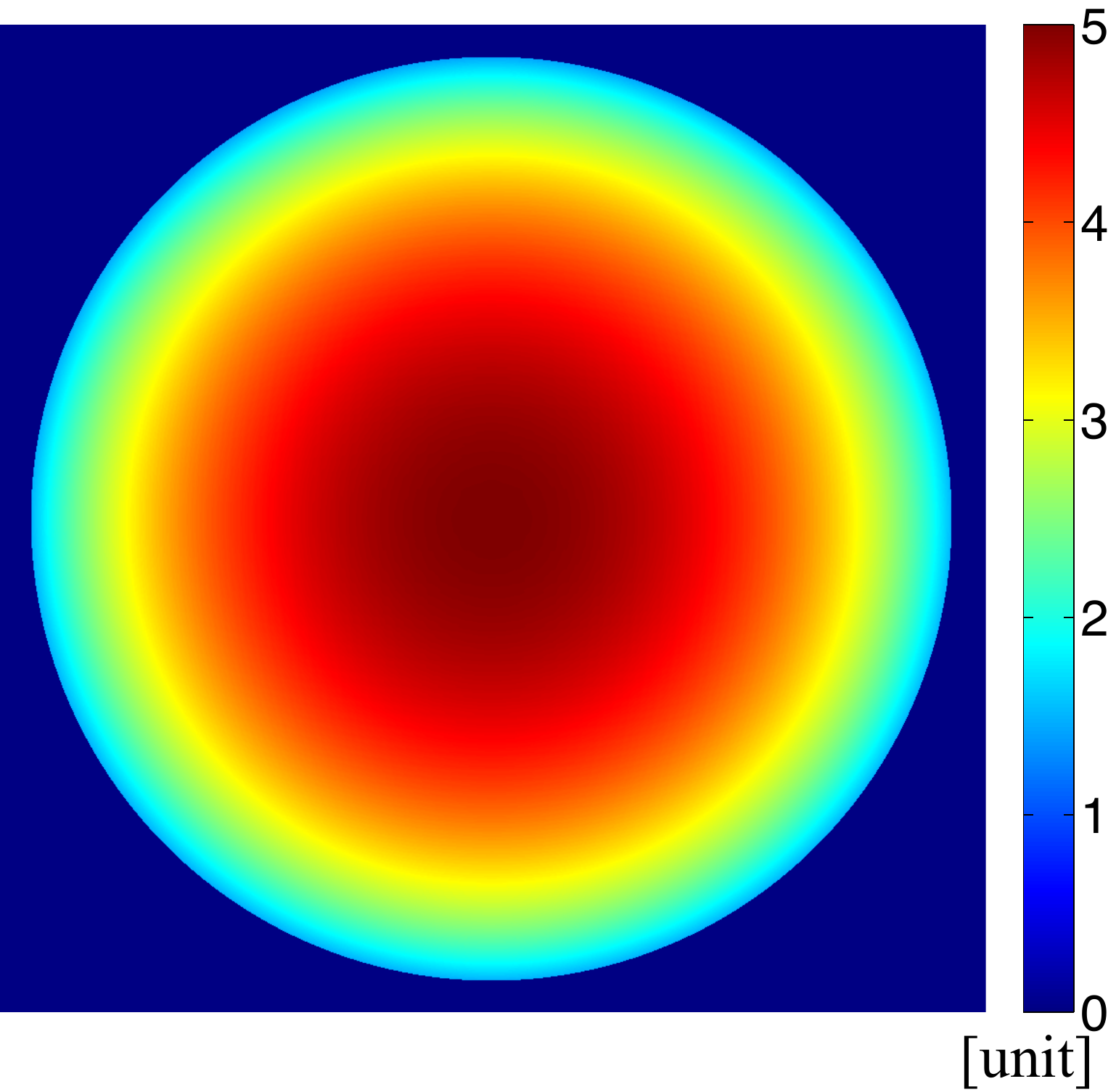}&
\includegraphics[width=\linewidth]{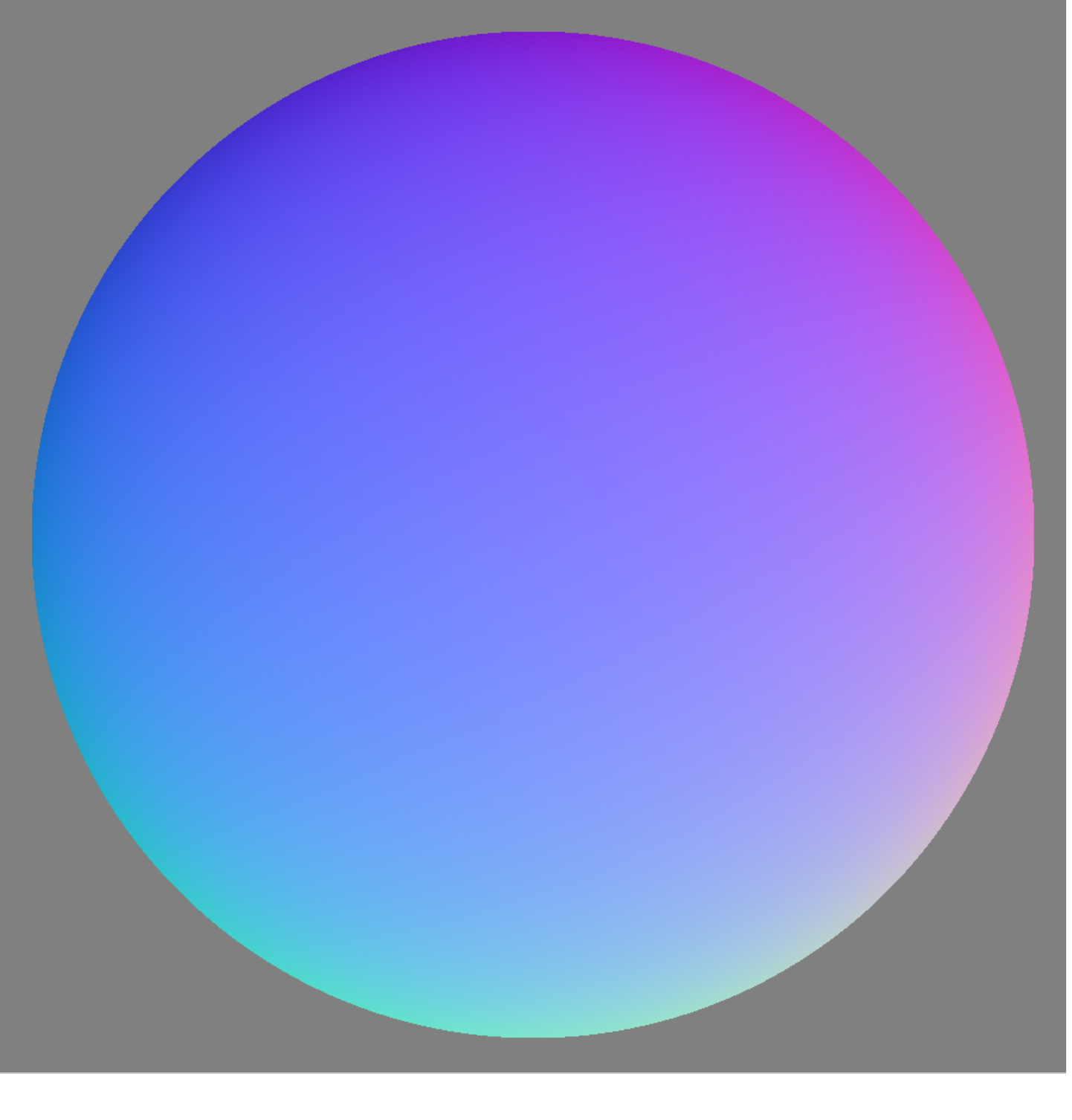}\\
\includegraphics[width=\linewidth]{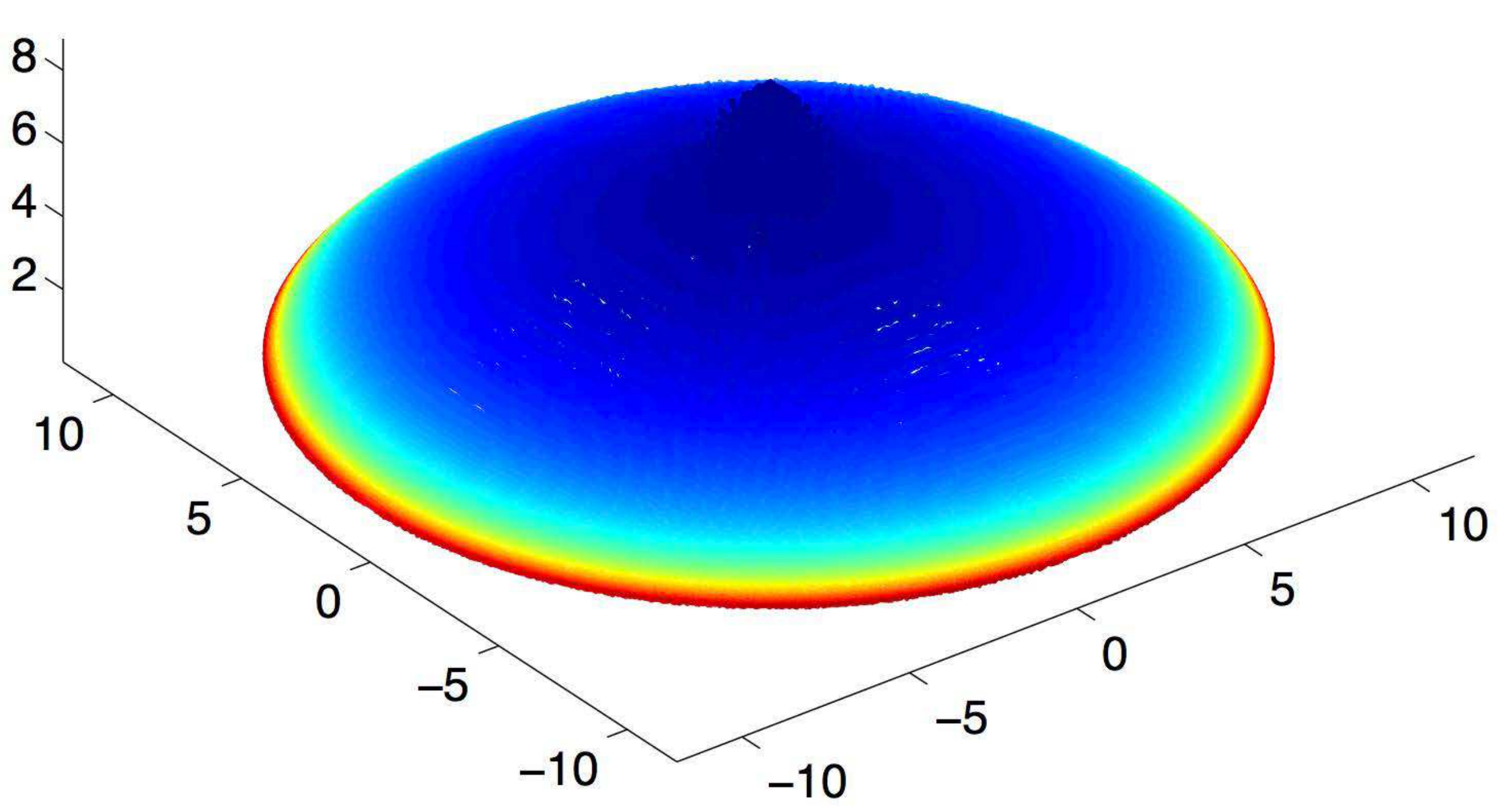} &
\includegraphics[width=\linewidth]{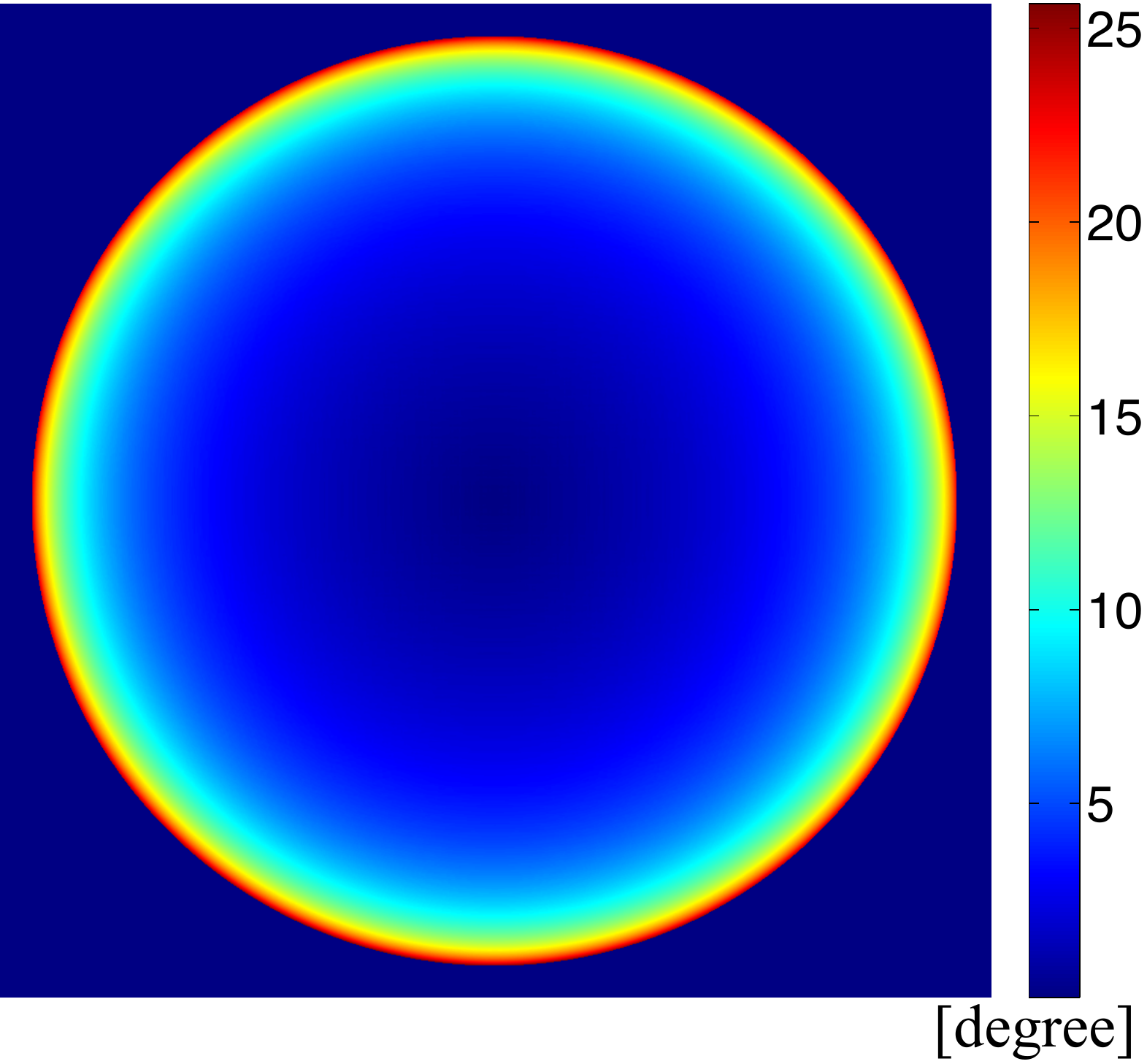} &
\includegraphics[width=\linewidth]{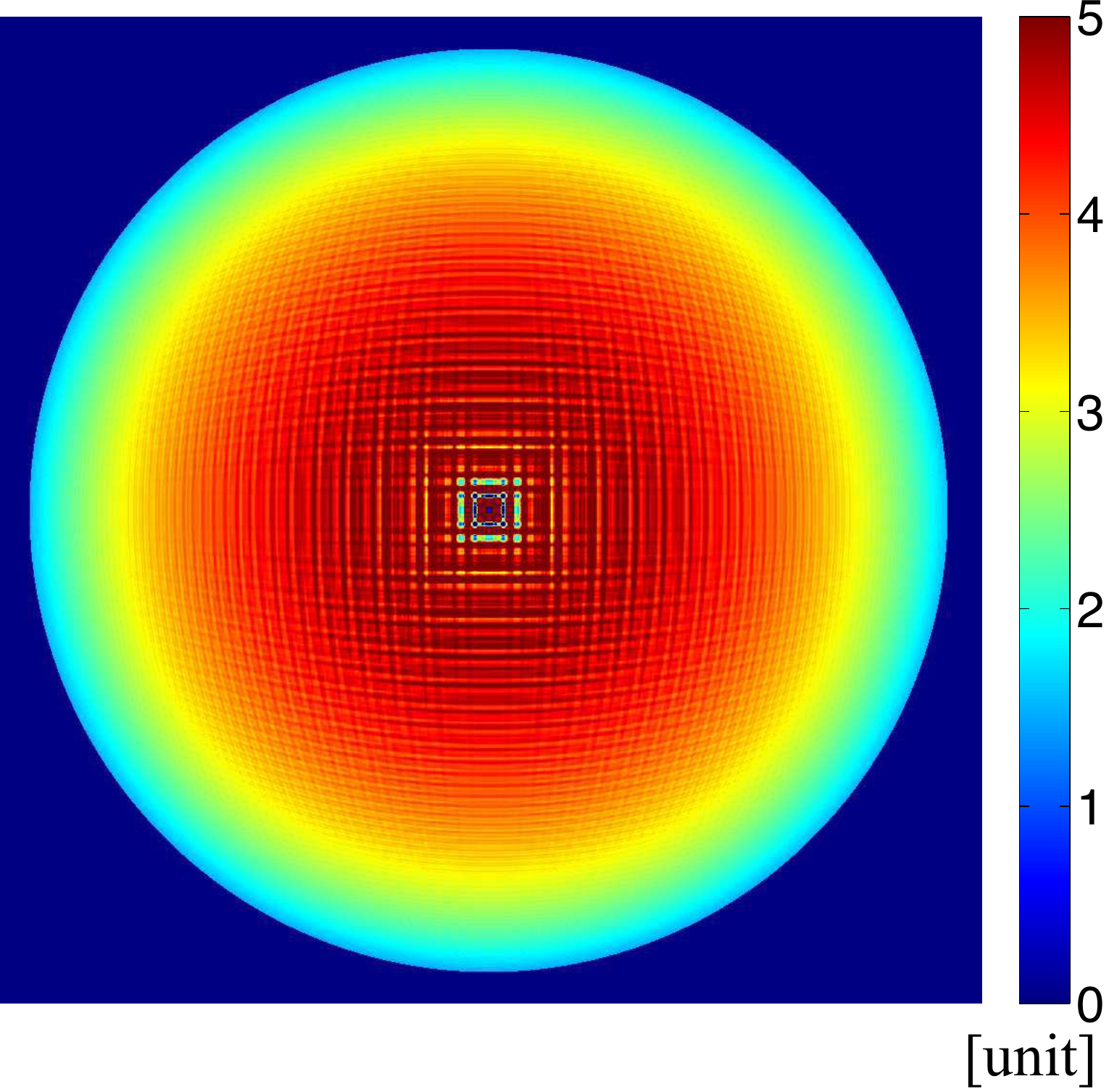} &
\includegraphics[width=\linewidth]{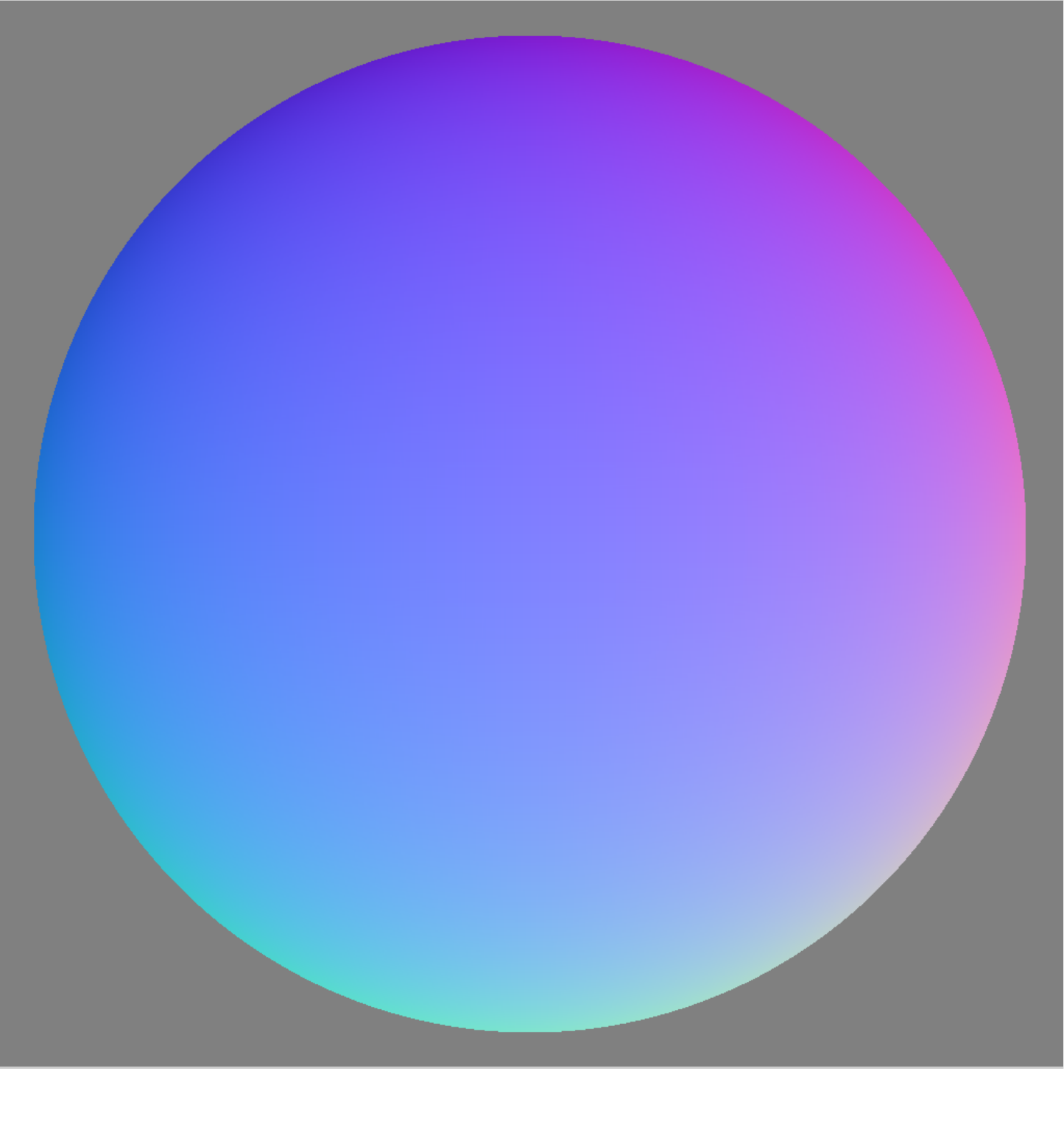}\\
\end{tabular}
\caption{Reconstruction of a synthetic semi-ellipsoid using the first method. First row: Ground truth. Second row: Our reconstructed results. First column: FEP cloud. Second column: angle between the PBCs in a pair. Third column: depth map. Fourth column: normal map.} 
\label{fig:synthetic_experiments}
\end{figure*}

It is also worth to note that the difference between the surface normals at the upper and lower surface points along the same light path will also affect the reconstruction accuracy under the single refraction approximation. If we increase the difference between the surface normals at $\mathbf{A}$ and $\mathbf{B}$ by rotating the lower surface around $\mathbf{B}$, $\mathbf{E}$ will get closer to $\mathbf{B}$, which indicates a smaller reconstruction error.

\section{Experimental Evaluation}
\label{sec:experiments}

We now demonstrate the effectiveness of our approach on synthetic and real objects. In the remainder of this section, we present both quantitative and qualitative reconstruction results. In the following, for the sake of clarity, we denote our general method that uses liquid to alter the incident light path as the first method, and the method that uses the object itself to alter the incident light path, which is tailored for thin transparent objects, as the second method.

\subsection{Synthetic Data}
\textbf{First method on a convex object}. For our synthetic experiments, we used~\emph{Pov-Ray} to simulate the entire experimental setup. First, we modeled a convex transparent object as a semi-ellipsoid with the following parametric equation

\begin{equation} \left\{
\begin{aligned}
 (\frac{x}{12.5})^2 + (\frac{y}{12.5})^2 + (\frac{z}{5})^2 = 1,  \\
 z>0. \\
\end{aligned}
\right.
\end{equation}
We further assumed the transparent ellipsoid has a refractive index $\lambda = 1.5$.\footnote{The transparent object can be inhomogeneous, namely the refractive index varies across the interior of the object.}  A reference plane displaying a set of thin stripe sweeping patterns was placed at two different positions. The size of the reference plane was $32 \times 32$ units in~\emph{Pov-Ray} environment and the thickness of the stripe was $\frac{1}{32}$ unit. A synthetic perspective camera with a resolution of $1024 \times 1024$ was used to capture the refraction of the reference pattern through the transparent object immersed in air ($\lambda = 1.0$) and liquid ($\lambda = 1.3$) respectively. We adopted the strategies described in Section~\ref{sec:correspondences} to obtain dense refraction correspondences. More than 700K refraction correspondences were used in our synthetic experiment.

We reconstructed a pair of PBCs for each FEP based on the retrieved refraction correspondences. The transparent surface was then recovered from the ray triangulation of PBC pairs. We also computed the surface normals from the PBC pairs and the refractive indices of the media. Fig.~\ref{fig:synthetic_experiments} depicts the reconstructed FEP cloud as well as surface normals. It also shows the depth map of the reconstructed object for accuracy evaluation\footnote{The depth map is defined as the $z$ component for each $3D$ point.}.

\begin{figure*}[htbp]
\centering
\tabcolsep=0.5cm
\begin{tabular}{
>{\centering\arraybackslash} m{0.32\textwidth}
>{\centering\arraybackslash} m{0.32\textwidth}}
\includegraphics[width=\linewidth]{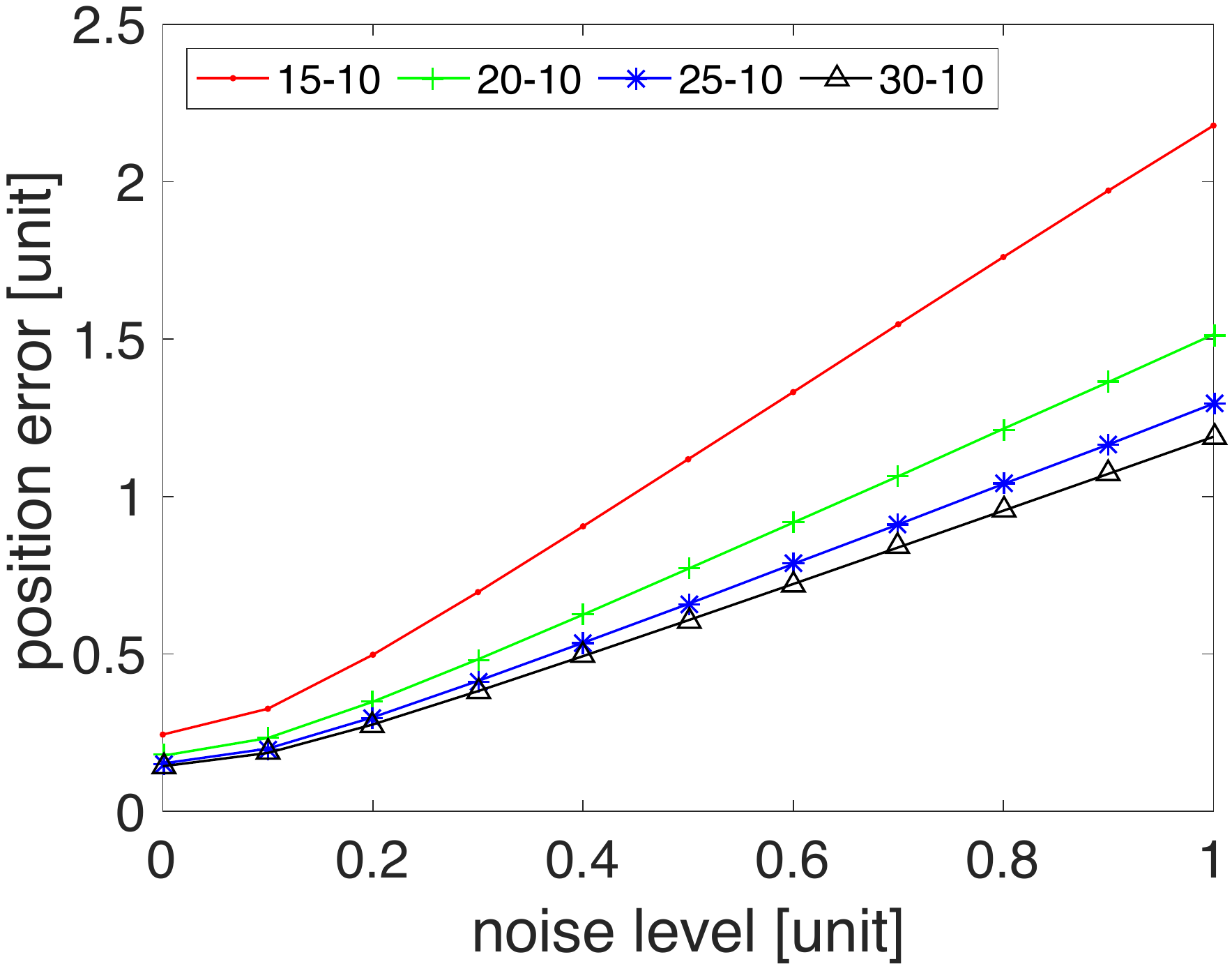} &
\includegraphics[width=\linewidth]{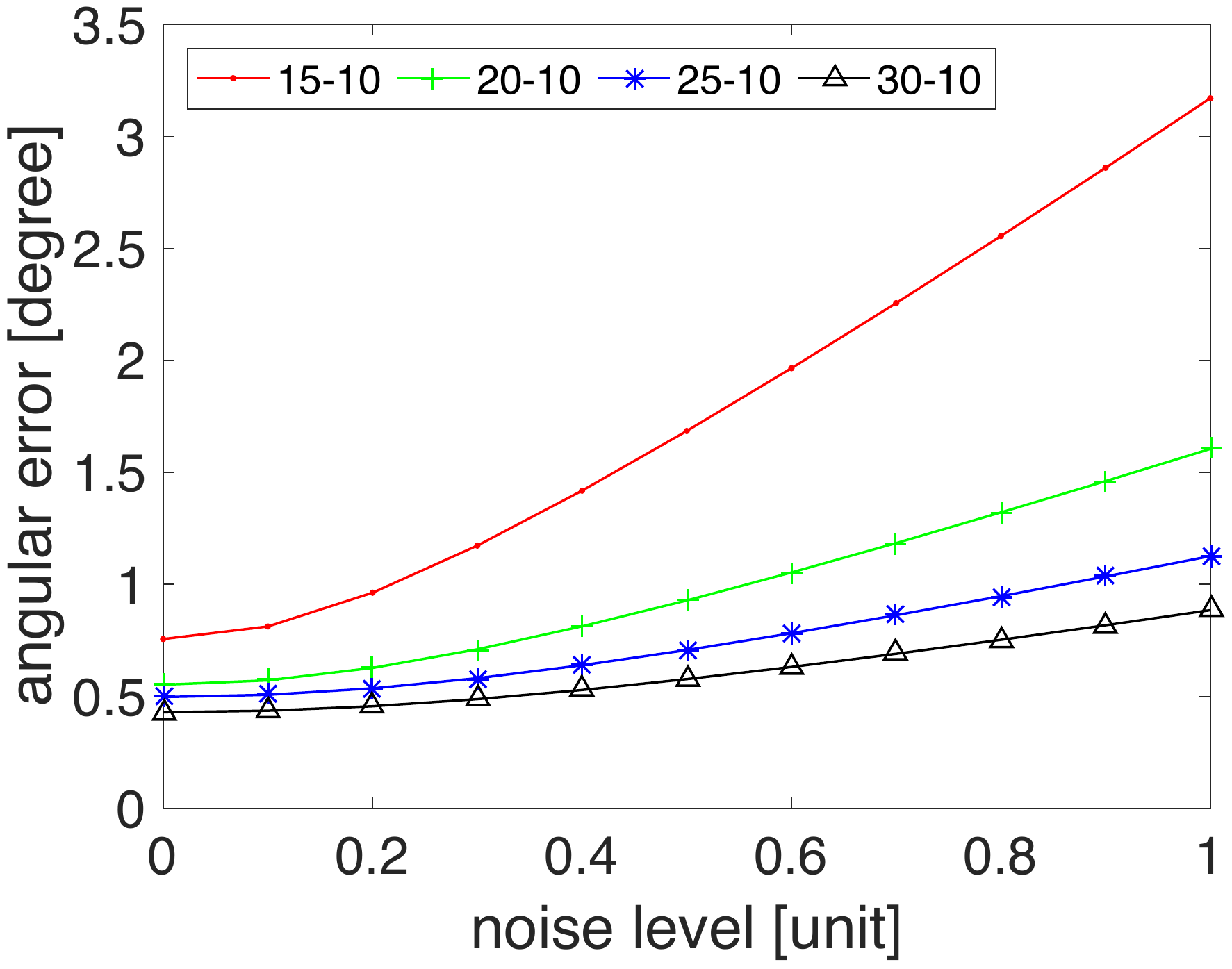}\\
(a) &
(b) \\
\includegraphics[width=\linewidth]{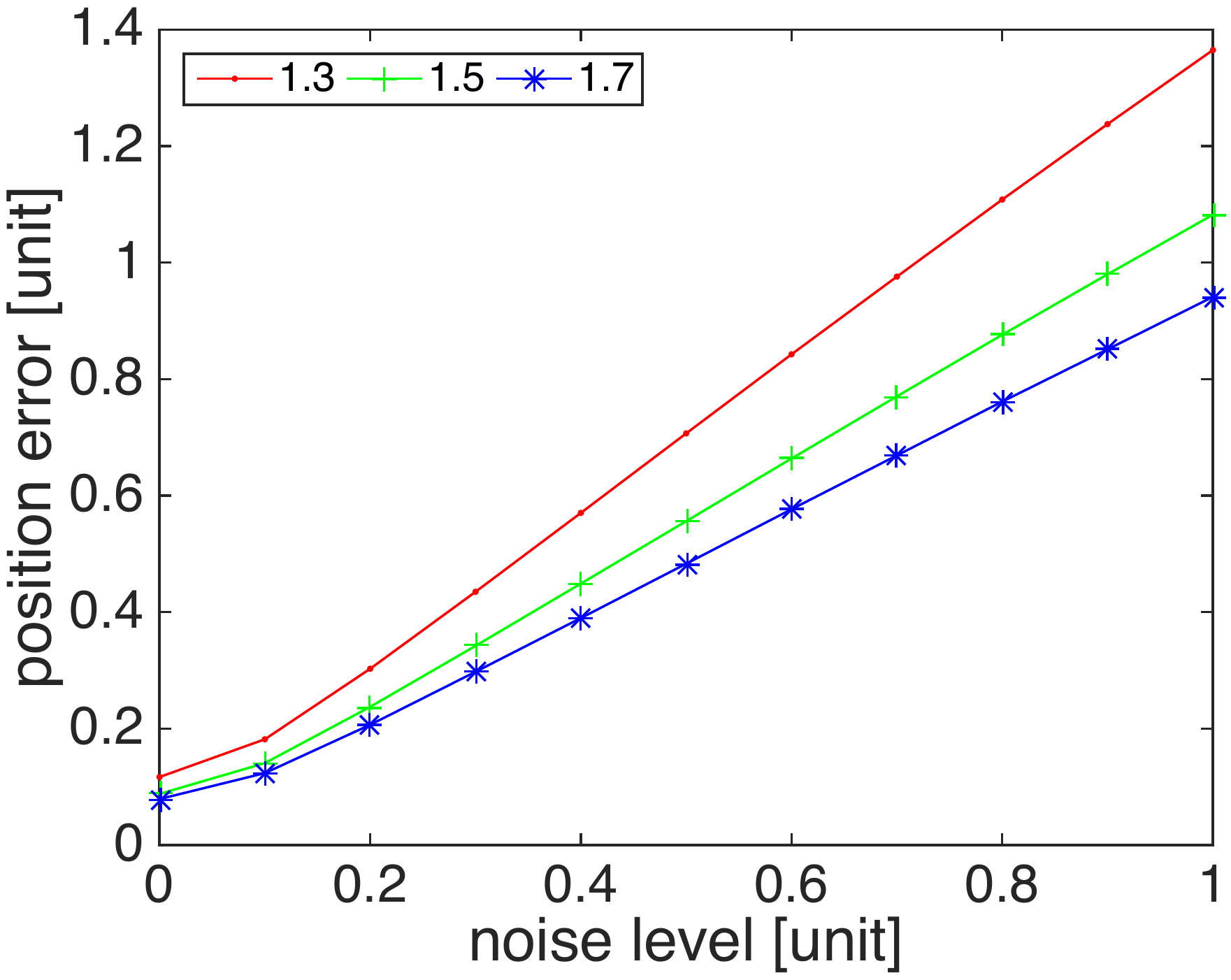}&
\includegraphics[width=\linewidth]{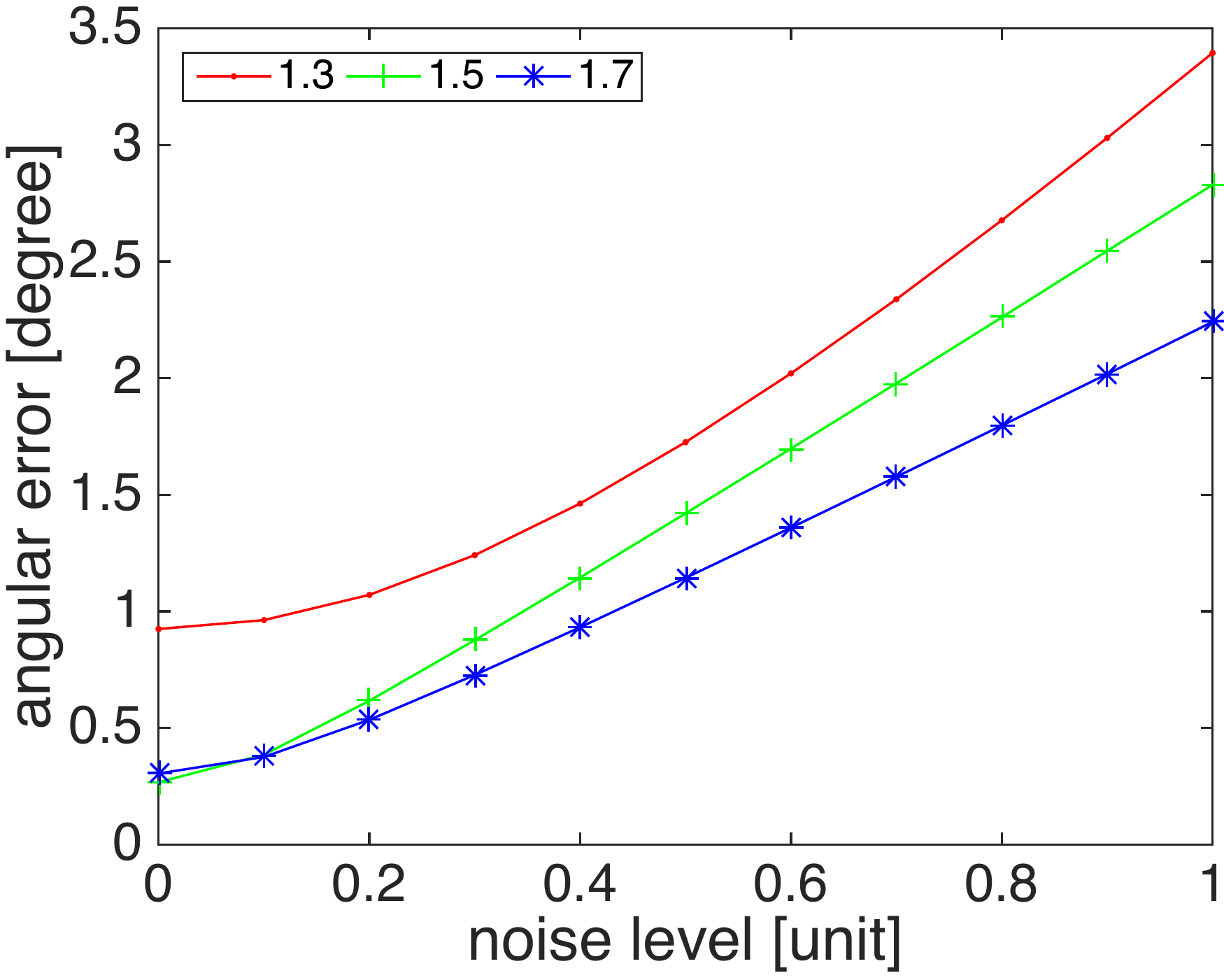}\\
(c) &
(d)\\
\end{tabular}
\caption{RMS errors for the positions and normals of the FEPs reconstructed using the first method. (a)-(b) show the RMS errors for the positions and normals, respectively, over 500 rounds with different random noise and relative distances between the two positions of the reference pattern. (c)-(d) show the RMS errors for the positions and normals, respectively, over 500 rounds with different random noise and refractive indices of the media.
} 
\label{fig:RMS_experiments}
\end{figure*}

\begin{figure*}[htbp]
\centering
\tabcolsep=0.2cm
\begin{tabular}{
>{\centering\arraybackslash} m{0.27\textwidth}
>{\centering\arraybackslash} m{0.210\textwidth}
>{\centering\arraybackslash} m{0.18\textwidth}
>{\centering\arraybackslash} m{0.210\textwidth}}
\includegraphics[width=\linewidth]{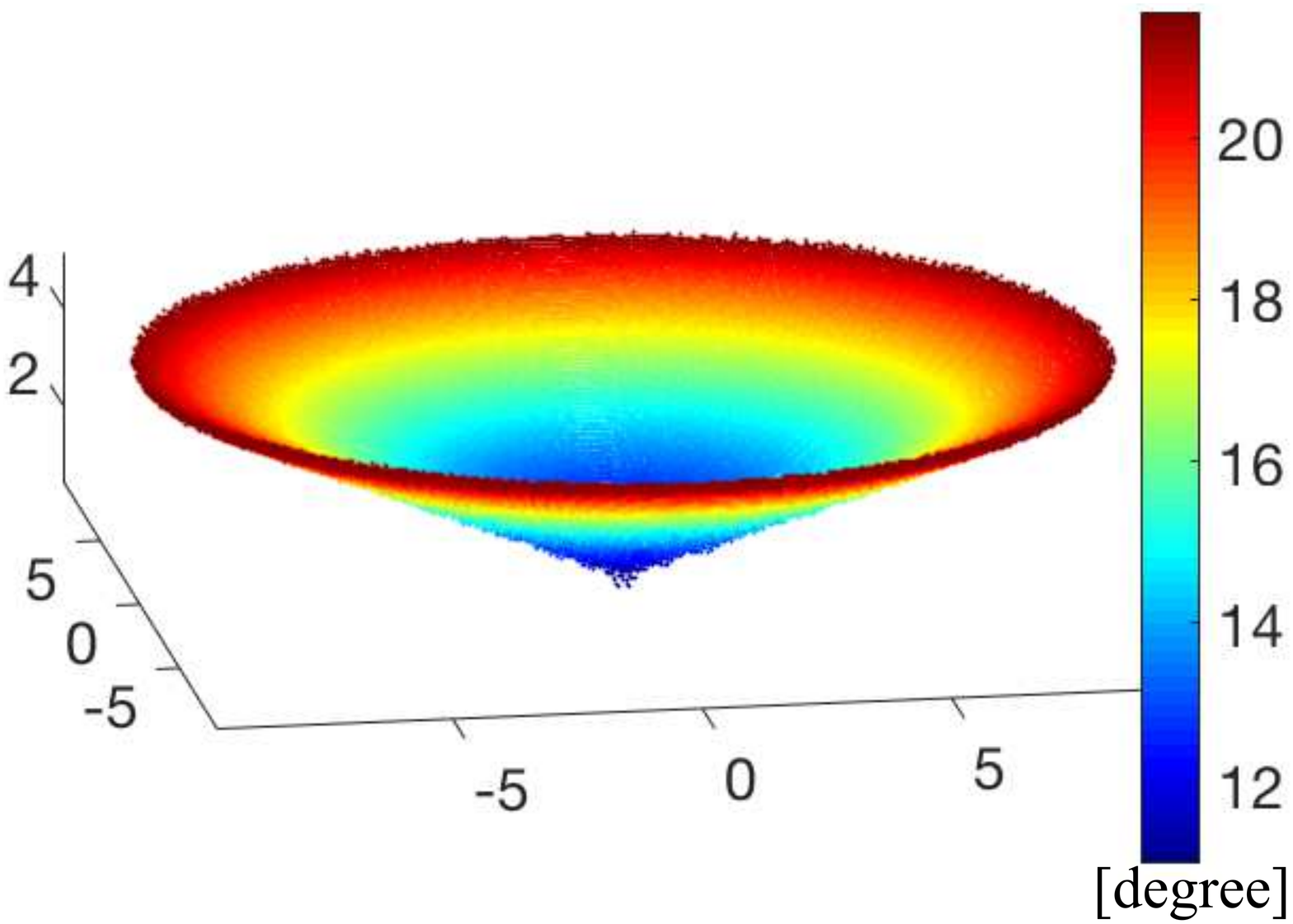}&
\includegraphics[width=\linewidth]{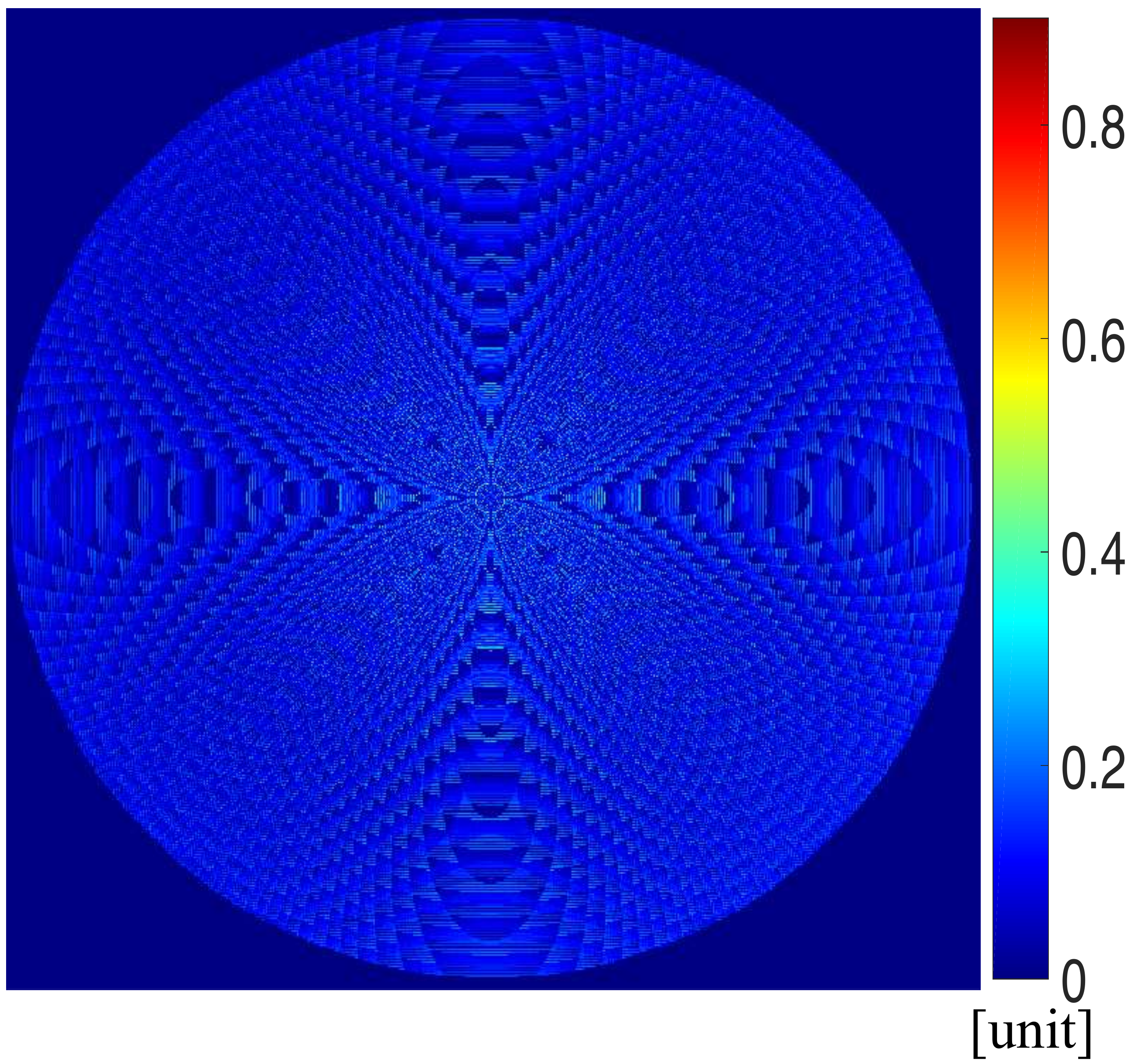}&
\includegraphics[width=\linewidth]{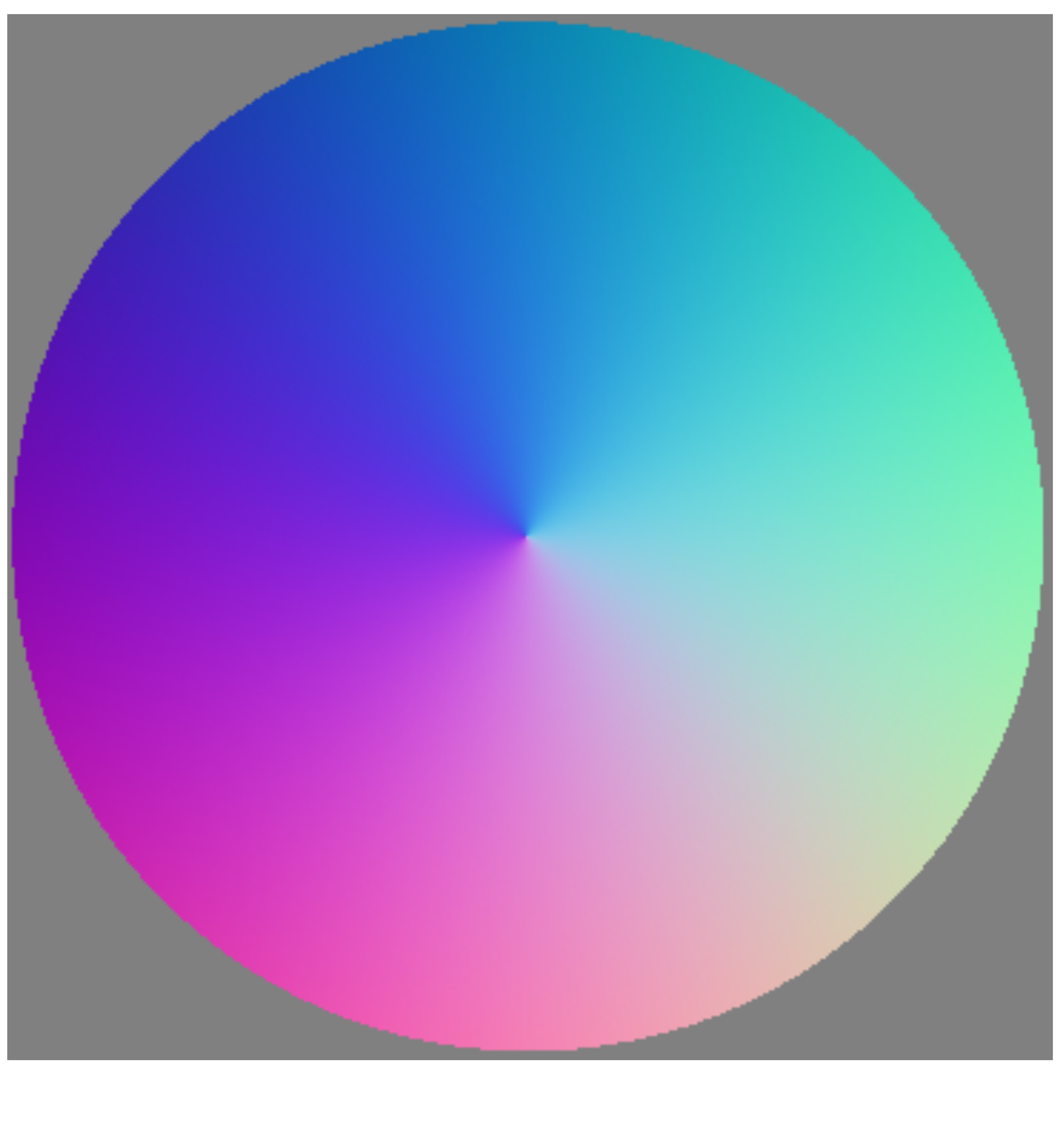}&
\includegraphics[width=\linewidth]{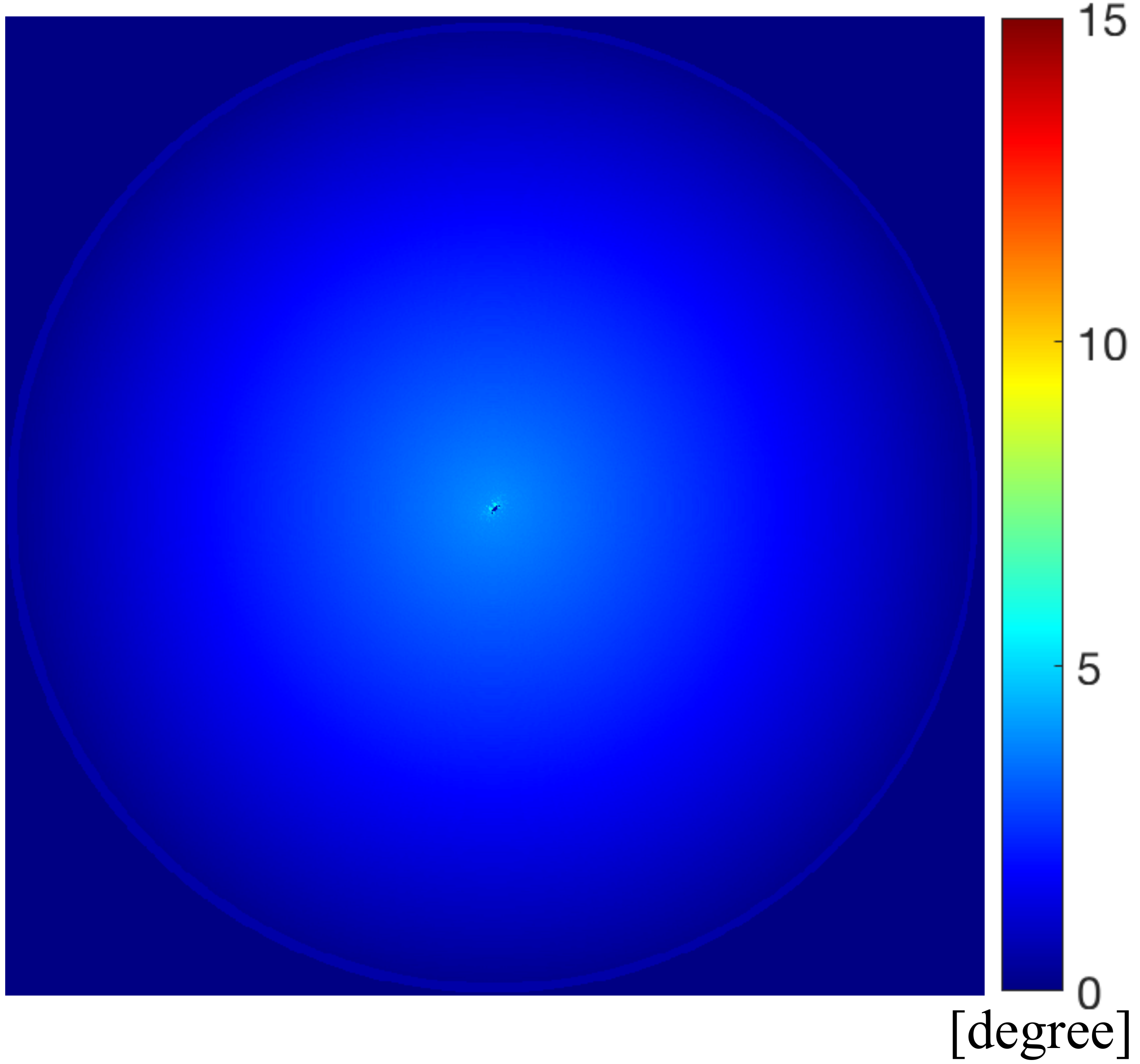}\\
\end{tabular}
\caption{Reconstruction of a concave transparent object using the first method. From left to right: the reconstructed FEP cloud (color coded by angle between the PBCs in a pair); error map between ground truth FEPs and reconstructed FEPs; reconstructed normal map; error map between ground truth normals and reconstructed normals.} 
\label{fig:synthetic_concave_cone}
\end{figure*}

In practice, reconstruction errors originate from the inaccuracy in finding the refraction correspondences on the reference patterns. Errors may increase as the relative distance between the two positions of the reference pattern decreases. We therefore carried out a joint analysis by adding $2D$ zero-mean Gaussian noise to the extracted dense correspondences on the reference pattern together with varying the relative distance between the two positions of the reference pattern. The noise level ranged from $0.1$ to $1.0$ unit. The relative distance between the positions of the reference pattern varied from $5$ to $20$ units. We fixed the first position of the reference pattern at $z=10$ in our experiment, and varied the second position of the pattern by placing it at $z = 15, 20, 25, 30$, respectively. The reconstruction accuracy was evaluated based on the root mean square (RMS) error between the ground truth surface and the reconstruction. We further computed the angular distances between our reconstructed normals and the ground truth normals computed from the analytical equation of the semi-ellipsoid. Fig.~\ref{fig:RMS_experiments}(a-b) show the RMS errors for the positions and normals of the reconstructed FEPs under different noise level and relative distances between the two positions of the reference pattern. It shows that the reconstruction errors decrease as the distance between the two positions of the reference pattern increases.

\begin{figure*}[htbp]
\centering
\tabcolsep=0.2cm
\begin{tabular}{
>{\centering\arraybackslash} m{0.01\textwidth}
>{\centering\arraybackslash} m{0.2\textwidth}
>{\centering\arraybackslash} m{0.225\textwidth}
>{\centering\arraybackslash} m{0.19\textwidth}
>{\centering\arraybackslash} m{0.26\textwidth}}

(a) &
\includegraphics[width=\linewidth]{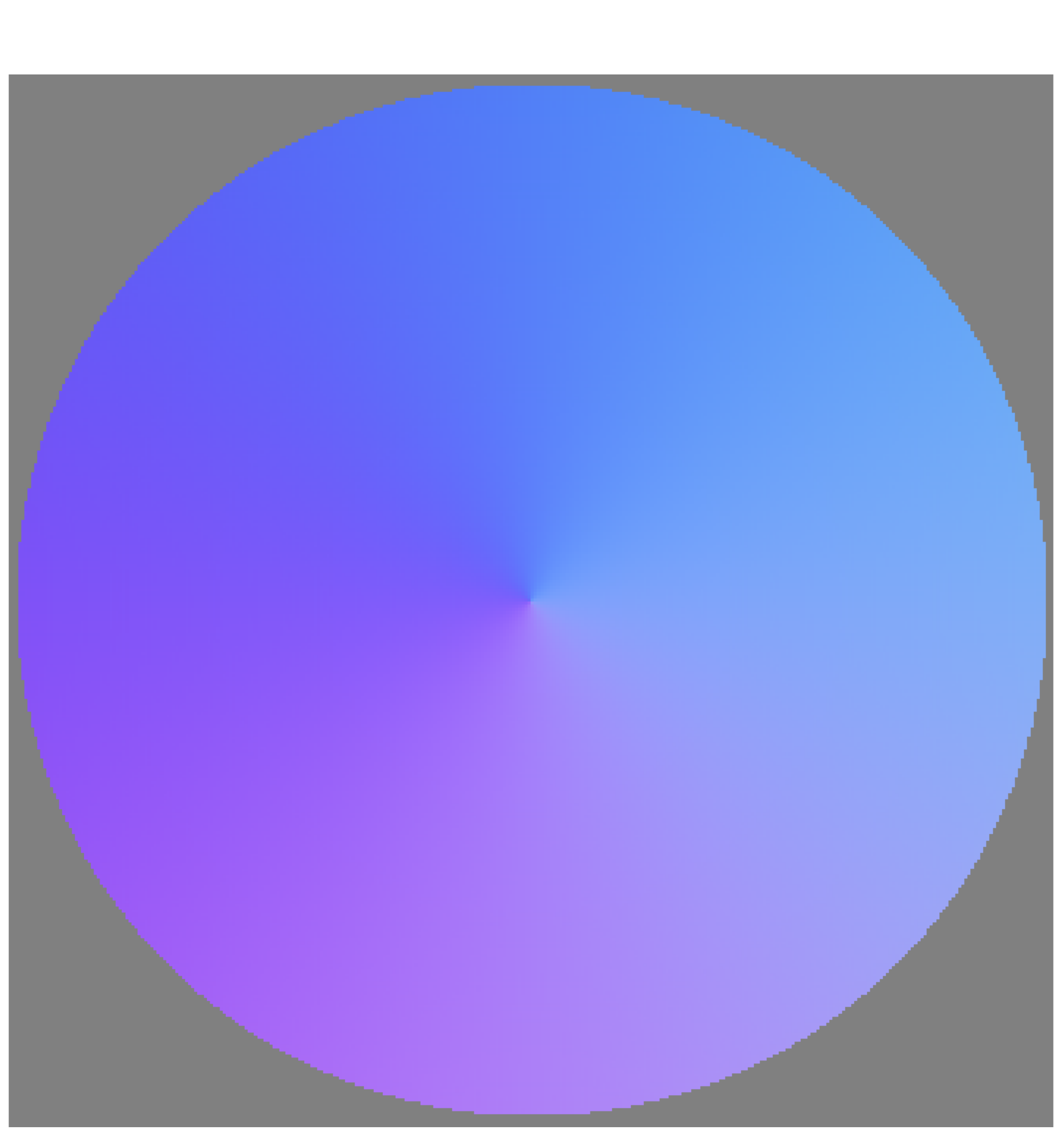}&
\includegraphics[width=\linewidth]{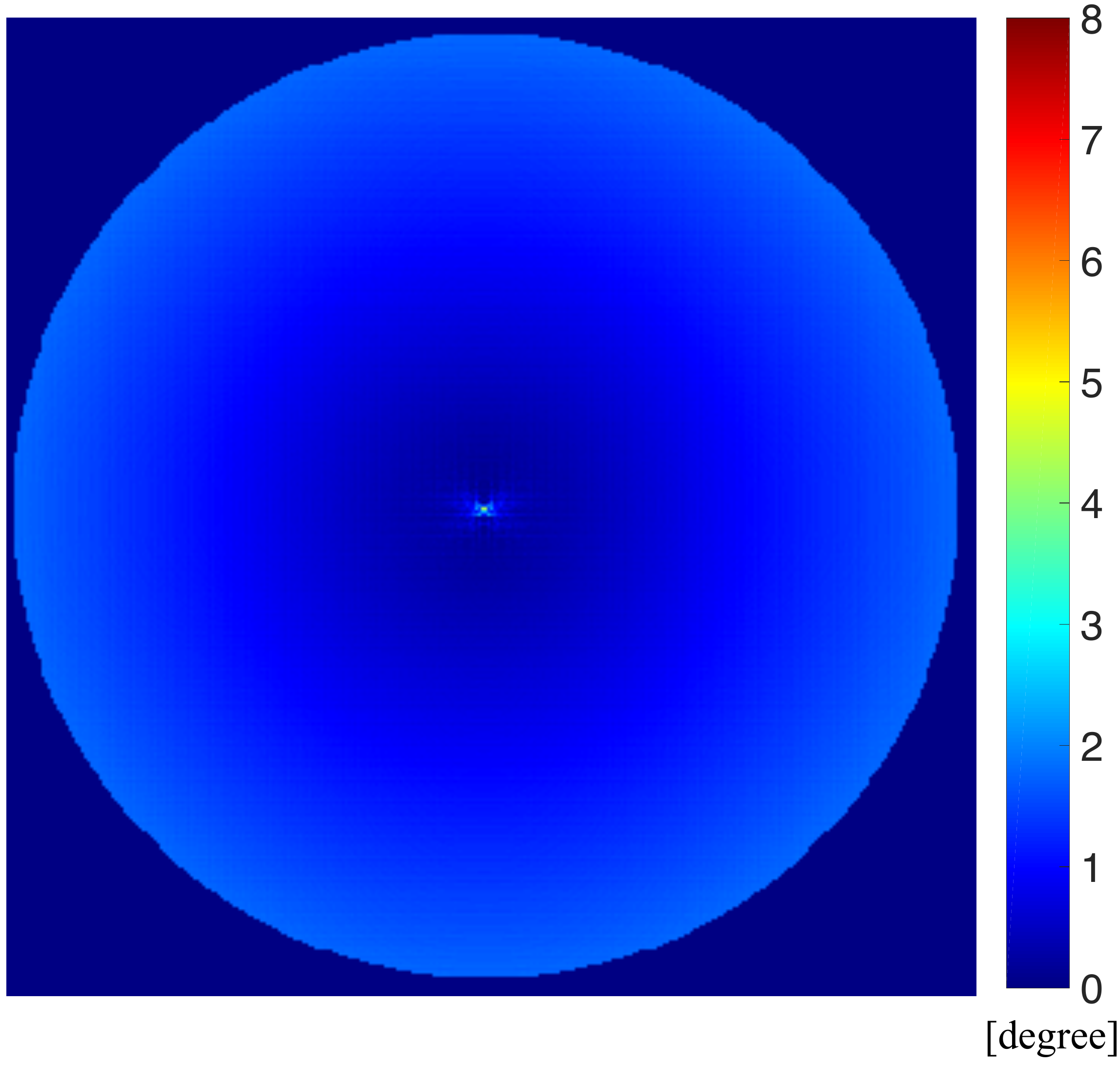}&
\includegraphics[width=\linewidth]{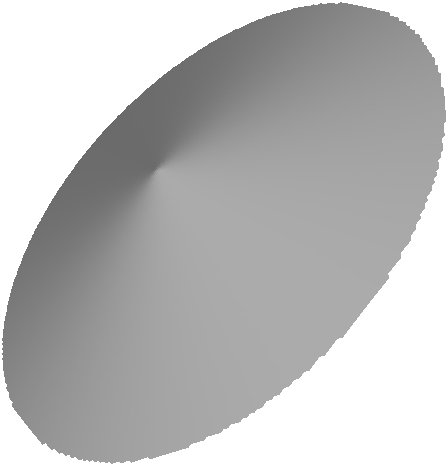}&
\includegraphics[width=\linewidth]{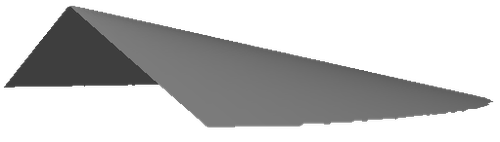}\\
(b) &
\includegraphics[width=\linewidth]{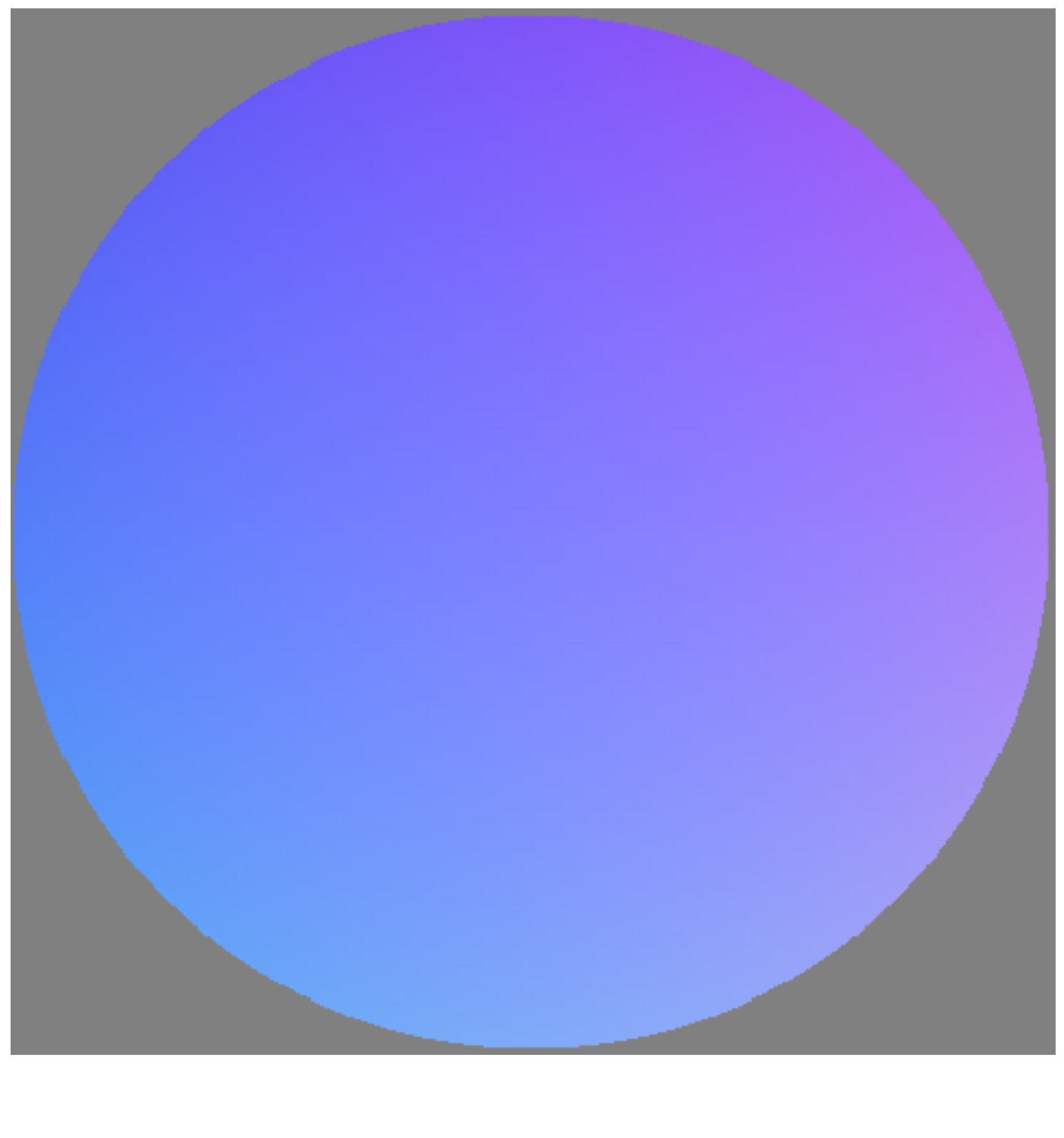}&
\includegraphics[width=\linewidth]{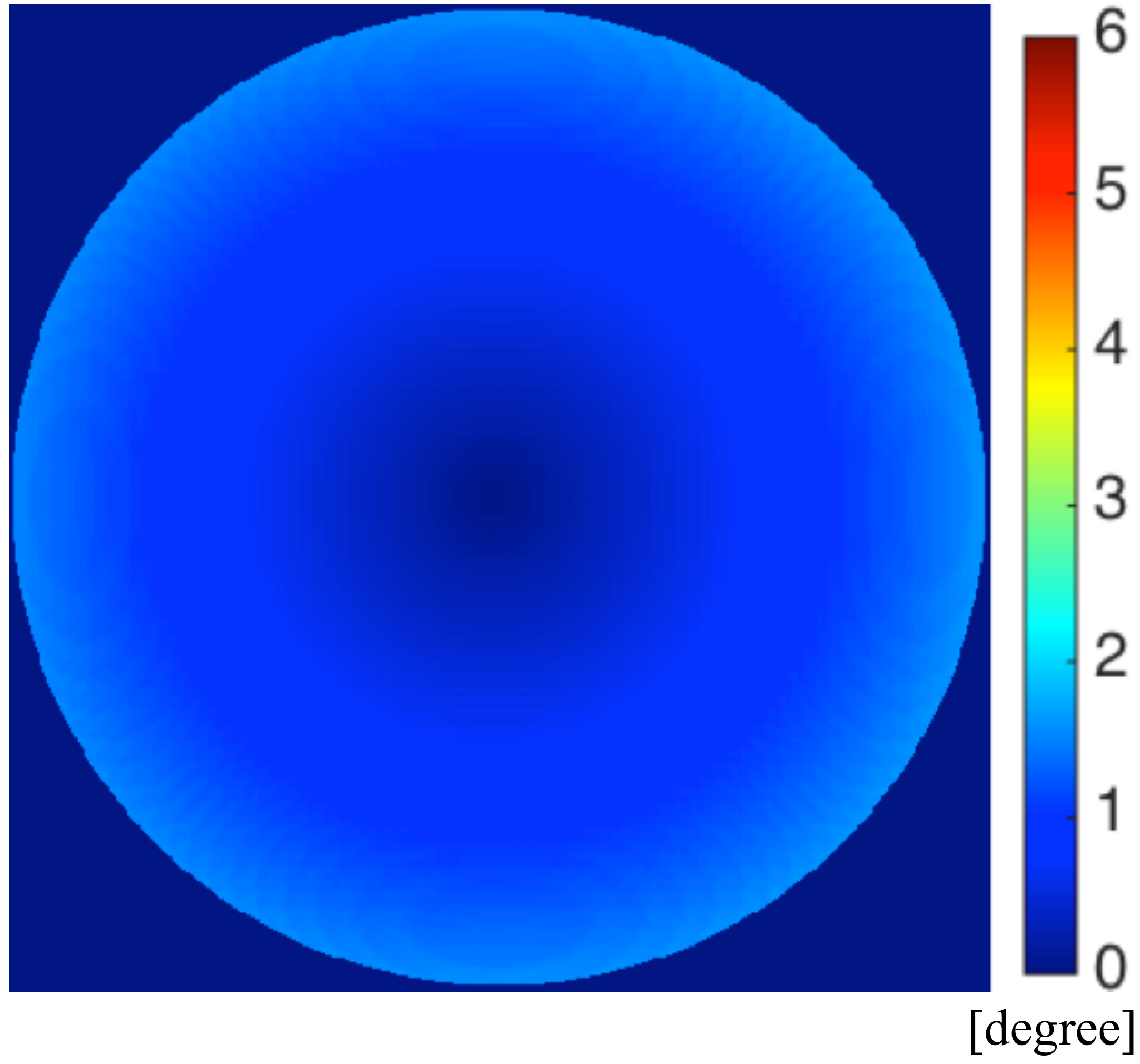}&
\includegraphics[width=\linewidth]{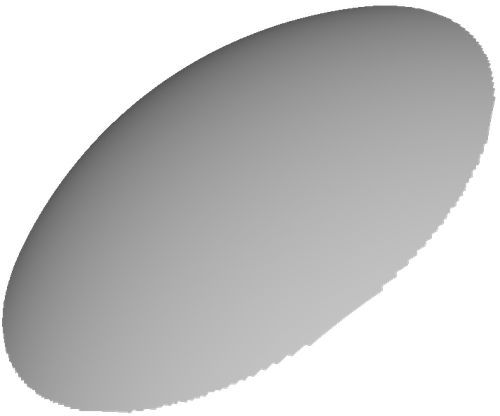}&
\includegraphics[width=\linewidth]{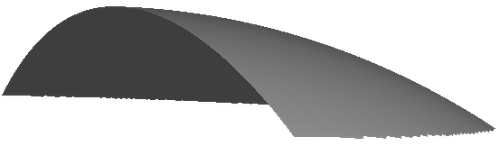}\\

\end{tabular}
\caption{Reconstruction of two synthetic thin objects using the second method: (a) thin cone; (b) spherical shell. First column: reconstructed angle map. Second column: error map between ground truth normals and reconstructed normals. Last two columns: two views of the reconstructed surface (the second view is a cross-section view).} 
\label{fig:synthetic_thin}
\end{figure*}

We further conducted an analysis on the reconstruction error with respect to the refractive index of the liquid medium. 
Two other media with different refractive indices were tested in the experiment, namely $\lambda=1.5$ and $\lambda=1.7$. The reference pattern was placed at $z = 6$ and $z = 10$, respectively. Fig.~\ref{fig:RMS_experiments}(c-d) show that the reconstruction errors  decrease as the refractive index of the medium increases.

\textbf{First method on a concave object}. We also evaluated our method on a concave surface with $\lambda = 1.7$, which was defined by the difference between a cylinder and a right circular cone. The radius and height of the cylinder were $5$ and $10$ respectively, while the height and radius of the cone were $4$ and $10$ respectively. The resulting shape was an object with a cylinder outer shape and a right circular cone inner shape. We reconstructed the inner shape of the object in this experiment by immersing the concave side of the object into water. Fig.~\ref{fig:synthetic_concave_cone} summarizes the results.  The RMS error for FEP was $0.141$ unit, and the RMS error for normal was $1.58^\circ$. It shows that our method can successfully reconstruct object with concavity as long as the light rays do not re-enter the object once they exit the object.

\textbf{Second method on a thin convex cone}.  Under the same synthetic environment as described above, we rendered a thin object to evaluate our approach in Section ~\ref{sec:thin_method}. We evaluated our thin transparent object reconstruction method on a right circular cone with $\lambda = 1.7$. Its height and radius were $1$ and $4$ respectively. 

\begin{figure}[htbp]
\begin{center}
\includegraphics[width=0.7\linewidth]{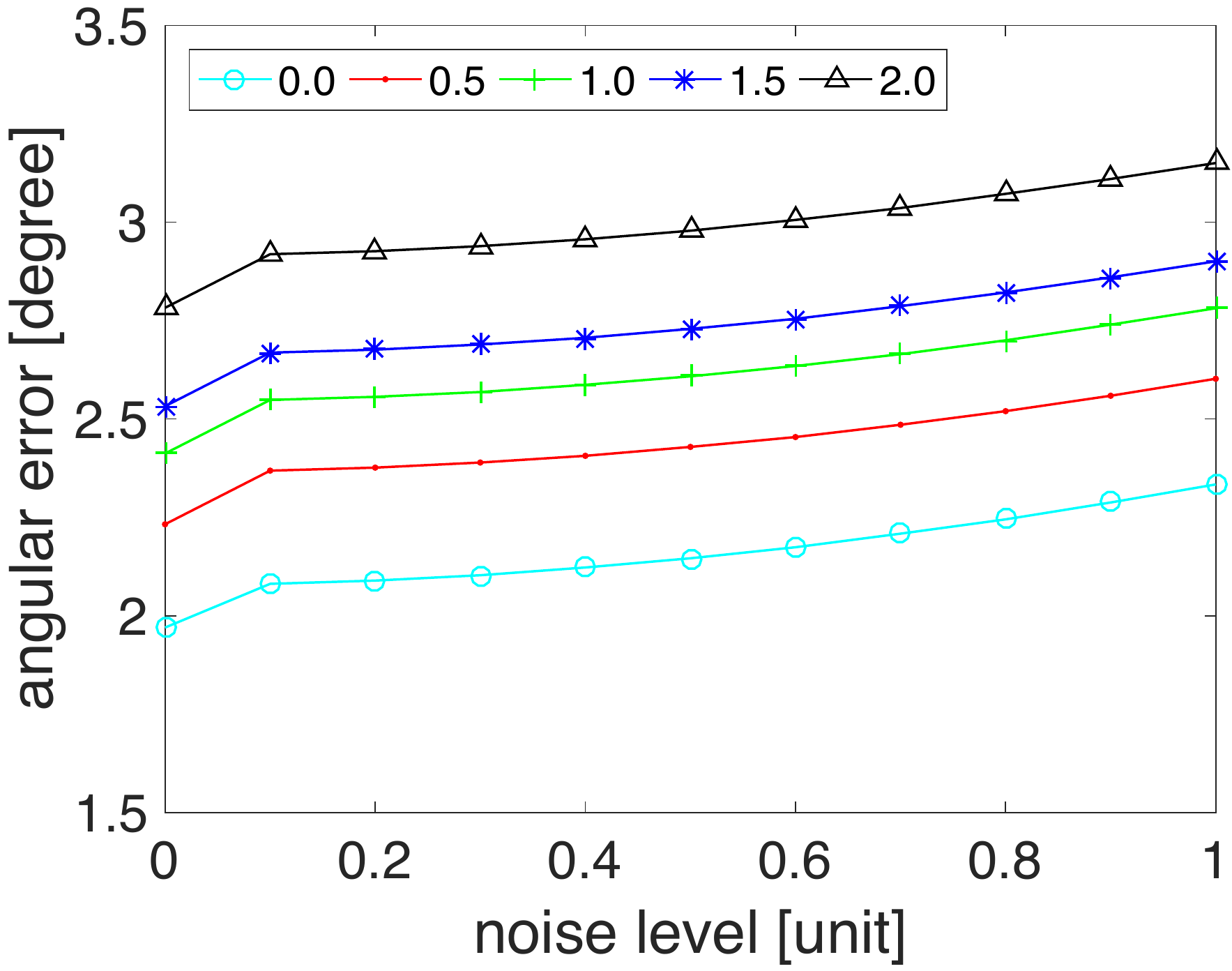}
\end{center}
   \caption{Error of surface normals recovered using the second method for a synthetic thin convex cone under different thicknesses. A cylinder was padded to the cone to change its thickness. The thickness values of the padded cylinder are shown in the legend.}
\label{fig:synthetic_thin_noise_cone}
\end{figure}

The reference plane was placed at $z = 20$ and $z = 30$ respectively. A synthetic perspective camera with a resolution of $1024 \times 1024$ was used to capture the image of the pattern directly and through the thin object, respectively. 
After we established four correspondences for each pixel, a PBC and a visual ray were reconstructed accordingly. The surface normals were then recovered using our method described in Section~\ref{sec:thin_method}. Other than the correspondence quality, the distance of the two positions of the pattern and the refractive index as discussed above, the accuracy of our thin transparent object reconstruction method will also be affected by the thickness of the objects. Hence we carried out a joint analysis by adding $2D$ zero-mean Gaussian noise to the extracted dense correspondences on the reference pattern together with varying thickness of the object. The noise level ranged from 0.1 to 1.0 units. The varying of the thickness was achieved by padding the cone with a cylinder of the same radius and setting the height of the cylinder to $h = 0.0, 0.5, 1.0, 2.0$, respectively. Fig.~\ref{fig:synthetic_thin}(a) shows the reconstruction result for $h = 0$, i.e., the cone without padding the cylinder. The errors induced by the single refraction approximation were small ($< 2^\circ$). The joint analysis results are presented in Fig.~\ref{fig:synthetic_thin_noise_cone}.  For a fixed thickness, with an increase of noise level, the RMS error of the estimated surface normals does not change a lot. This demonstrates the robustness of our approach. It can also be seen that the errors decrease with thickness of the object.

\textbf{Second method on a spherical shell}. 
We constructed a thin transparent spherical shell with $\lambda=1.7$ by subtracting a solid transparent sphere defined by (\ref{eq:thin_obj_a}) from another solid transparent sphere defined by (\ref{eq:thin_obj_b}).

\begin{equation}
 (\frac{x}{10})^2 + (\frac{y}{10})^2 + (\frac{z - s}{10})^2 = 1, s \in\{1,2,3,4,5\}\\
\label{eq:thin_obj_a}
\end{equation}

\begin{equation}
 (\frac{x}{10})^2 + (\frac{y}{10})^2 + (\frac{z}{10})^2 = 1
\label{eq:thin_obj_b}
\end{equation}

\begin{figure}[htbp]
\begin{center}
\includegraphics[width=0.7\linewidth]{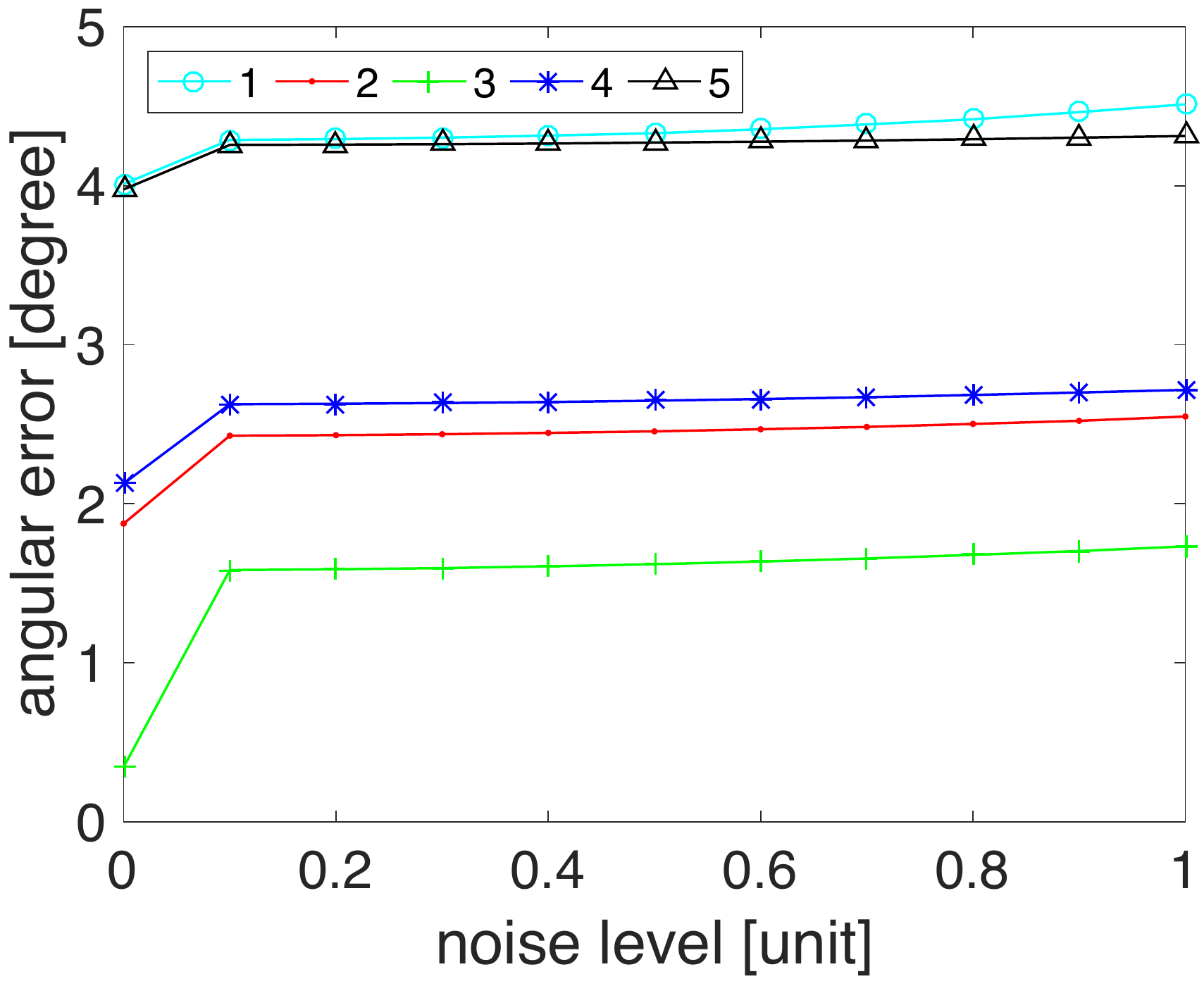}
\end{center}
   \caption{Error of surface normals recovered using the second method for a synthetic thin spherical shell under different thicknesses. The thickness values are shown in the legend, which are the distances between the two sphere centers defined by $s$ in (\ref{eq:thin_obj_a}).}
\label{fig:synthetic_thin_noise}
\end{figure}

Similar joint analysis as before was also conducted for this object. The varying of the thickness was achieved by setting different $s$ value in (\ref{eq:thin_obj_a}), which specifies the distance between the two sphere centers. Fig.~\ref{fig:synthetic_thin}(b) shows the reconstruction result for $s = 3$. Fig.~\ref{fig:synthetic_thin_noise} depicts the results of the joint analysis. For this object, the error does not keep decreasing with its thickness. The error decreases with the thickness at first, but then it starts to increase after some particular thickness (e.g., 3 units in our experiment). For the spherical shell, with the decrease of its thickness, the difference between the normals at the upper and lower surface points will also decrease. When the object gets too thin, the normals at the upper and lower surface points along the each light path tend to become parallel. In this case, the single refraction assumption is no longer applicable.

\begin{figure*}[htbp]
\centering
\tabcolsep=0.1cm
\begin{tabular}{
>{\centering\arraybackslash} m{0.3\textwidth}
>{\centering\arraybackslash} m{0.215\textwidth}
>{\centering\arraybackslash} m{0.22\textwidth}
>{\centering\arraybackslash} m{0.17\textwidth}}
\includegraphics[width=\linewidth]{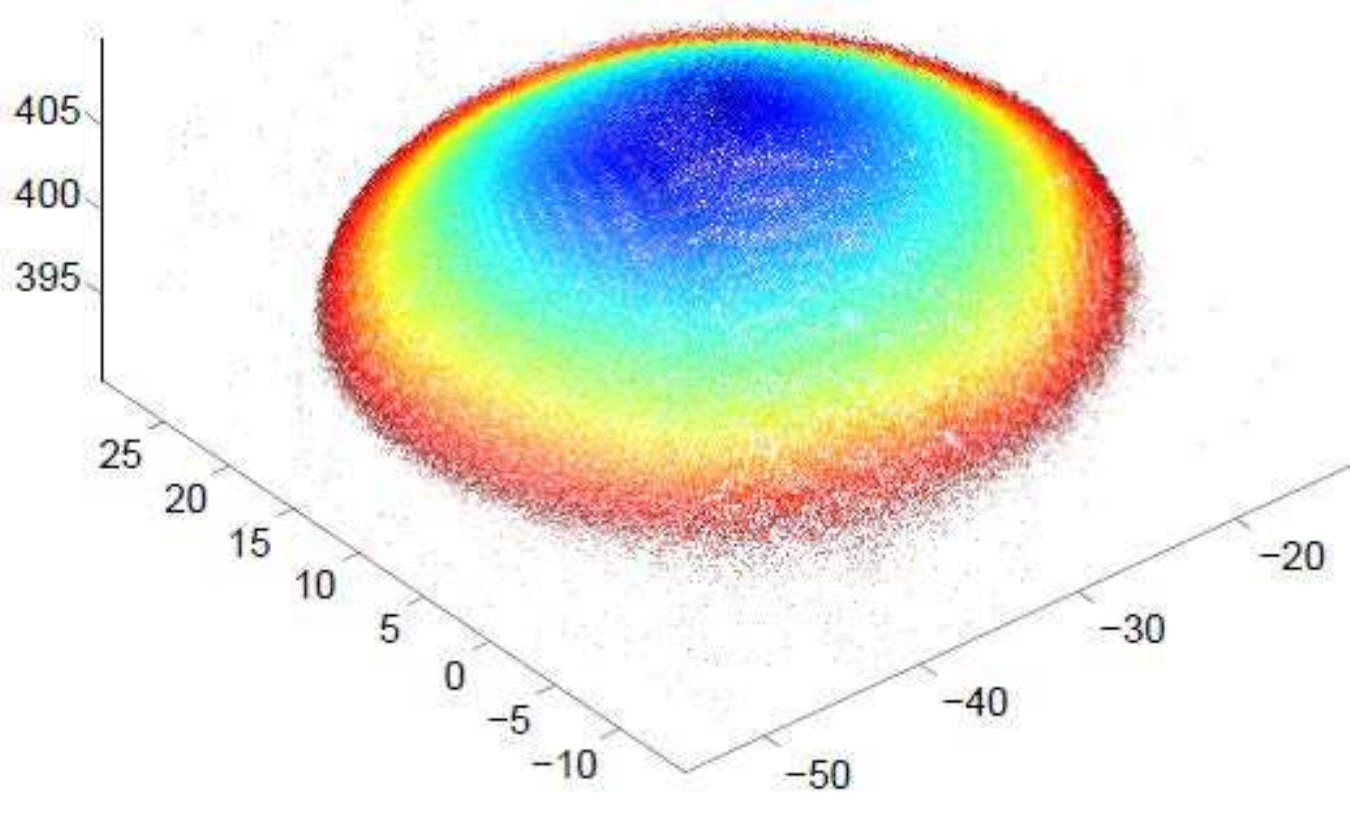} &
\includegraphics[width=\linewidth]{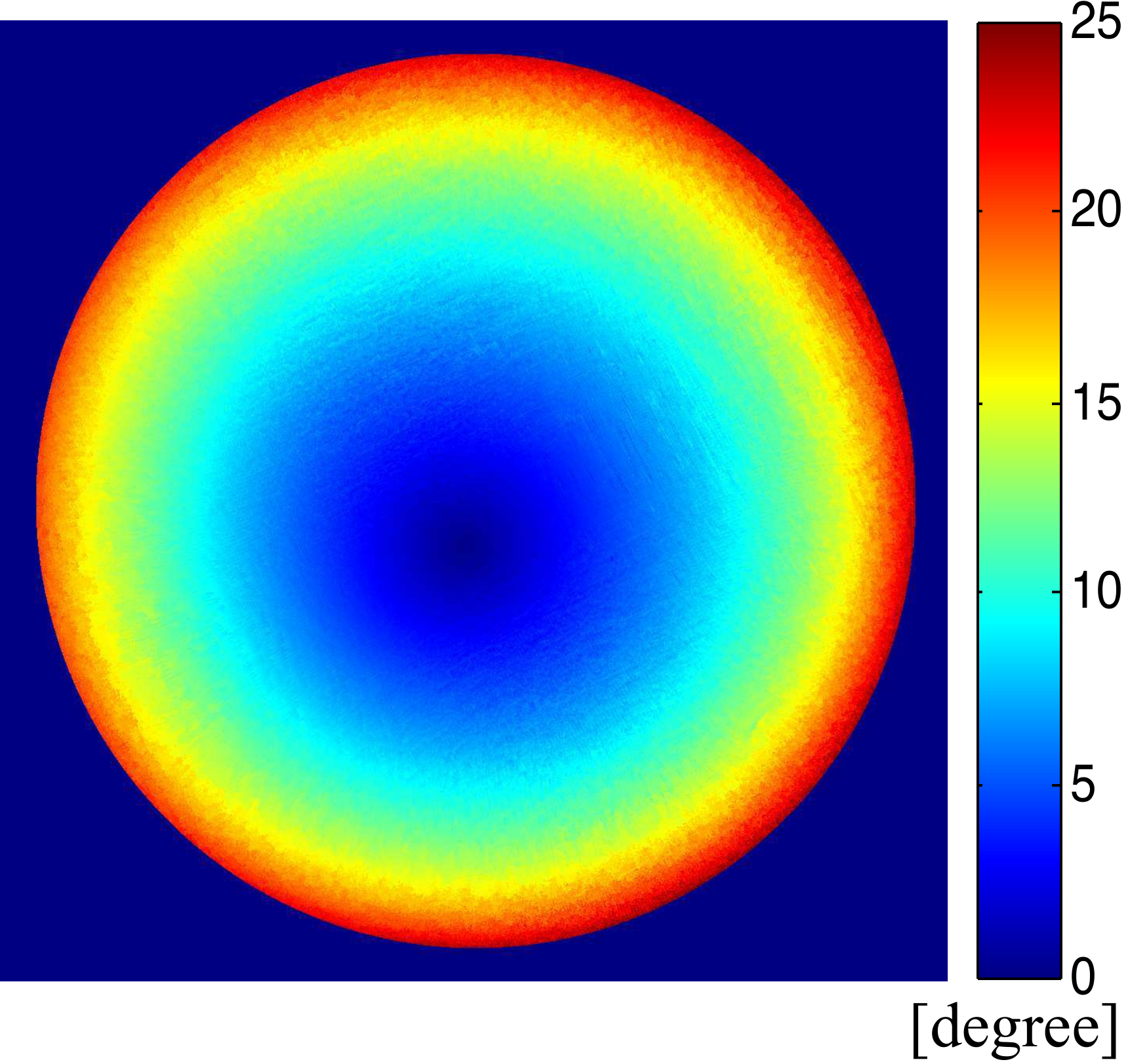}&
\includegraphics[width=\linewidth]{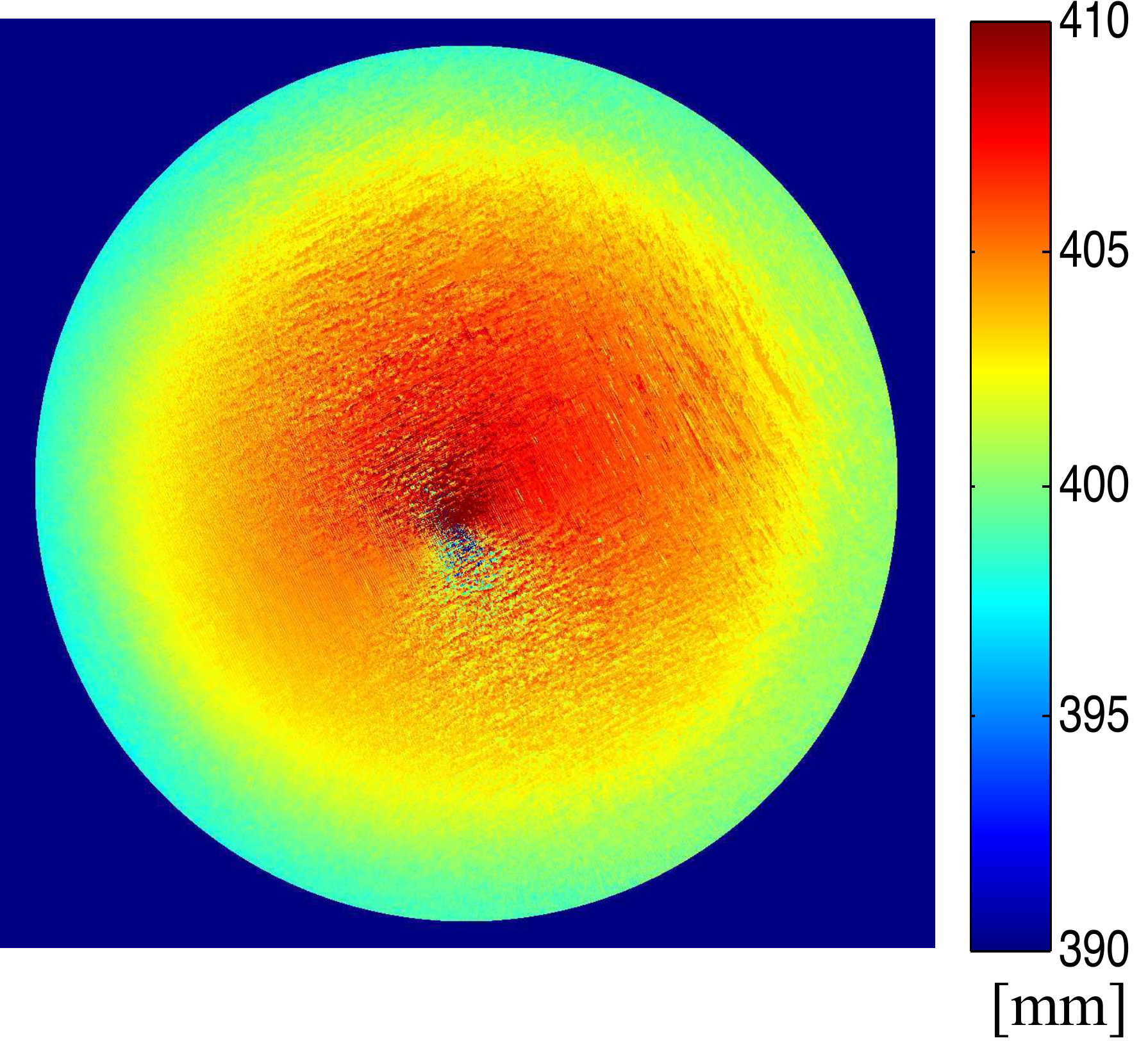}&
\includegraphics[width=\linewidth]{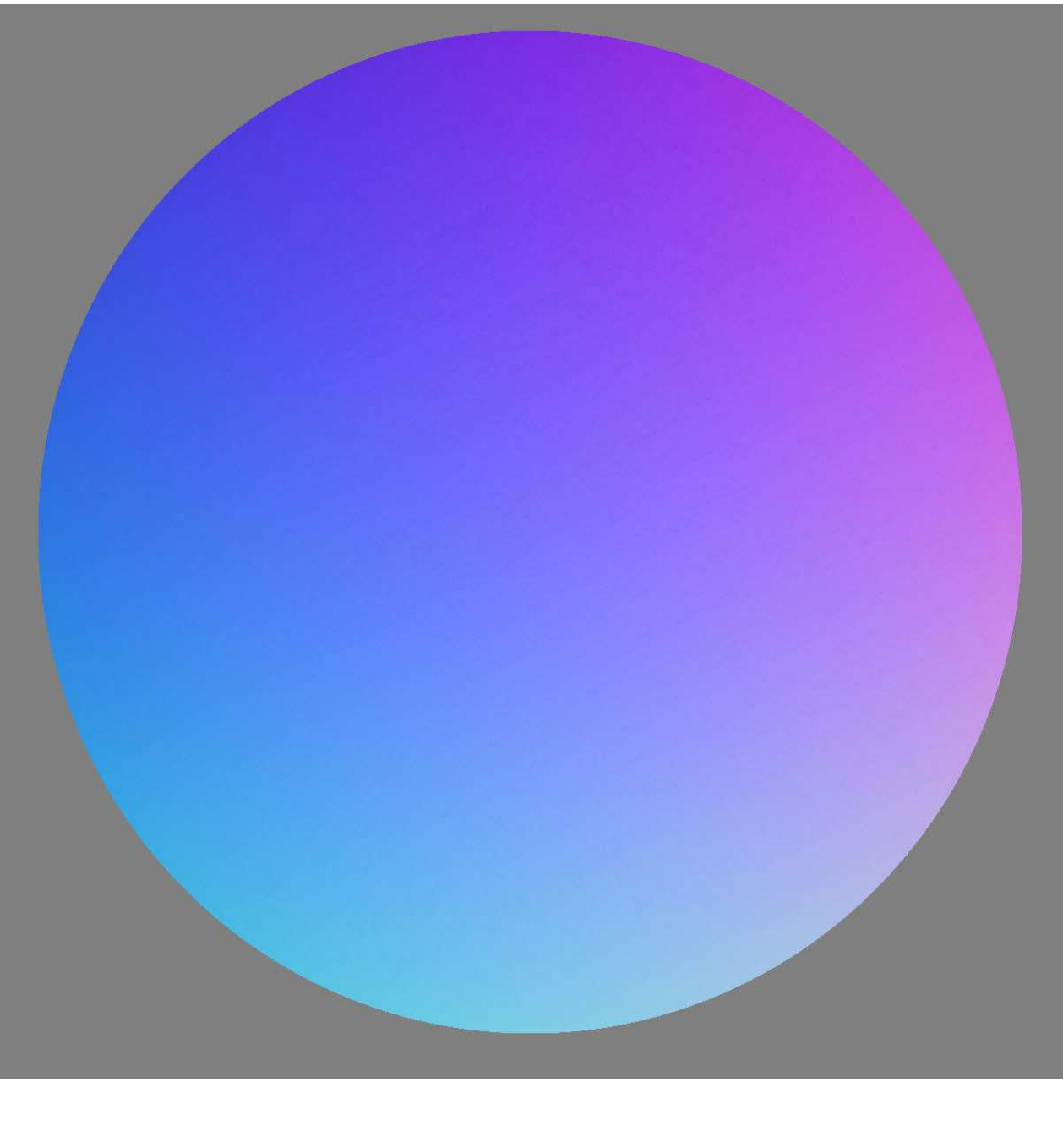}\\
\includegraphics[width=\linewidth]{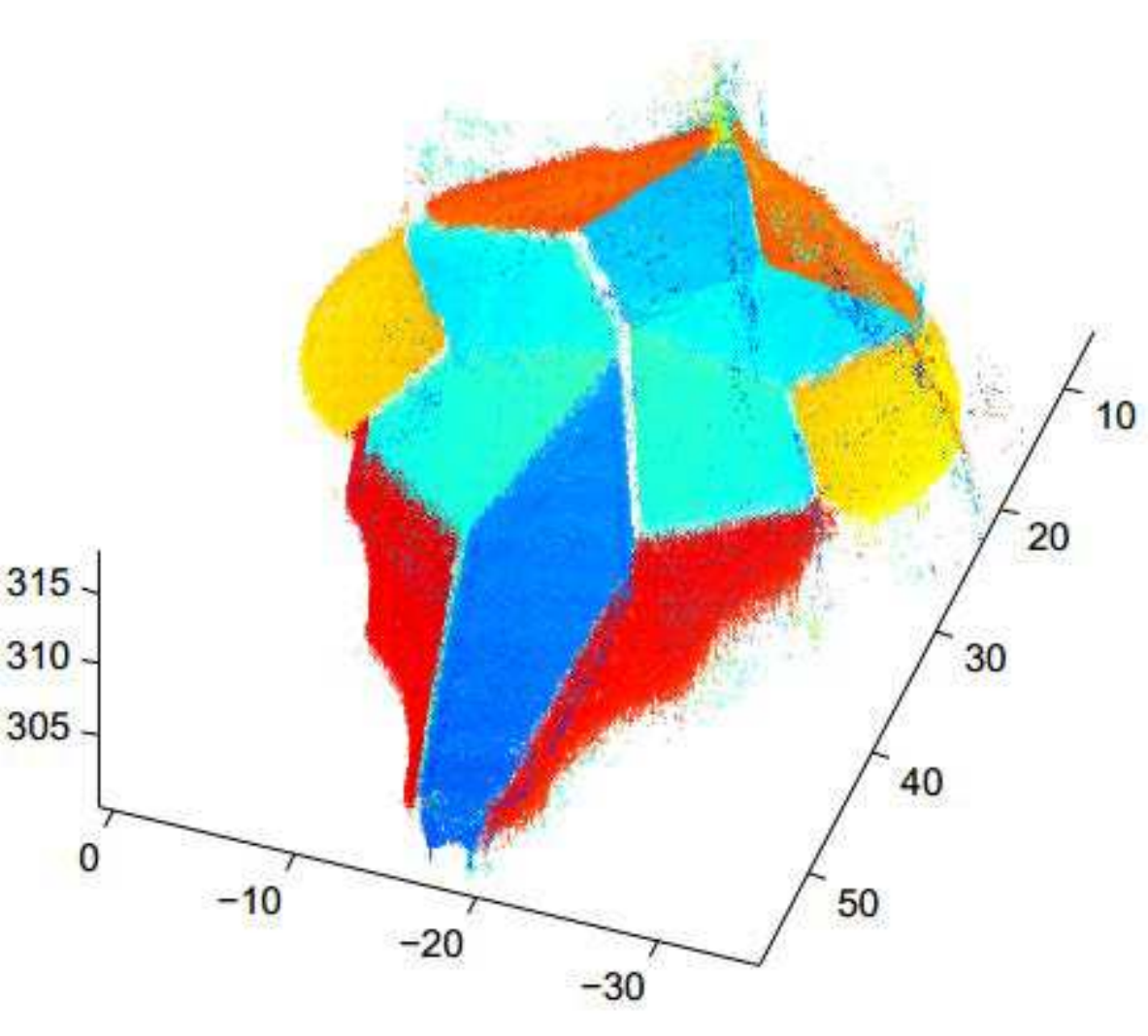} &
\includegraphics[width=\linewidth]{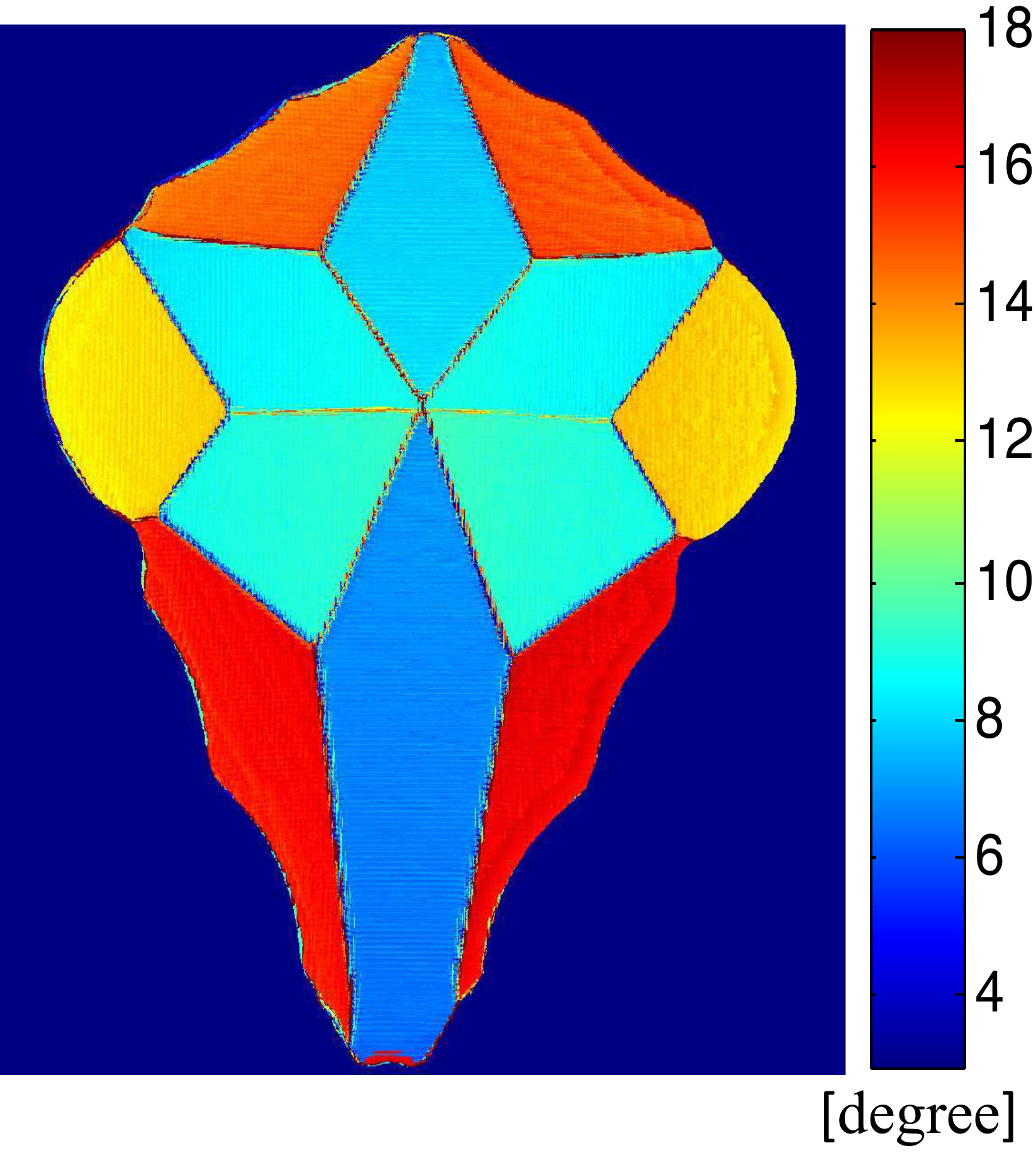} &
\includegraphics[width=\linewidth]{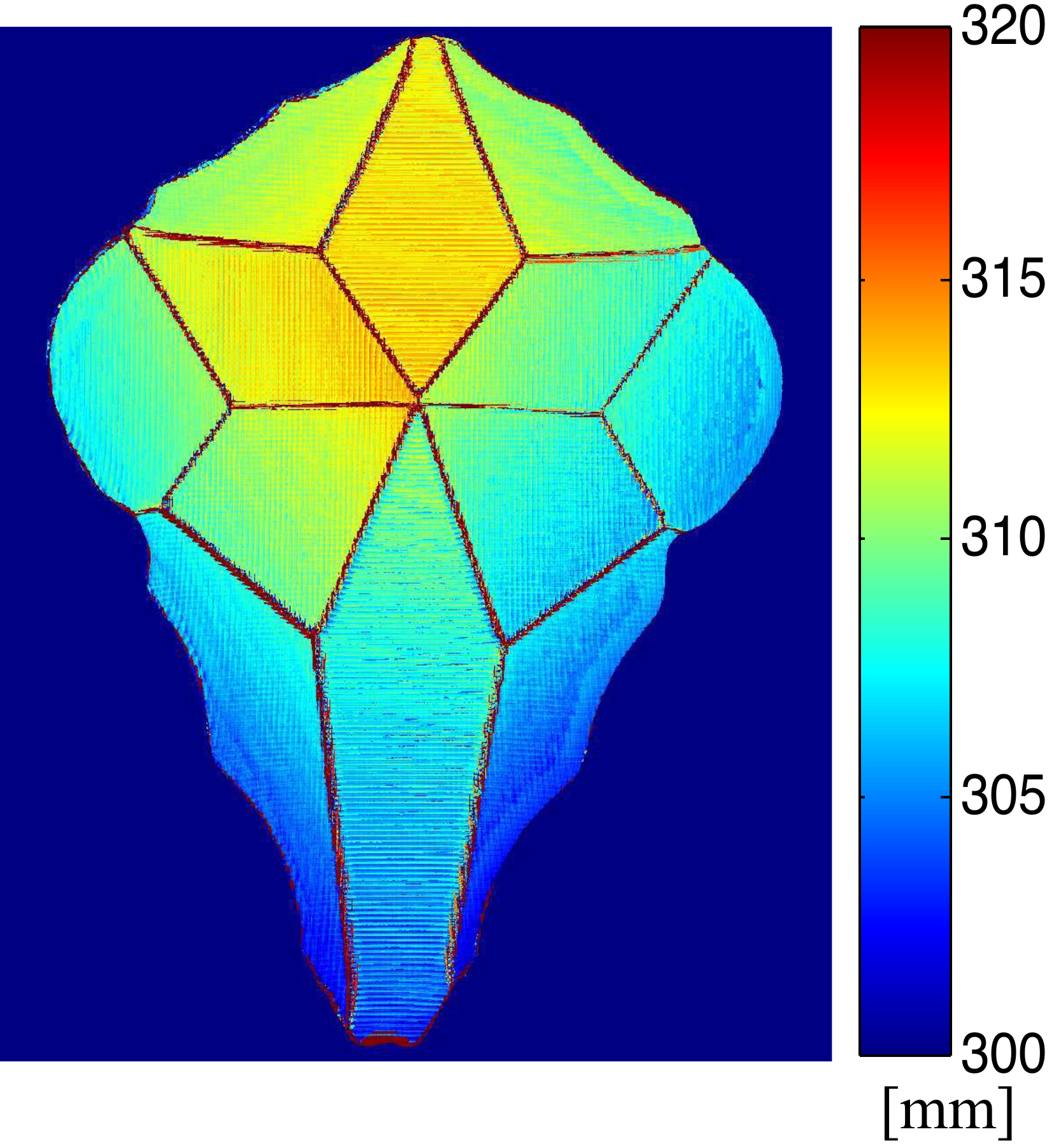} &
\includegraphics[width=\linewidth]{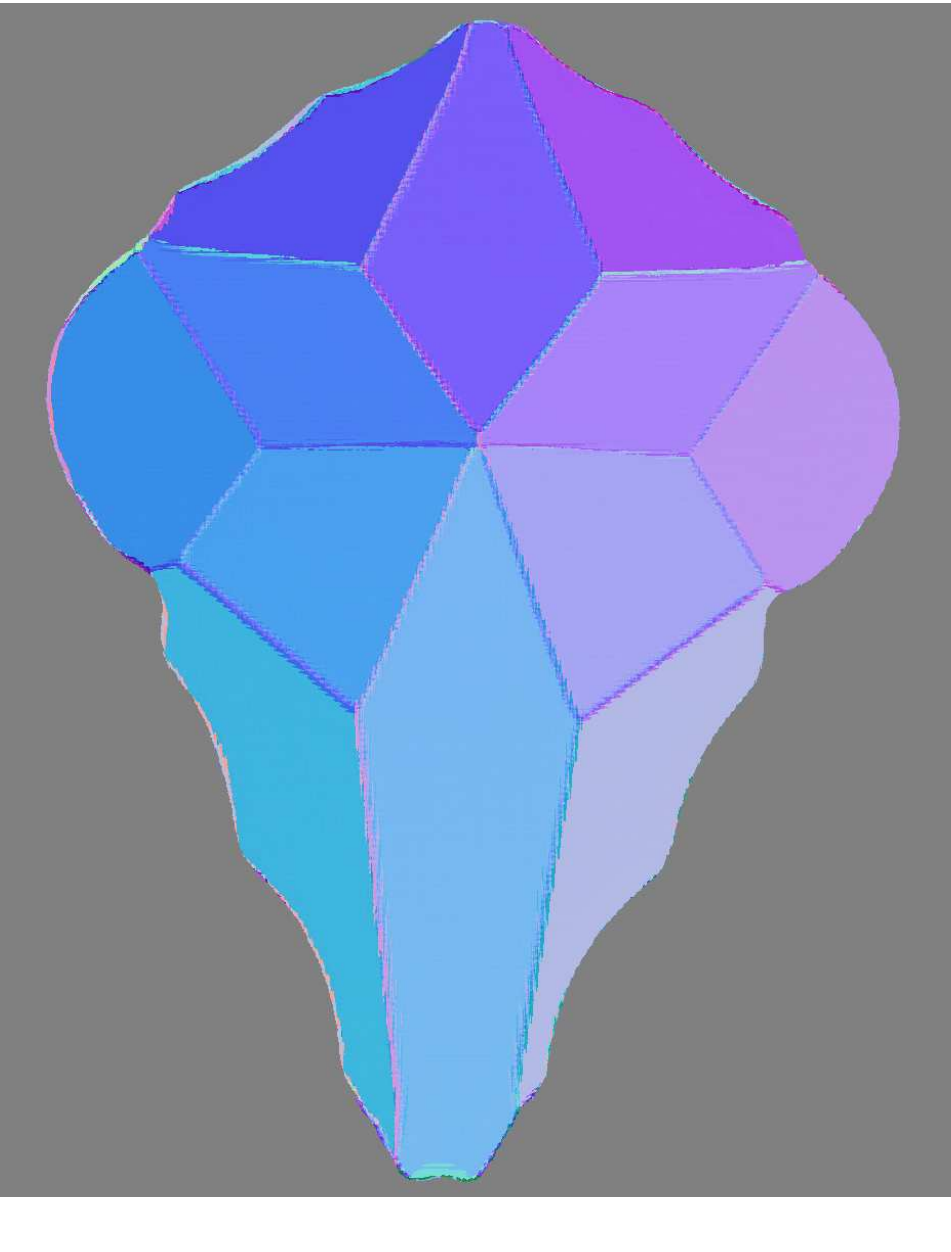}\\

\end{tabular}
\caption{Reconstruction results of the first method on real data. First row: $hemisphere$ reconstruction results. Second row: $ornament$ reconstruction results. The first column shows the reconstructed FEPs; the second column shows the angle between each PBC pairs; the third column shows the depth map; the fourth column shows the reconstructed normal map.} 
\label{fig:real_experiments}
\end{figure*}

\begin{figure*}[htbp]
\centering
\tabcolsep=0.3cm
\begin{tabular}{
>{\centering\arraybackslash} m{0.19\textwidth}
>{\centering\arraybackslash} m{0.25\textwidth}
>{\centering\arraybackslash} m{0.19\textwidth}
>{\centering\arraybackslash} m{0.18\textwidth}}
\includegraphics[width=\linewidth]{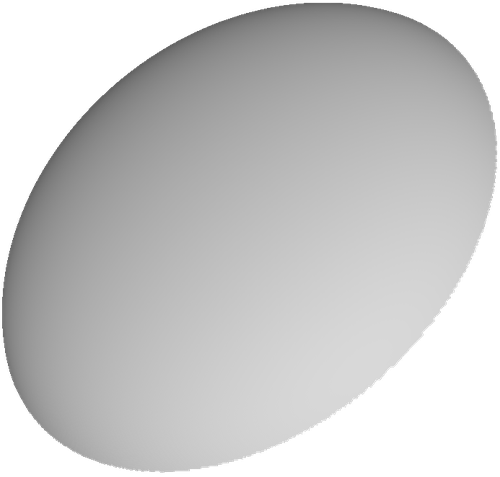}&
\includegraphics[width=\linewidth]{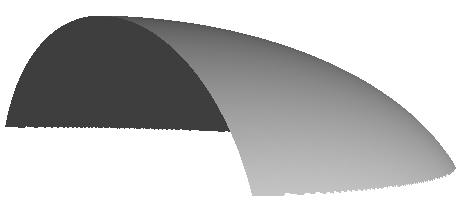}&
\includegraphics[width=\linewidth]{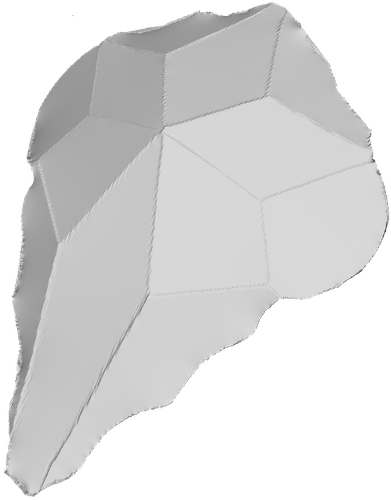}&
\includegraphics[width=\linewidth]{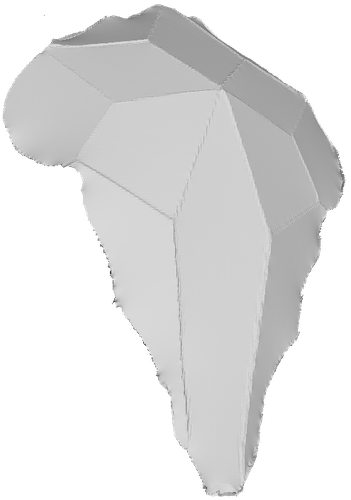}\\
\end{tabular}
\caption{Two views of the reconstructed surface using the first method. First two columns: $hemisphere$ reconstruction results (the second view is a cross-section view). Last two columns: $ornament$ reconstruction results. Note the shown surfaces are the surfaces touching the water in the experiment.}
\label{fig:real_experiments_mesh}
\end{figure*}

\subsection{Real Data}
\begin{figure*}[htbp]
\centering
\tabcolsep=0.1cm
\begin{tabular}{
>{\centering\arraybackslash} m{0.15\textwidth}
>{\centering\arraybackslash} m{0.26\textwidth}
>{\centering\arraybackslash} m{0.18\textwidth}
>{\centering\arraybackslash} m{0.19\textwidth}
>{\centering\arraybackslash} m{0.125\textwidth}}
\includegraphics[width=\linewidth]{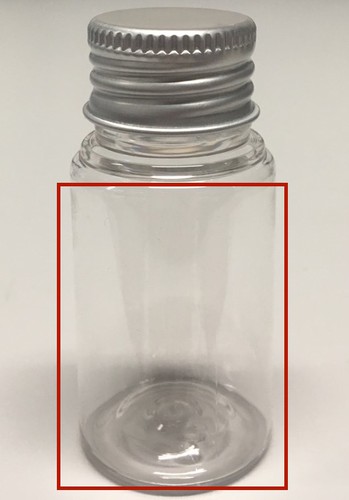}&
\includegraphics[width=\linewidth]{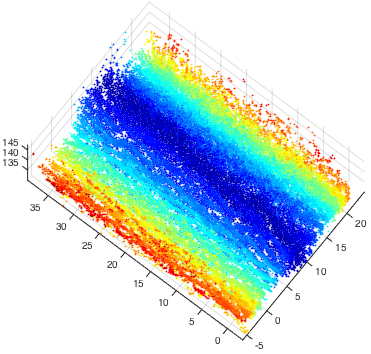}&
\includegraphics[width=\linewidth]{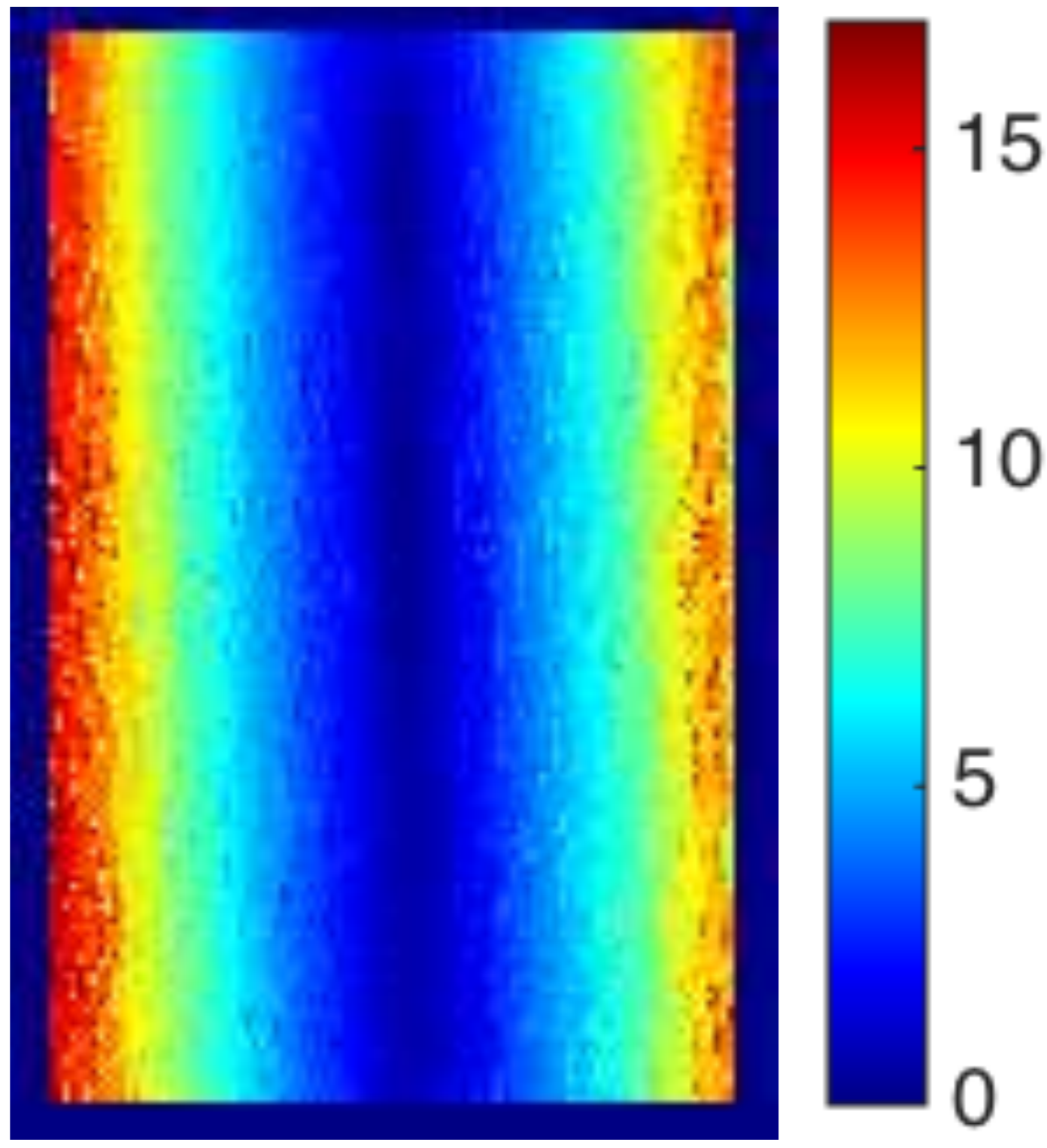}&
\includegraphics[width=\linewidth]{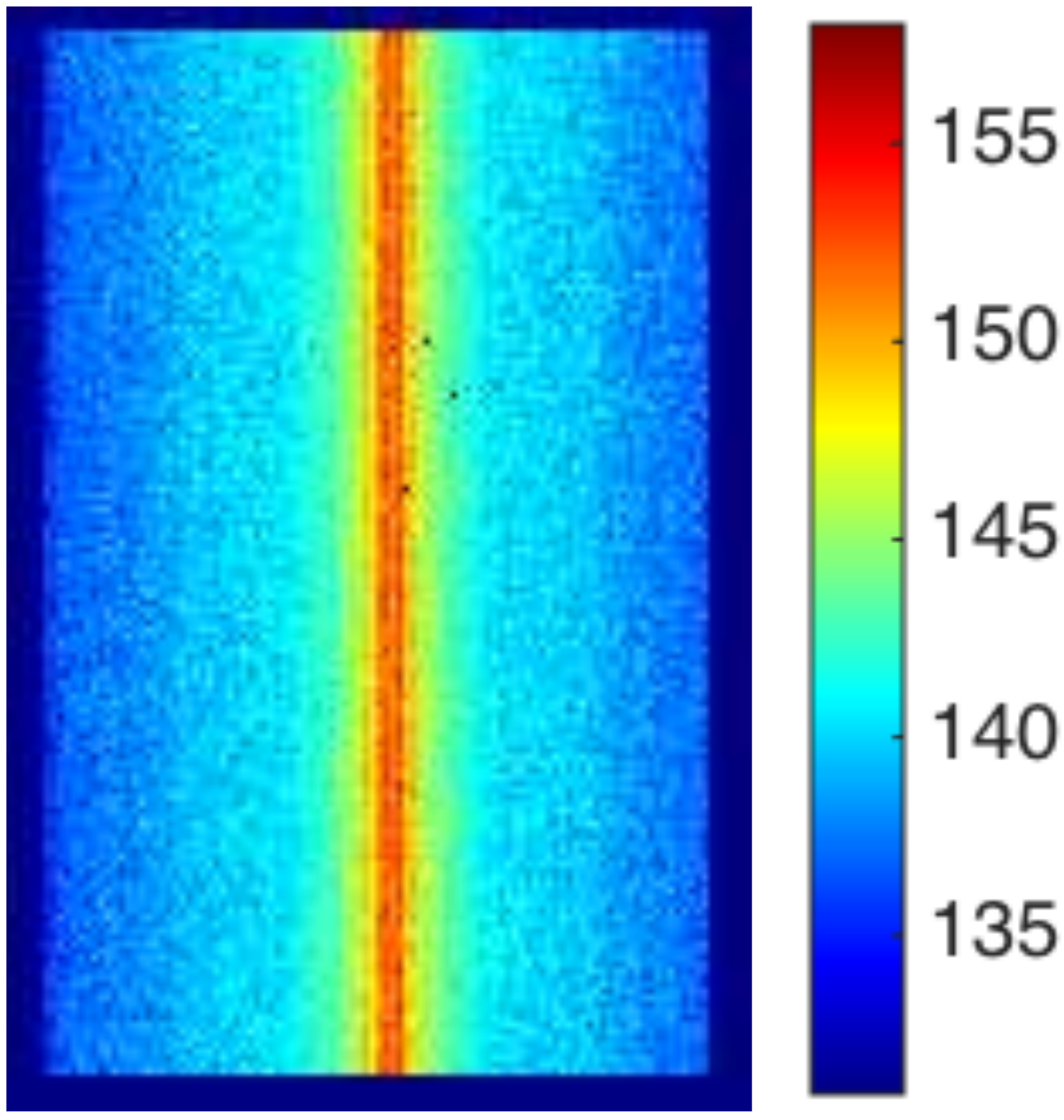}&
\includegraphics[width=\linewidth]{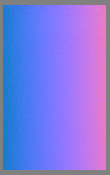}
\end{tabular}
\caption{Reconstruction result for $bottle$ using the first method. Left to right: real object (red box highlights the region for reconstruction); reconstructed FEPs; angle between the PBCs in a pair; estimated depth map; reconstructed normal map.} 
\label{fig:real_bottle}
\end{figure*}

To evaluate the accuracy of our \textbf{first method on real data}, we performed experiments on a smooth glass \emph{hemisphere}, a diamond-shape  \emph{ornament} with piecewise planar surfaces (see Fig.~\ref{fig:setup}), and a small \emph{bottle} (see Fig.~\ref{fig:real_bottle}). We acquired images with a $Canon$ $EOS$ $40D$ camera equipped with a $24$ $mm$ lens and used a $9.7$-$inch$ iPad with a resolution of $2048\times1536$ as the  reference plane. We displayed stripe patterns on the iPad for extracting the dense refraction correspondences using the strategy in Section~\ref{sec:correspondences}. In order to reconstruct PBCs, the reference plane was placed at two different positions in a water tank. Under each position, we first took one set of images of the sweeping stripe patterns refracted by the object directly. We then filled the tank with water, having a refractive index $\lambda=1.33$,  to alter the PBCs and took another set of images. In brief, four sets of images with a resolution of $3888\times2592$ were captured for each object. This yielded dense correspondences (see Table~\ref{tab:statistic}). The poses of the reference plane relative to the camera were calibrated with Matlab Calibration Toolbox~\cite{MatlabCalibrationToolBox}.

\begin{table}
\renewcommand{\arraystretch}{1.3}
\caption{Statistics for our real experiment with the first method. We show the number of captured encoding pattern images, refraction correspondences, reconstructed FEPs, and reconstructed normals for our dense reconstructions of $hemisphere$, $ornament$ and $bottle$, respectively.}
\label{tab:statistic}
\begin{center}
\begin{tabular}{l c c c}
\hline
  & $hemisphere$ & $ornament$ & $bottle$\\
\hline
Images & 2,800 & 2,200  & 2,100\\
Corres. & 1,180,300 & 546,173 & 483,052\\
FEPs & 1,115,748 & 519,162 & 471,874\\
Normals & 1,115,748 & 519,162 & 471,874\\
\hline
\end{tabular}
\end{center}
\end{table}

\begin{table}
\renewcommand{\arraystretch}{1.3}
\caption{Reconstruction errors of $hemisphere$ using the first method. The position error is defined as the difference between the distance from the fitted center to each FEP and the length of fitted radius. The normal error is defined as the angle between the ray from the fitted center to each FEP and the reconstructed normal for each FEP.}
\label{tab:hemi}
\begin{center}
\begin{tabular}{l c c}
\hline
  & Position (mm) & Normal (degree) \\
\hline
Mean error& 0.5903 & 6.9665\\
Median error& 0.4179 & 6.9215 \\
\hline
\end{tabular}
\end{center}
\end{table}

\begin{table*}[htbp]
\renewcommand{\arraystretch}{1.3}
\caption{Reconstruction errors of $ornament$ using the first method. Left figure: shows the labels for each facet of the ornament. Right table: shows the various error metric used in the evaluation for $ornament$. Due to its piecewise property, we fitted each facet using RANSAC with an inlier threshold of $0.5 mm$ and then measured the distances from the FEPs to the fitted plane and also the angle differences between the reconstructed normals of each facet region and the fitted facet normal.}
\label{tab:diamond}
\begin{center}
\begin{tabular}{>{\centering\arraybackslash} m{0.125\textwidth} l c c c c c c c c}
\cline{2-8}
\multirow{5}{*}{\includegraphics[width=\linewidth]{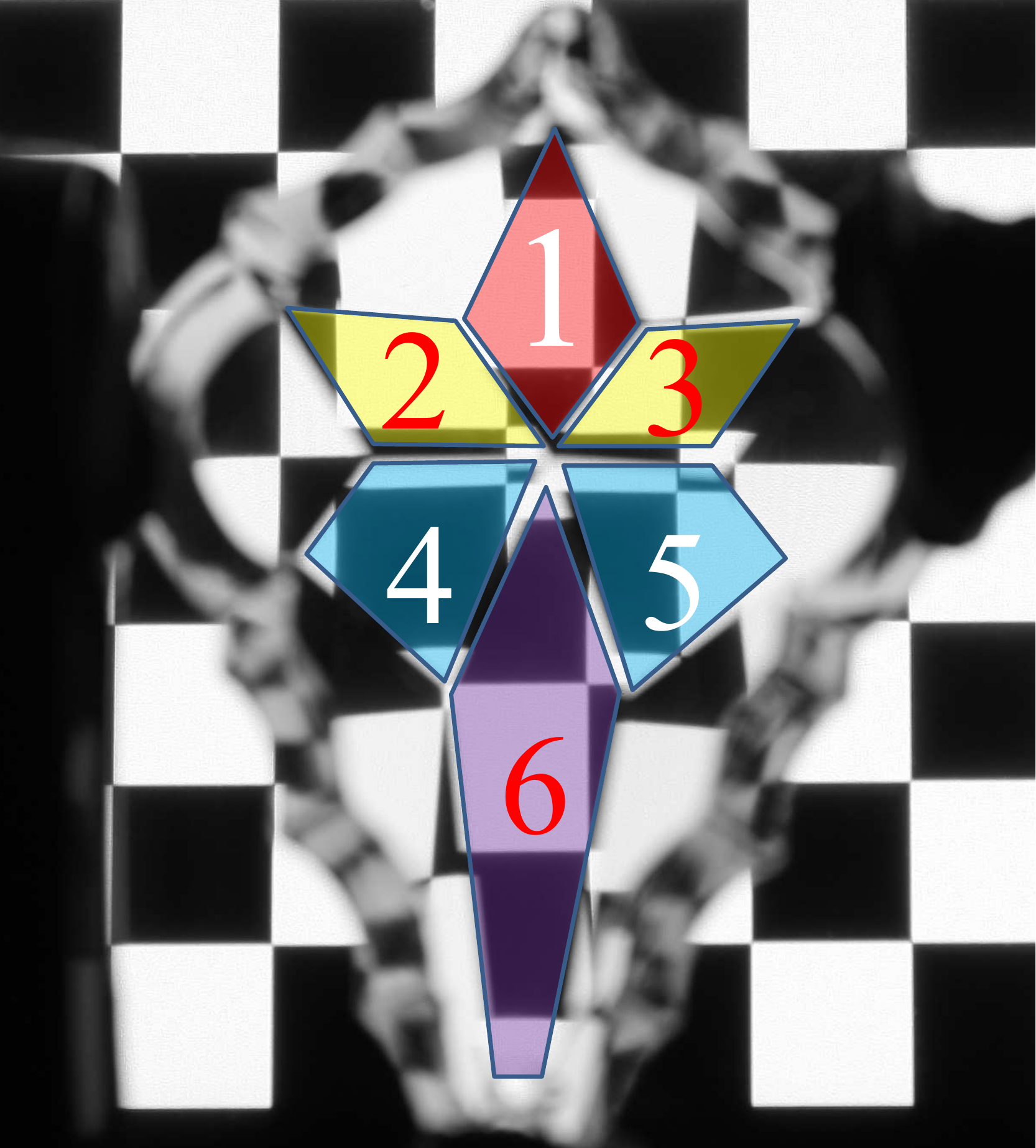}}&Facet label & 1 & 2 & 3 & 4 & 5 & 6\\
\cline{2-8}
&Mean normal error (degree) & 6.2654 & 9.9585 & 7.5905 & 9.6871 & 3.6591 & 6.8677\\
&Median normal error (degree) & 6.1677 & 9.5906 & 7.5109 & 9.7511 & 3.4741 & 6.7908\\
&Mean position error (mm) & 0.7250 & 0.6814 & 0.6675 & 0.6767 & 0.5881 & 1.0442\\
&Median position error (mm) & 0.6108 & 0.5945 & 0.5755 & 0.5721 & 0.5133 & 0.6333\\
&RANSAC position inliers (\%) & 40.33 & 42.97 & 44.06 & 43.38 & 49.02 & 40.64\\
\cline{2-8}
\end{tabular}
\end{center}
\end{table*}

\begin{figure}[htbp]
\centering
\tabcolsep=0.3cm
\begin{tabular}{
>{\centering\arraybackslash} m{0.18\textwidth}
>{\centering\arraybackslash} m{0.18\textwidth}}
\includegraphics[width=\linewidth]{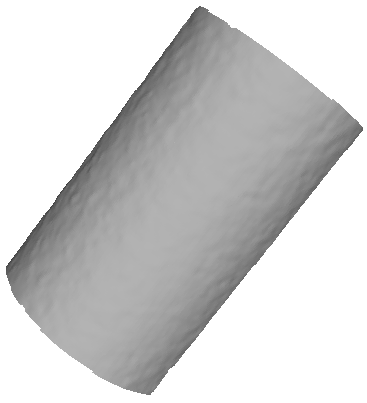}&
\includegraphics[width=\linewidth]{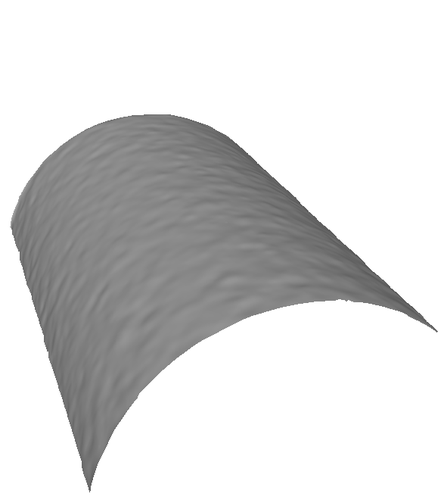}\\
\end{tabular}
\caption{Two views of the  \emph{bottle} reconstructed using the first method.} 
\label{fig:real_bottle_mesh}
\end{figure}

\begin{table}
\renewcommand{\arraystretch}{1.3}
\caption{Reconstruction errors of $bottle$ using the first method. The position error is defined as the difference between the distance from the fitted cylinder axis to each FEP and the length of fitted radius. The normal error is defined as the angle between the normal computed analytically from the fitted cylinder and the reconstructed one for each FEP.}
\label{tab:bottle}
\begin{center}
\begin{tabular}{l c c}
\hline
  & Position (mm) & Normal (degree) \\
\hline
Mean error& 0.6356 & 6.1256\\
Median error& 0.6278 & 5.9183 \\
\hline
\end{tabular}
\end{center}
\end{table}

A pair of PBCs were reconstructed from the extracted refraction correspondences for each 
image point. These PBCs were triangulated to give an estimate of the FEP.
We treated those reconstructed FEPs with a small PBC angle ($< 1^\circ$),  or out of the depth range between the camera and reference plane as noise points. The normal for each FEP was then recovered with the knowledge of refractive indices $1.0$ and $1.33$ for air and water respectively. In Fig.~\ref{fig:real_experiments}, we show our reconstructed $3D$ FEP cloud, angles between the PBCs in a pair, depth map, and surface normals for $hemisphere$ and $ornament$ respectively. Note that large reconstruction errors occur in regions with small PBC angles. We also employed the integration method proposed by Xie {\em et al.} in \cite{xie2014cvpr} to generate surface meshes with our recovered normals. These meshes are shown in Fig.~\ref{fig:real_experiments_mesh}.

Since no ground truth was available, a sphere was fitted to the FEP cloud to evaluate the reconstruction accuracy for the $hemisphere$. We compared the fitted radius with the physical measurement, which were $26.95$ $mm$ and $27.99$ $mm$, respectively. The error was $1.04$ $mm$ (i.e., $3.7\%$ compared with the measurement). Table~\ref{tab:hemi} shows the reconstruction errors of the $hemisphere$ compared against the fitted sphere. The mean and median FEP position errors were $< 0.6$ $mm$, and the mean and median normal errors were $< 7.0^{\circ}$. This shows a high accuracy of the reconstruction. In order to evaluate the reconstruction of the $ornament$, we first used RANSAC~\cite{Fishchler1981} to fit a plane for each facet. The reconstruction error for each facet was measured by the distances from the reconstructed FEPs to the fitted plane, as well as the angles between the reconstructed normals and the normal of the fitted plane. The results shown in Table~\ref{tab:diamond} suggest that our proposed approach can accurately reconstruct the piecewise planar $ornament$. The mean and median FEP position errors were $< 1.0$ $mm$ \footnote{Except facet 6 with a mean of $1.0442$.} and the mean and median normal errors were $< 10.0^{\circ}$.

\begin{figure}[tbp]
\begin{center}
\includegraphics[width=0.65\linewidth]{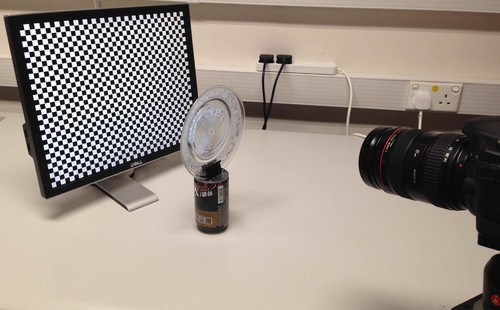}
\end{center}
   \caption{Real reconstruction setup for thin transparent objects.}
\label{fig:real_thin_setup}
\end{figure}

\begin{figure*}[htbp]
\centering
\tabcolsep=0.2cm
\begin{tabular}{
>{\centering\arraybackslash} m{0.2\textwidth}
>{\centering\arraybackslash} m{0.21\textwidth}
>{\centering\arraybackslash} m{0.245\textwidth}
>{\centering\arraybackslash} m{0.205\textwidth}}
\includegraphics[width=\linewidth]{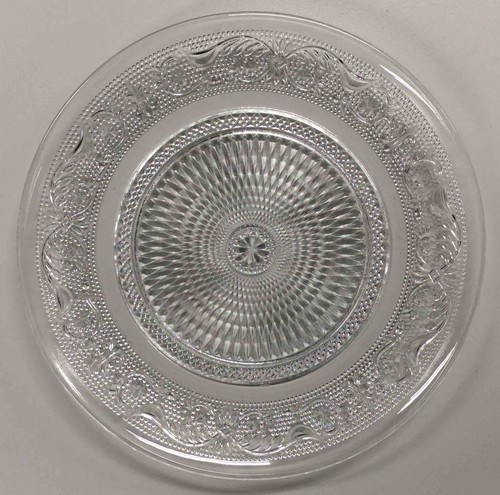} &
\includegraphics[width=\linewidth]{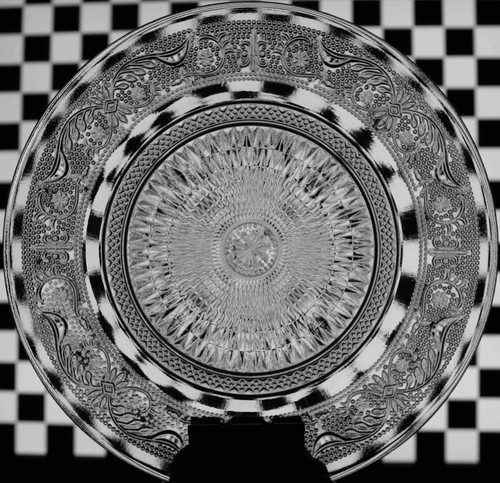} &
\includegraphics[width=\linewidth]{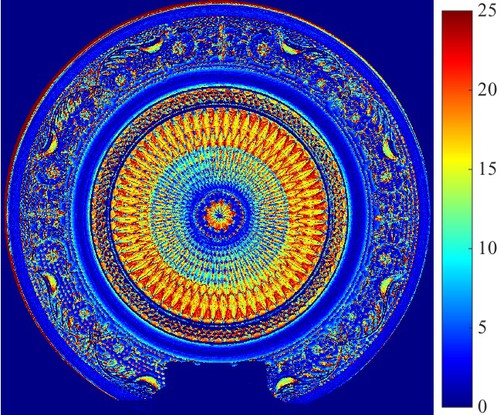}&
\includegraphics[width=\linewidth]{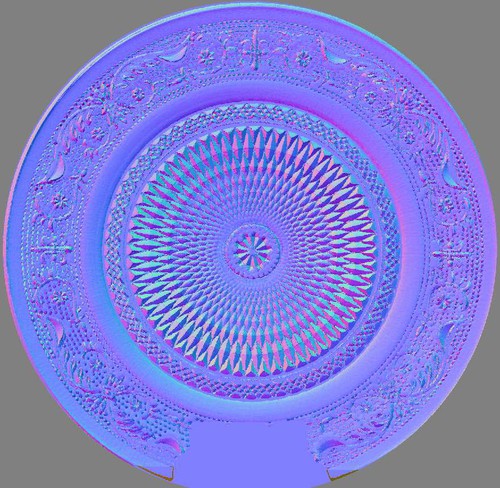}\\
\includegraphics[width=\linewidth]{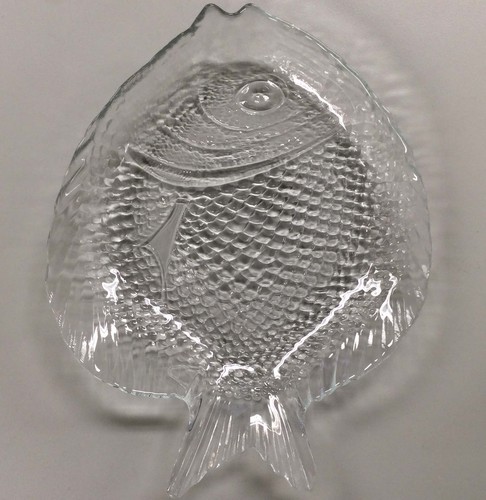} &
\includegraphics[width=\linewidth]{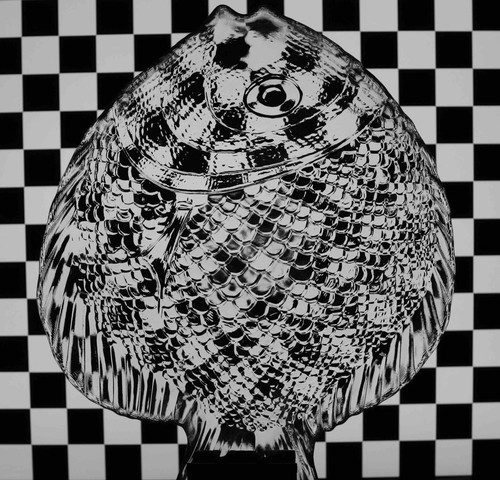} &
\includegraphics[width=\linewidth]{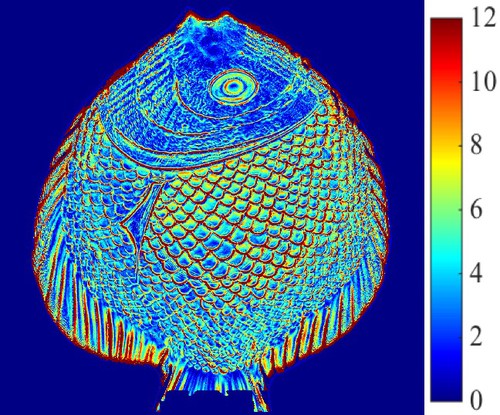}&
\includegraphics[width=\linewidth]{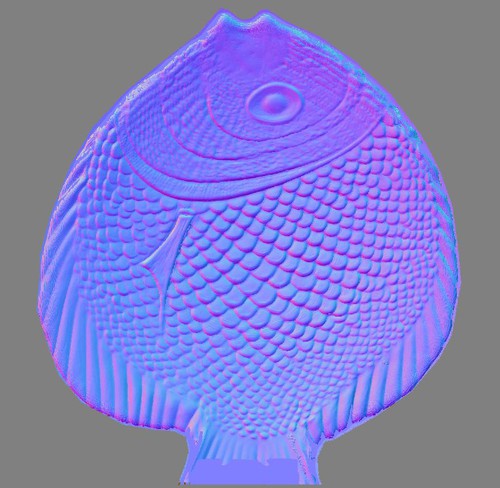}\\
\end{tabular}
\caption{Reconstruction results of the second method on real data. First row: $circular\;plate$ reconstruction results. Second row: $fish\;plate$ reconstruction results. The first column shows the real objects under room illumination; the second column shows the refraction of the pattern caused by the objects; the third column shows the angle between the visual ray and PBC in a pair; the last column shows the reconstructed normal map.} 
\label{fig:real_experiments_thin}
\end{figure*}

\begin{figure*}[htbp]
\centering
\tabcolsep=0.2cm
\begin{tabular}{
>{\centering\arraybackslash} m{0.24\textwidth}
>{\centering\arraybackslash} m{0.21\textwidth}
>{\centering\arraybackslash} m{0.22\textwidth}
>{\centering\arraybackslash} m{0.22\textwidth}}
\includegraphics[width=\linewidth]{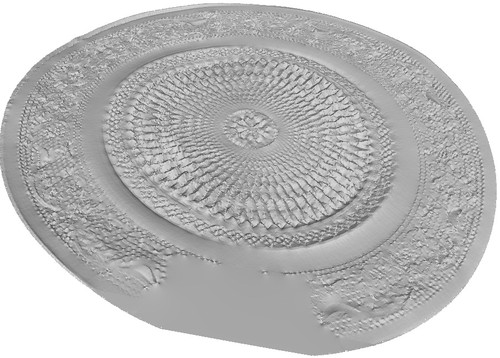}&
\includegraphics[width=\linewidth]{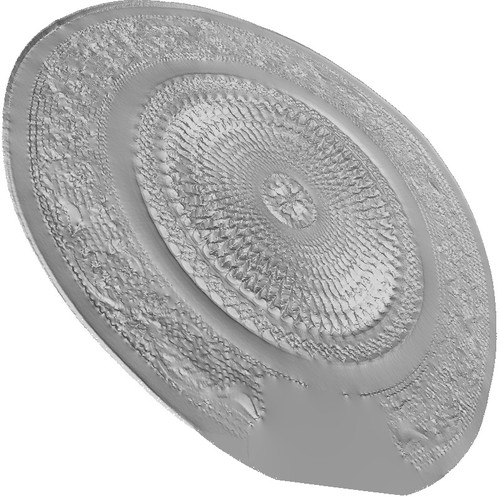}&
\includegraphics[width=\linewidth]{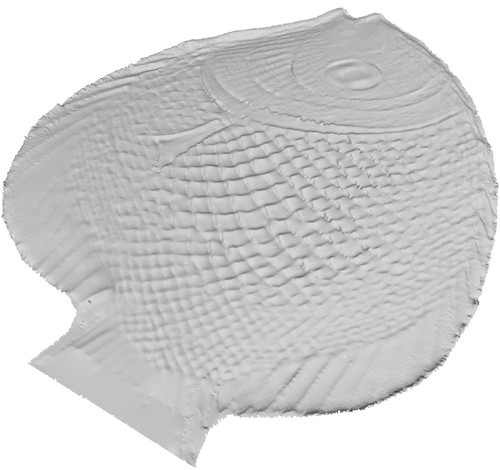}&
\includegraphics[width=\linewidth]{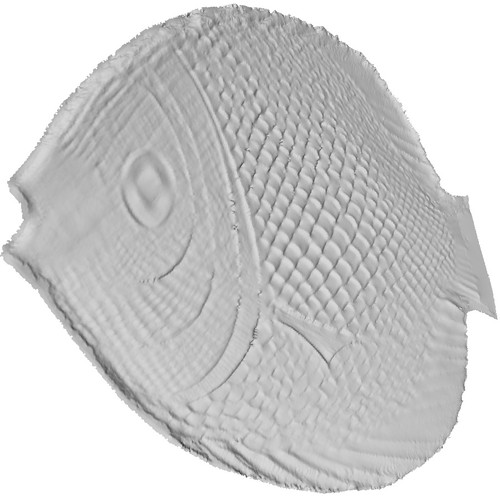}\\
\end{tabular}
\caption{Two views of the surface reconstructed using the second method. First two columns: $circular\;plate$ reconstruction results. Last two columns: $fish\;plate$ reconstruction results.} 
\label{fig:real_experiments_thin_mesh}
\end{figure*}

Besides, we also validated our approach on a hollow object by reconstructing a small transparent bottle. The reconstruction result is presented in Fig.~\ref{fig:real_bottle} and the reconstructed surface mesh is shown in Fig.~\ref{fig:real_bottle_mesh}. The measured height and radius were $55.65$ $mm$ and $14.18$ $mm$ respectively, and the body part (red box region in Fig.~\ref{fig:real_bottle}) of the bottle is $44.73$ $mm$ in height.  We fitted a cylinder to the reconstructed point cloud using MLESAC \cite{Torr2000cviu}. We set the point-to-cylinder distance threshold to $0.5$ $mm$ during fitting, and $41.17\%$ of reconstructed FEPs are inliers. The fitted height and radius were $46.26$ $mm$ and $15.32$ $mm$ respectively. Both the height error and radius error were $< 2$ $mm$. The position error and normal error are shown in Table~\ref{tab:bottle}.

To evaluate our \textbf{second method on real data}, we applied our method on two thin glass plates, namely a $circular\;plate{}$ and a $fish\;plate$ (see the first column in Fig.~\ref{fig:real_experiments_thin}). Since the objects were considered thin enough ($\approx 0.3$ $cm$), compared with the size of the objects ($circular\;plate$ : diameter = $17.5$ $cm$,  $fish\;plate$: $25.6$ $cm$ $\times$ $20.7$ $cm$) and the distance between the camera and the objects ($\approx 50$ $cm$), the light path displacements inside the objects could be ignored. Fig.~\ref{fig:real_thin_setup} shows our real setup for thin object recovery. We used a $19$-$inch$ LCD display with a resolution of $1280 \times 1024$ as the reference plane. Similar to the experiments done before, we captured four sets of images to establish refraction correspondences for each image point by arranging the display in two different positions. Differently, it was not necessary to immerse the thin object partially in a liquid any more. The simplified setup largely reduced the efforts in taking images.

We first captured an image sequence of the moving stripe on the reference plane, and then put the object in front of the camera to take another sequence. To take the third sequence, we moved the the reference plane to another position while keeping the camera and object stationary. We then removed the object and took the last sequence.  After reconstructing a PBC and visual ray for each observed surface point, the surface normals were recovered with the known refractive index of glass ($\lambda = 1.52$). Fig.~\ref{fig:real_experiments_thin} shows the recovered normal map. The angles between rays in a pair are larger for regions with more details as these regions are less planar. The recovered normal maps were consistent with the real objects. The reconstructed surfaces are shown in Fig.~\ref{fig:real_experiments_thin_mesh}, which can correctly show the details of the real objects. 

\section{Conclusions}
\label{sec:conclusion}
In this paper, we develop a fixed viewpoint approach to dense surface reconstruction of transparent objects. We introduce a simple setup that allows us to alter the incident light paths by immersing the object partially in a liquid, while keeping the rest of the light paths fixed as light rays travel through the object. This greatly simplifies the problem by making it not necessary to model the complex interactions of light inside the object, and allows the object surface to be recovered by triangulating the incident light paths. Our approach can handle transparent objects with a relatively complex structure, with an unknown and inhomogeneous refractive index. The only assumption to the objects is that the light paths should not re-enter the liquid medium once they enter the object. If the refractive index of the liquid is known a priori, our method can also recover the surface normal at each reconstructed surface point. Besides, for thin transparent objects, we show that the acquisition setup can be simplified by adopting a single refraction approximation. Experimental results demonstrate both the feasibility and robustness of our methods.

\section*{Acknowledgement}
This project is supported by a grant from the Research Grant Council of the Hong Kong (SAR), China, under the project HKU 718113E.

\bibliographystyle{spmpsci}      %
\bibliography{reference}   %

\end{document}